\documentclass[twocolumn,aps,pre,showpacs,showkey,superscriptaddress,floatfix]{revtex4-1}

\usepackage{amsmath,amssymb,amsthm}

\usepackage{subfigure}

\usepackage{graphicx}% Include figure files
\usepackage{dcolumn}% Align table columns on decimal point
\usepackage{bm}% bold math

\usepackage{enumitem}

\newcommand{\PCM}{p^{\rm{CM}}}
\newcommand{\PA}{p^{\rm{A}}}
\newcommand{\JU}{u}
\newcommand{\JD}{l}
\newcommand{\pA}{\pi^{\rm{A}}}
\newcommand{\PM}{p^{\rm{M}}}
\newcommand{\pM}{\pi^{\rm{M}}}

\begin{document}

\title{
Dynamical systems on large networks with predator-prey interactions are stable and exhibit oscillations }

\author{Andrea Marcello Mambuca}
 \email{andreamambuca@gmail.com}
\affiliation{%
 Department of Mathematics, King’s College London,
 Strand, London, WC2R 2LS, United Kingdom
}%
 \author{Chiara Cammarota}
  \email{chiara.cammarota@uniroma1.it}
 \affiliation{Dipartimento di Fisica, Sapienza Università di Roma, P.le A. Moro 5, 00185 Rome, Italy}
 \affiliation{%
 Department of Mathematics, King’s College London,
 Strand, London, WC2R 2LS, United Kingdom
}%
 \author{Izaak Neri}
  \email{izaak.neri@kcl.ac.uk}
 \affiliation{%
 Department of Mathematics, King’s College London,
 Strand, London, WC2R 2LS, United Kingdom
}%

\date{\today}

\begin{abstract}
We analyse the  stability of  linear   dynamical systems defined on sparse, random graphs with predator-prey, competitive, and mutualistic interactions. These systems are aimed at modelling the stability of fixed points in large systems defined on  complex networks, such as,  ecosystems consisting of a large number of species that interact through a food-web.
We 
develop an exact theory for the spectral distribution and the leading eigenvalue of the corresponding sparse Jacobian matrices. 
This theory reveals that the nature of local interactions have a strong influence on system's  stability.   We show that, in general,  linear dynamical systems defined on random graphs with a prescribed degree distribution of unbounded support  are unstable if they are large enough, implying a tradeoff between stability and diversity.   Remarkably,  in contrast to the generic case,   antagonistic systems   that only contain interactions of the predator-prey type  can be stable in the infinite size limit.     This qualitatively new feature for antagonistic systems is accompanied by a peculiar oscillatory behaviour of the dynamical response of the system after a perturbation, when the mean degree of the graph is small enough. Moreover, for  antagonistic systems we also find that there exist a dynamical phase transition and critical mean degree above which the response becomes non-oscillatory.
\end{abstract}

\maketitle

\section{\label{sec:Intro} Introduction:}
Complex systems  consist of a large number of components interacting through  a  network \cite{newman2010networks, dorogovtsev2013evolution}, as it is for instance the case for neural networks \cite{brunel2000dynamics, bullmore2009complex, sporns2010networks}, ecosystems and food-webs \cite{dunne2002food, bascompte2009disentangling}, financial and economic markets \cite{Onnela2003,Boginski2005,easley_kleinberg_2010,Tse2010,DebtRank2012,moran2019may}, and gene-regulatory networks \cite{guo2021exploring}. 
In all these fields it is crucial to understand how the static and dynamic properties of a complex system  depend on the properties of the underlying  network \cite{newman2010networks, dorogovtsev2013evolution}.  

In this paper, we are interested in two main aspects.  First, we focus on how the {\it linear stability} of fixed points in  complex systems depends on the network topology and  the properties of the interactions between the system constituents.  Second, we discuss how  network topology and the properties of the interactions determine the dynamical response of a large system to an external perturbation, interestingly giving rise to peculiar oscillatory patterns in some specific cases.

It has been speculated that the first question on the linear stability of fixed points in complex systems  is relevant to understand, among others, the resilience of ecosystems to external perturbations to species abundances \cite{allesina2012, mougi2012diversity, mougi2014stability, allesina2015predicting, coyte2015ecology},  the onset of chaos in random neural networks \cite{chaos, kadmon2015transition}, systemic risk in financial markets \cite{SystemicRisk,Sandhue1501495,Bardoscia2017,mambuca2018SystemicRisk, moran2019may}, and homeostasis of protein concentrations in cells \cite{guo2021exploring}.   In ecology numerical simulations and full dynamical solutions of ecosystems models show that fixed points, and their linear stability features, control the large time evolution for a large portion of the model parameters~\cite{bunin2017ecological, barbier2018generic, roy2019numerical}.   Note that there exists other approaches to address stability of complex systems.  In particular, recent studies show that in the context of ecology {\it structural stability}, linked to the existence (feasibility) of stationary points and referring to their sensitivity to changes in the ecological parameters, should be primarily looked at 
\cite{rossberg2013food, rohr2014structural, rossberg2017structural, dougoud2018feasibility, o2019metacommunity}.   
Although this paper mainly focuses on linear stability, we will show at the end of the paper that our result on system stability also apply to structural stability.

The second question on the nature of the dynamical response of a complex system to external perturbations is relevant to understand, among others, how the  patterns of brain activity that emerge in  response  to an external perturbation depend on     the structure of  cortical networks \cite{ Massimini2228, rogasch2013assessing, kadmon2015transition}, or how ecosystems'  response to external perturbations depends on the graph structure and type of interactions of the underlying food-web \cite{neubert1997alternatives, fox2000population, koons2005transient, arnoldi2018ecosystems}. 
More specifically, it has been shown that, when perturbed, the response 
of  a large complex system, such as a cortical network can be oscillatory \cite{Massimini2228, rogasch2013assessing}. Evidences of oscillations in population abundances of ecological systems (emerging either as stationary phenomenon or in relaxational dynamics after perturbation) and studies of their underlying mechanisms have a long history \cite{ volterra1926fluctuations, vandermeer1994qualitative, huisman2001biological, vandermeer2004coupled,  stone2007chaotic, arnoldi2018ecosystems}.
Despite many steps forward in both fields \cite{brunel2000dynamics, hermann2012heterogeneous, del2013synchronization, bimbard2016instability}, the origin of such oscillatory behaviour  in brain responses after solicitation or in ecological assemblies, especially in the limit of large ecosystems,  is not  fully understood.
In the present work we show that oscillatory dynamical patterns may arise with high probability in large complex systems for specific type of interactions and network structure that we are able to identify.

In order to address these questions, we perform a linear stability analysis of a set of coupled differential equations of the form 
\begin{equation}
\partial_t \vec{x}= \vec{f}(\vec{x}), \label{eq:ODEs}
\end{equation}  
where $\vec{x}(t) = (x_1(t), x_2(t),\ldots, x_N(t))^{\rm T}\in \mathbb{R}^N$ is a column vector that describes the state of the system at a time $t$, and where $\vec{f}:\mathbb{R}^N\rightarrow \mathbb{R}^N$ is an arbitrary function.      
In the vicinity of a fixed point $\vec{x}^\ast$, defined by the condition
\begin{equation}
\vec{f}(\vec{x}^\ast) = 0,
\end{equation}
the dynamics given by Eq.~(\ref{eq:ODEs}) is well approximated by a set of  linear, coupled differential equations  of the form \cite{grobman1959homeomorphism, hartman1960lemma}
\begin{equation}
 \partial_t \vec{y}= -\textbf{d}\, \vec{y}(t) +  \textbf{A}\vec{y}(t),  \label{eq:ODEs_linearised}
\end{equation}
where  the vector
\begin{equation}
 \vec{y} =  \vec{x}-\vec{x}^\ast
\end{equation}
denotes the deviation from the fixed point $\vec{x}^\ast$, and where 
\begin{equation}
\partial_j f_k(\vec{x}^\ast) = -d_j\,  \delta_{j,k} + A_{kj} \label{eq:Jacob}
\end{equation}
 is the Jacobian of $\vec{f}$ at the fixed point.   In Eqs.~(\ref{eq:ODEs_linearised}) and~(\ref{eq:Jacob}), we have conveniently expressed the Jacobian as the sum of a diagonal matrix $\textbf{d}$ with diagonal elements $d_j$, and a {\it coupling matrix} $\mathbf{A}$ with elements representing interspecies coupling strengths $A_{kj}$, so that $ A_{jj}=0$, for all $ j$. We write $\delta_{j,k}$ for the Kronecker delta function. If the   parameters $d_j$ are positive, then they represent decay rates that describe how fast $\vec{y}$ relaxes to the stationary state   $\vec{y}=0$  in the absence of  coupling ($A_{kj}=0$).       
 Equations of the form (\ref{eq:ODEs})-(\ref{eq:ODEs_linearised}) are used to model, among others,  the dynamics of neural networks \cite{chaos,PhysRevE.91.012820,kadmon2015transition,amir2016non} and ecosystems \cite{allesina2012,grilli2016modularity, gibbs2018effect} in the vicinity of some fixed point $\vec{x}^\ast$.

 If the coupling strengths $A_{kj}$ are small enough, then  the fixed point $\vec{x}^{\ast}$ 
 is stable since
\begin{equation}
\lim_{t\rightarrow \infty}|\vec{y}(t)| = 0
\end{equation}
for all initial states $\vec{y}(0)$,  where $|\cdot|$ is the norm of a vector.   On the other hand, large  $A_{kj}$ can destabilize the fixed point giving 
\begin{equation}
 \lim_{t\rightarrow \infty}|\vec{y}(t)| = \infty
 \end{equation}
 for all initial states $\vec{y}(0)$, although in this limiting case the Eq.~(\ref{eq:ODEs_linearised}) does not approximate well Eq.~(\ref{eq:ODEs}) and one should focus on the nonlinear dynamics.  

To quantitatively determine the stability of the system, we define the leading eigenvalue $\lambda_1$ of the matrix $\mathbf{A}$ as the eigenvalue that has the largest real part.
If there exist  several eigenvalues with the same real part, for example because $\lambda_1$ has a nonzero imaginary part, then we choose $\lambda_1$ to be the eigenvalue with the largest imaginary part. 
 For instance under the assumption that $d_j=d$ for all $j$, if the real part ${\rm Re}(\lambda_1(\mathbf{A}))$ of $\lambda_1$ satisfies
 \begin{equation}
{\rm Re}(\lambda_1(\mathbf{A})) < d \label{eq:crit1}, 
\end{equation}
then  the fixed point $\vec{y}=0$ is {\it stable} 
since the matrix $\textbf{M}=-\textbf{d}+\textbf{A}$ has leading eigenvalue with negative real part, 
while if 
 \begin{equation}
{\rm Re}(\lambda_1(\mathbf{A})) >d, \label{eq:crit2}
\end{equation}
$\textbf{M}=-\textbf{d}+\textbf{A}$ has leading eigenvalue with positive real part,
then the fixed point is {\it unstable}, as we detail in Appendix~\ref{App:Oscil}.

Besides  system stability, we are also interested in how large dynamical systems respond to external perturbations.  
As we discuss in  Appendix~\ref{App:Oscil},   if
 \begin{equation}
{\rm Im}(\lambda_1(\mathbf{A}))  = 0  \label{eq:crit3}
\end{equation} 
then the response of  $\vec{y}$ is {\it nonoscillatory}, while if 
 \begin{equation}
{\rm Im}(\lambda_1(\mathbf{A})) > 0  \label{eq:crit4}
\end{equation}
then the response  is  {\it oscillatory}.  
In particular, the imaginary part of $\lambda_1$ determines the frequency of oscillations of the slowest mode when the system is stable, and of the fastest destabilizing mode when the system is unstable.    

An important prediction in the theory of complex systems  is that a linear system of randomly interacting components is unstable when $N$ is large enough \cite{gardner1970connectance, may1972will}.    Indeed, under the assumption that the   $A_{ij}$ can be well approximated by independent and identically distributed (i.i.d.)~random variables drawn from a distribution $p_A$ with 
finite second moment $\sigma^2$, as proposed in a pioneering work by R. M. May \cite{may1972will}, it holds that for $N$ large enough \cite{tao2012topics} 
\begin{equation}
\lambda_1 = \sigma \sqrt{N} .  \label{eq:MayLambda1}
\end{equation} 
Hence,  for  $N<N^\ast= (d/\sigma)^2$ the system is stable, while  for $N>N^\ast$ the system is unstable. In the context of ecology, this result seems to imply that  the biodiversity of an ecosystem is constrained by system stability, standing at the core of the diversity-stability debate \cite{mccann2000diversity, Allesina2015, biroli2018marginally}.

The model of R.M.~May exhibits an unavoidable tradeoff between linear stability of fixed points and diversity  because  the real part ${\rm Re}(\lambda_1)$ of the leading eigenvalue diverges as a function of $N$.     If on the other hand,  ${\rm Re}(\lambda_1)$  converges to a finite value as a function of $N$, then fixed points can be stable even in the limit of infinitely large system size $N$ if $d$ in our example is sufficiently large, yet finite.   On these grounds, considering random matrix models with distribution of the nonzero elements $A_{ij}$ that does not depend on $N$,  we  define the following two classes of stability: 
\begin{enumerate}[label=(\roman*)]
\item   {\it Size-dependent stability}:   the leading eigenvalue ${\rm Re}(\lambda_1)$ diverges as a function of $N$.   Therefore, there exists a critical value $N^\ast$ above which models are unstable and  biodiversity is constrained by system stability.   
\item {\it Absolute stability}:   the leading eigenvalue ${\rm Re}(\lambda_1)$ converges to a finite value as a function of $N$.   Therefore, a finite value of $d$ can stabilise the fixed point, even in  the limit of $N\gg 1$.    Note that, absolutely stable models can be constructed from models with size-dependent stability by simply rescaling the entries of the matrix  $\mathbf{A}$  with $N$, which guarantees  that the matrix norm  $\|\mathbf{A}\| = {\rm sup}\left\{\|\mathbf{A}\vec{x}\|: \vec{x}\in\mathbb{R}^N \ {\rm with} \ \|\vec{x}\|=1\right\}$   is finite in the limit of $N\gg 1$ and thus also the leading eigenvalue is finite.    For example, if we consider random matrices $\mathbf{A}$ with entries that are i.i.d.~random variables with a variance $\sigma^2/N$, {\it i.e.} a model where coupling strengths $A_{kj}$ tend to zero for large $N$, then the leading eigenvalue ${\rm Re}(\lambda_1)$ converges to $\sigma$.  
However such rescaling is introducing a dependence on $N$ of the  matrix elements $A_{ij}$, and therefore in the following we will call absolutely stable only those models for which fixed points can be stabilised by a finite value of $d$ without  rescaling of the matrix entries with $N$.
\end{enumerate}

From a network perspective, May's model \cite{may1972will} can be described in terms of a dense graph, which can represent special cases of well-mixed ecosystem.
However, usually the constituents of real-world systems interact through specific preferential interaction whose structure is better described by large, complex networks where nodes are not all linked with all the others. It is therefore interesting  to understand the stability of dynamical systems defined on  infinitely large, sparse, random graphs with edges that are characterised by random weights.  Under these assumptions, interspecies coupling strenghts can still be approximated by random variables but it can be studied how the non trivial graph structure may affect system's stability.      
        
So far, the leading eigenvalue  of adjacency matrices of sparse graphs has been studied on nondirected graphs with symmetric couplings and on directed graphs, for which the couplings are unidirectional.  
For symmetric matrices,  the leading eigenvalue ${\rm Re}(\lambda_1)$ scales as $O(\sqrt{k_{\rm max}})$ \cite{krivelevich2003largest, chung2004spectra, susca2019top}, where $k_{\rm max}$ is the maximal degree of the graph.    On the other hand,  for adjacency matrices of  random, directed graphs \cite{neri2016eigenvalue, tarnowski2020universal, neri2019spectral} ${\rm Re}(\lambda_1)$ scales as $O(\tilde{c})$, where $\tilde{c}$ is the mean number of links pointing outward, also known as mean outdegree.   Hence,  in the limit $N\rightarrow \infty$, linear dynamical systems on random directed graphs require a finite mean outdegree to guarantee absolute stability,  while models defined on nondirected graphs need that the maximal degree is finite, which is a stronger requirement.      
This result gives a first example of how the nature of the interactions between single components play a major role on the stability property of the system granting the absolute stability of models on directed graphs that have the same structure and statistical properties as  models on undirected graphs that are not absolutely stable.

Symmetric and unidirectional couplings are often not realistic types of interactions for modelling  real world systems, such as, ecosystems \cite{allesina2012, barbier2018generic} and neural networks \cite{marti2018correlations}.    In general, interactions between the constituents of complex systems are bidirectional and nonsymmetric.   For example, the trophic interactions between species in ecosystems can be of  the {\it predator-prey}, {\it competitive} or {\it mutualistic} type (see Fig.\ref{fig:model}). Starting from this observation, Refs.~\cite{allesina2012, mougi2012diversity, mougi2014stability} have considered how predator-prey, competitive, and mutualistic interactions affect the stability of ecosystems defined on dense graphs.   Although predator prey interactions tend to decrease the real part of the leading eigenvalue  ${\rm Re}(\lambda_1)$ by a constant prefactor, it does  not  alter  the scaling of ${\rm Re}(\lambda_1)$ as $\sqrt{N}$.   
Hence, in the case of dense graphs, stability depends on system sizes in all cases studied unless absolute stability is restored in the trivial sense by rescaling the interaction variables.

Inspired by these studies on complex ecosystems \cite{Allesina2008, allesina2012, brauer2012mathematical, allesina2015predicting,barbier2018generic},  we analyse in this paper the linear  stability of fixed points of dynamical systems defined on  large sparse random graphs with   predator-prey,  competitive or  mutualistic interactions.  Such dynamical systems serve as models for    large   ecosystems    defined  on foodwebs \cite{bascompte2009disentangling, james2015constructing}
described as
sparse, nondirected, random graphs   that have a prescribed degree distribution \cite{molloy1995critical, molloy_reed_1998, newman2001random, newman2010networks, dorogovtsev2010lectures} and 
random interactions.       We obtain the typical leading eigenvalue for these graphs with  the cavity method \cite{cavity_nonH, NeriMetz2012, Izaak_Metz_2019}, which is applied to specific cases as an illustration, but also 
holds for general nonsymmetric and bidirectional interactions.  
Using this theory, we  determine linear stability of fixed points in an  infinitely large ecosystem, characterised by random interactions on a sparse random graph.
We find that, unlike in the dense cases and at variance with what generally expected, for sparse graphs the type of interactions can strongly affect stability, leading in some models to absolute stability, instead of size-dependent stability. 
The result also applies to other kind of system stability, such as structural stability and feasibility of ecosystems, as discussed in the last section.
We also determine  how the dynamics in the vicinity of a fixed point can strongly depend on the nature of the interactions and the network topology. 

The paper is organized as follows:  In Sec.~\ref{sec:Model}, we introduce the random matrix models we study in this paper, namely, antagonistic and mixture random matrices defined on sparse random graphs.   In Sec.~\ref{sec:MainResults}, we present the main results for  the stability and dynamics of infinitely large ecosystems defined on sparse graphs, which are based on exact results for the leading eigenvalue of the random matrix models defined in the previous section.   In Sec.~\ref{sec:Theory},   we present the theory we use to derive   the main results for the leading eigenvalue of antagonistic and mixture matrices on random graphs.   In Sec.~\ref{sec:BoundaryResults}, we present    results for the spectra of antagonistic and mixture matrices, which extend the results for the leading eigenvalue discussed in Sec.~\ref{sec:MainResults}.     In Sec.~\ref{sec:networkTop}, we analyse how graph topology affects the stability of large ecosystems, and in Sec.~\ref{sec:disc}, we present a discussion of the results in this paper and their broader application in the ecological context.

\subsection{Notation}
We denote real or complex numbers with a regular, serif font, e.g., $a$, $b$, vectors with an arrow, e.g., $\vec{a}$, $\vec{b}$, matrices of size $2\times 2$ with a sans serif font, e.g., $\mathsf{a}$, $\mathsf{b}$, and matrices of size $N\times N$ in bold, e.g.,  $\mathbf{a}$, $\mathbf{b}$.     Notation wise, we do not make a distinction between  deterministic numbers and random variables.   If $u$ is a random variable drawn from a distribution $p$, then we write $\langle u\rangle_p$ for its average.     We use $u^\ast$ to denote the typical value of a random variable $u$, i.e., $u^\ast = {\rm argmax}\: p(u)$.  If $z = x + {\rm i}y$ is a complex number, then we denote its complex conjugate by $\overline{z} = x - {\rm i}y$.   We denote the real part of a complex number $z$ by ${\rm Re}(z)$ and we denote its imaginary part by ${\rm Im}(z)$.  The complex conjugate of a vector is denoted by $\vec{a}^\dagger$, $\vec{b}^\dagger$.

\section{\label{sec:Model} Model definitions} 
The models we study in this paper generalize the random matrix models for complex ecosystems with all-to-all interactions  defined in Ref.~\cite{allesina2012} and the models for complex systems  on random graphs with unidirectional interactions  studied in Refs.~\cite{neri2016eigenvalue, Izaak_Metz_2019, metz2020localization}.   

\begin{figure}[htbp]
        \includegraphics[width=0.5\textwidth]{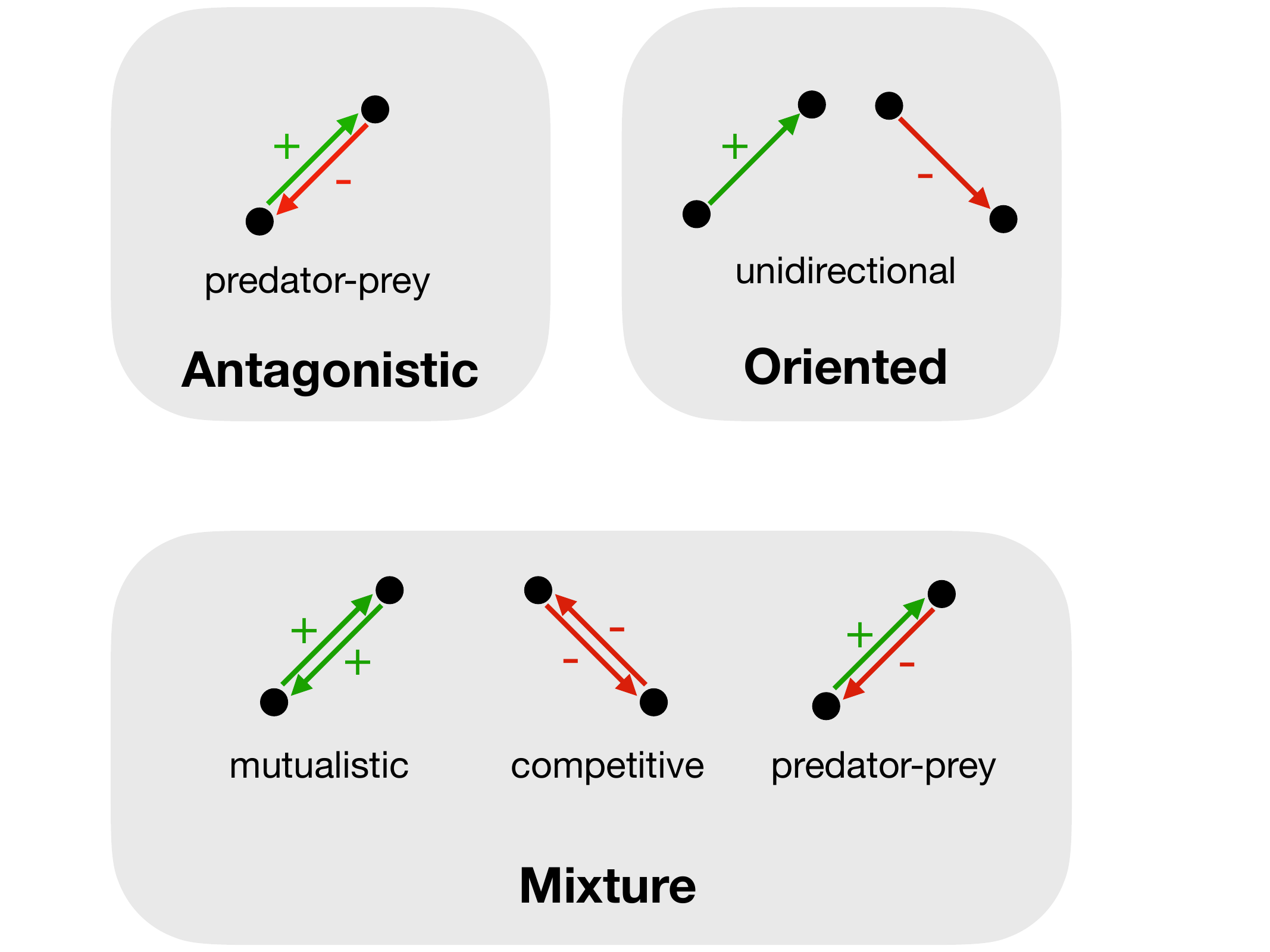}
         \caption{Sketch of the interactions in the two main models,  the antagonistic and mixture model,  that we study in this paper, as well as, in the oriented ensemble studied before in Refs.~\cite{neri2016eigenvalue, Izaak_Metz_2019, metz2020localization}.
         }
 \label{fig:model}
\end{figure}

We first define a general model for  a dynamical system defined on a network irrespectively of the specific choice for the interactions between the nodes of the network.    Subsequently, we will  focus on random matrix models that represent dynamical systems with predator-prey interactions only or  a combination of predator-prey, mutualistic, and competitive interactions. We call the former {\it antagonistic} ensemble, following Refs.~\cite{allesina2012, Cicuta}, and the latter the {\it mixture} ensemble; see Fig.~\ref{fig:model} for an illustration of these two models.

\subsection{General model}\label{sec:generalModel}
For simplicity, in Eq.~(\ref{eq:Jacob}) we consider that all diagonal entries of $\mathbf{d}$ are all the same, {\it i.e.}, $d_j=d$ for all $j$.  
This is because we want to understand how the properties of the interactions influence system stability, which will not be directly affected by having different values $d_j$.

The interspecies coupling strengths are modelled by random matrices 
$\mathbf{A}$ with entries 
\begin{equation}
A_{ij}=C_{ij}J_{ij}, \hspace{1cm} i,j \in \left\{1,\ldots,N\right\}, \label{eq:ADef}
\end{equation} 
 where $\mathbf{C}=\{C_{ij}\}$ is the   adjacency matrix of a nondirected, random graph with a prescribed degree distribution $p_{\rm deg}(k)$ \cite{molloy1995critical, molloy_reed_1998, newman2001random, newman2010networks, dorogovtsev2010lectures}   with mean degree 
 \begin{equation}
    c = \sum^{\infty}_{k=0}p_{\rm deg}(k)\:k,
\end{equation}
and where the pairs $(J_{ij}, J_{ji})$ are i.i.d.~real-valued random variables drawn from a probability distribution $p(\JU,\JD)$ that is symmetric in its arguments, 
 \begin{equation}
 p(\JU,\JD) = p(\JD,\JU) ,\quad \JU,\JD\in\mathbb{R}. \label{eq:puv}
 \end{equation}

 So far, in the research literature, the spectral properties of matrices of the form given by Eq.~(\ref{eq:ADef}) have been discussed in the cases of symmetric random matrices \cite{Abou_Chacra_1973, cugliandolo,fyodmirl,  cavity, LocTreeLike, Kabashima_2010, susca2019top}, for which
 \begin{equation}
    p(\JU,\JD) = p^{\rm S}(\JU,\JD) = \tilde{p}(\JU) \delta(\JU-\JD), \label{eq:sym}
 \end{equation}
 or for  oriented random matrices \cite{cavity_nonH,  neri2016eigenvalue, Izaak_Metz_2019, tarnowski2020universal, neri2019spectral}, for which 
 \begin{equation}
    p(\JU,\JD) = p^{\rm O}(\JU,\JD) = \frac{1}{2}\tilde{p}(\JU) \delta(\JD) + \frac{1}{2}\tilde{p}(\JD) \delta(\JU),  \label{eq:oriented} 
 \end{equation}
 where  $\tilde{p}$ is a probability density generally supported on $\left(-\infty,\infty \right)$.  In the oriented case, interactions are unidirectional, as illustrated in Fig.~\ref{fig:model}.   We write $\delta(x)$ for the Dirac distribution.
 
In this paper, we extend these studies and present a theory for deriving the spectral properties of  random matrices of the type given by Eq.~(\ref{eq:ADef}) with $J_{ij}$ and $C_{ij}$ as described.   We apply this theory to two types of ensembles that have not been discussed before in the literature, namely, antagonistic random matrices, for which  $p(\JU,\JD)$ represents predator-prey interactions, and mixture matrices, for which $p(\JU,\JD)$ represent a mixture of  predator-prey, competitive, and mutualistic interactions.  We  discuss these two cases in the next two subsections, respectively.

 \subsection{Antagonistic matrices} \label{subsec:AntagModel}
Antagonistic coupling matrices are defined in terms of the sign-constraint of antagonistic interactions:
\begin{eqnarray}
J_{ij}J_{ji}<0 \label{eq:predPrey}
 \end{eqnarray}
 for each pair of indices $i\neq j$.  Note that antagonistic matrices are  not antisymmetric matrices as in general $J_{ij} \neq -J_{ji}$.    In the context of ecosystems, they represent   predator-prey interactions.
 %Without loss of generality 
 This condition could be realised by the following $p(\JU,\JD)$
  \begin{align}\label{eq:Antagonisticp}
     &p(\JU,\JD) = \PA(\JU,\JD)= \nonumber\\
     &=\frac{1}{2}\tilde{p}(\left| \JU \right|)\tilde{p}(\left|\JD\right|)\left[\theta(\JU)\theta(-\JD)+\theta(-\JU)\theta(\JD)  \right],
      \end{align}  
 where 
 \begin{eqnarray}
 \theta(x) = \left\{\begin{array}{ccc}0&& x< 0,  \\ 1 && x\geq 0,\end{array}\right.
 \end{eqnarray} 
 is the Heaviside function, and where  $\tilde{p}$  for the antagonistic ensemble is a probability density supported on $\left[0,\infty \right)$.

 \subsection{Mixture matrices}\label{subsec:mix}
The mixture ensemble consists of a mixture of predator-prey interactions, for which Eq.~(\ref{eq:predPrey}) holds, mutualistic interactions,  which determine the following coupling strengths
\begin{eqnarray}
J_{ij}>0\  {\rm and} \ J_{ji}>0, \label{eq:mutual}
 \end{eqnarray}
 and competitive interactions, for which corresponding coupling strengths are 
 \begin{eqnarray}
J_{ij}<0\  {\rm and} \ J_{ji}<0. \label{eq:mutual}
 \end{eqnarray}

For mixture matrices the distribution $p$ reads
 \begin{eqnarray}
    \lefteqn{ p(\JU,\JD)=\PM(\JU,\JD)=}&& \nonumber\\
 &=&\pA\PA(\JU,\JD)+\left(1-\pA\right) \PCM(\JU,\JD),\label{eq:MixtureP}
 \end{eqnarray} 
 where $\pA\in  \left[0,1\right]$, $\PA$ is the distribution defined in Eq.~\eqref{eq:Antagonisticp} that describes predator-prey interactions, and $\PCM$ is the distribution
 \begin{align}
     &\PCM(\JU,\JD)= \nonumber\\
     &=\tilde{p}(\left| \JU \right|)\tilde{p}(\left|\JD\right|) \left[\pM\theta(\JU)\theta(\JD)+\left(1-\pM \right)\theta(-\JU)\theta(-\JD)  \right]\label{eq:MixturepCM}
 \end{align}  
that describes both mutualistic and competitive interactions  with $\pM\in  \left[0,1\right]$. 
Hence, for mixture matrices the couple $(\JU,\JD)$ is with probability $\pA$ a predator-prey-like interaction, while it is mutualistic with probability $(1-\pA)\pM$  or  competitive with probability $(1-\pA)(1-\pM)$.  

\subsection{Two examples: Model A and Model B} \label{sec:relEx}
In the numerical examples of this paper, we  consider that $\tilde{p}$ is a uniform distribution with unit {second moment}, namely,  
 \begin{equation}
     \tilde{p}(x) = \frac{1}{b}\left[1-\theta(x-b)\right], \quad x\geq0, \label{eq:uniform}
 \end{equation}
 with $b=\sqrt{3}$.

We set $\pi_M = 0.5$ so that on average couplings are balanced, i.e., 
 \begin{equation}
 \langle u\rangle = \langle l\rangle = 0, \label{eq:balanced}
\end{equation} 
%which is a reasonable assumption for neural networks \cite{PhysRevLett.97.188104}, 
even though the results of paper also hold for $\pi_M \neq 0.5$.   
With this choice the variance of the coupling strengths ($u$ and $l$) is one.
 
We often consider random matrices defined on Erd\H{o}s-R\'{e}nyi graphs \cite{ER, bollobas2001random}.   For an Erd\H{o}s-R\'{e}nyi graph the $C_{ij}$ are i.i.d.~random variables that  are  with a probability $c/(N-1)$   equal to one and  with a probability $1-c/(N-1)$ equal to zero.      The  Erd\H{o}s-R\'{e}nyi ensemble is also a random graph with the prescribed degree distribution 
 \begin{equation}
p_{\rm deg}(k) = \left(\begin{array}{c}  N-1 \\ k \end{array}\right) \left(\frac{c}{N-1}\right)^k \left(1-\frac{c}{N-1}\right)^{N-1-k}.
 \end{equation}
 In the limit $N\rightarrow \infty$, the degree distribution is Poissonian  with mean degree $c$, i.e., 
\begin{equation}
p_{\rm deg}(k) = \frac{c^k}{k!}e^{-c} \label{eq:poisson}. 
\end{equation}

We define two ensembles of reference that we will often consider in the numerical examples and that  we call  Model A and Model B.   Model A is an antagonistic, random matrix defined on an Erd\H{o}s-R\'{e}nyi graph  with  $\tilde{p}$ given by Eq.~(\ref{eq:uniform}).   Model B, is a mixture random matrix defined on an Erd\H{o}s-R\'{e}nyi graph with $\pi^{\rm A} = 0.9$, ${\pi^{\rm M} = 0.5}$ and  $\tilde{p}$ given by Eq.~(\ref{eq:uniform}).   Since $\pi^{\rm A} = 0.9$, most of the interactions in Model B are predator-prey interactions.     Hence,  the difference between Model A and Model B is that all interactions in Model A are predator-prey interactions, while Model B contains a small fraction of mutualistic and competitive interactions.

\section{\label{sec:MainResults} Main Results}
In this section, we present the main results of this study. In particular we compare 
the linear stability of antagonistic systems with systems that contain a mixture of interactions, and 
we discuss the peculiar nature of their dynamics in the vicinity of a fixed point.       We consider cases for which the degree distribution $p_{\rm deg}(k)$ has unbounded support  so that with probability one  the norm $\|\mathbf{A}\|$    diverges as a function of $N$.      
It is natural to expect that if the norm diverges, then  also  ${\rm Re}(\lambda_1)$  diverges as a function of $N$ implying that the stability of the fixed point depends on the system size.     However, as  we will show this is not always the case and whether the stability of fixed points  is size-dependent or not size-dependent  follows from the nature of the interactions.
\subsection{\label{susec:ST}Infinitely large antagonistic systems  are absolutely stable} 
The first main result of this paper  is that infinitely large systems with   predator-prey interactions  are significantly more stable than their counterparts  that contain mutualistic and competitive interactions.    Indeed, we show  that in general, the stability of   fixed points of dynamical systems on random graphs is dependent on the system size, and for large enough $N$ fixed points are unstable, implying a tradeoff between stability and diversity.   An exception on this generic behaviour is when all interactions are of the predator-prey type.   In this case, fixed points are absolutely stable.    

\begin{figure}[htbp]
        \includegraphics[width=0.5\textwidth]{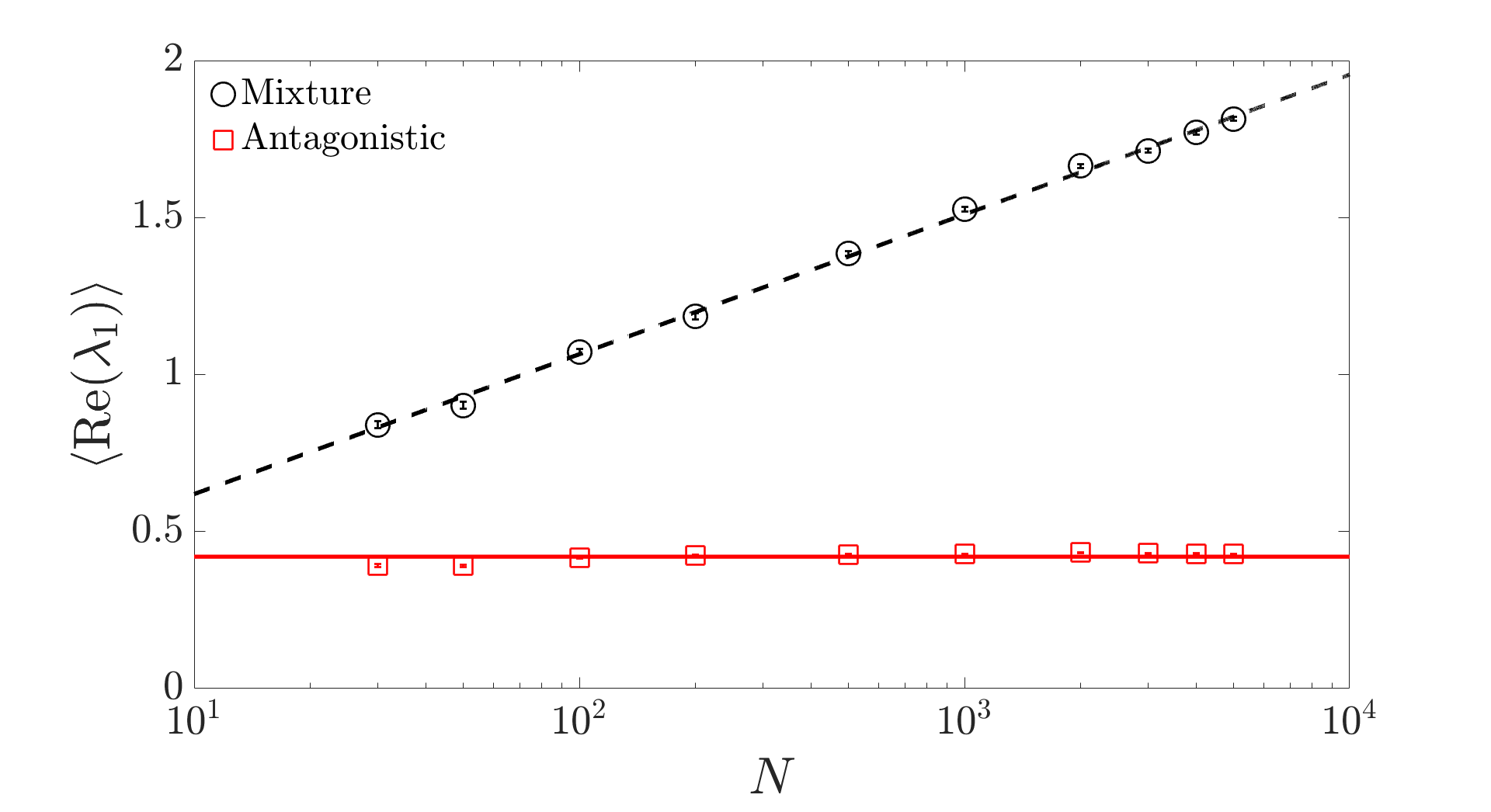}
         \caption{
          Real part $\langle {\rm Re}[\lambda_1]\rangle$ of the mean value of the leading eigenvalue $\lambda_1$ as a function of $N$ for antagonistic matrices (Model A) and mixture matrices (Model B) on  Erd\H{o}s-R\'{e}nyi graphs  with mean degree $c=4$.  Markers are  sample means of $\lambda_1$ obtained from directly diagonalizing $10^3$ matrices. The continuous red line is the typical value  of ${\rm Re}(\lambda^\ast_1)$  obtained with the cavity method, see  Sec.~\ref{sec:Theory}, and the black dashed line is obtained from fitting the function $a\log(N)+b$ to the data.  
         }
 \label{fig:ReLambda1VSN}
\end{figure}

This result follows from an analysis of the mean value 
\begin{equation}
\langle{\rm Re}(\lambda_1)\rangle  = \langle{\rm Re}(\lambda_1(\mathbf{A})) \rangle_{p(\mathbf{A})}
\end{equation}
of the leading eigenvalue $\lambda_1$ as a function of $N$.    Figure \ref{fig:ReLambda1VSN} shows that  for antagonistic random matrices 
$\langle{\rm Re}[\lambda_1]\rangle$ converges for large $N$ to  a finite number, while  for mixture matrices $\langle {\rm Re}[\lambda_1]\rangle$ diverges as a function of $N$.     Since stability is granted whenever ${\rm Re}[\lambda_1]<d$, this result tells that mixture models are unstable when $N$ is large enough, while antagonistic ones can be stable for $d$ sufficiently large but finite, even when $N$ infinitely large.  
The results for mixture matrices have been obtained using $\pi^{\rm A}=0.9$
for  Model B,  indicating that a small  fraction of competitive or mutualistic interactions  are sufficient to render a system always unstable in the infinite size limit.

  The results presented in Figure \ref{fig:ReLambda1VSN} are obtained by using two independent methods.  First, we  diagonalize  $10^3$  matrices sampled from  Model A (antagonistic) and Model B (mixture) and  compute the sample means of the leading eigenvalue.   For mixture matrices  $\langle {\rm Re}[\lambda_1]\rangle$  diverges logarithmically in $N$, while for antagonistic random matrices $\langle {\rm Re}[\lambda_1]\rangle$ is more or less independent of $N$.     Second, we compute  the typical value $\lambda^\ast_1$  of the leading eigenvalue in the limit $N\rightarrow \infty$  with the cavity method, which is an exact mathematical method for the spectral properties of sparse nonHermitian matrices; we  explain the cavity method in full detail in the next section.   For antagonistic matrices, the cavity method provides   a finite value that is confirmed by the numerical diagonalization results, while for mixture matrices we obtain that $\lambda^\ast_1$ is infinitely large in the infinite $N$ limit, which is also in agreement with the logarithmic divergence of $\lambda^\ast_1$ with $N$ observed in the direct diagonalization results.

\begin{figure}[htbp]
        \includegraphics[width=0.5\textwidth]{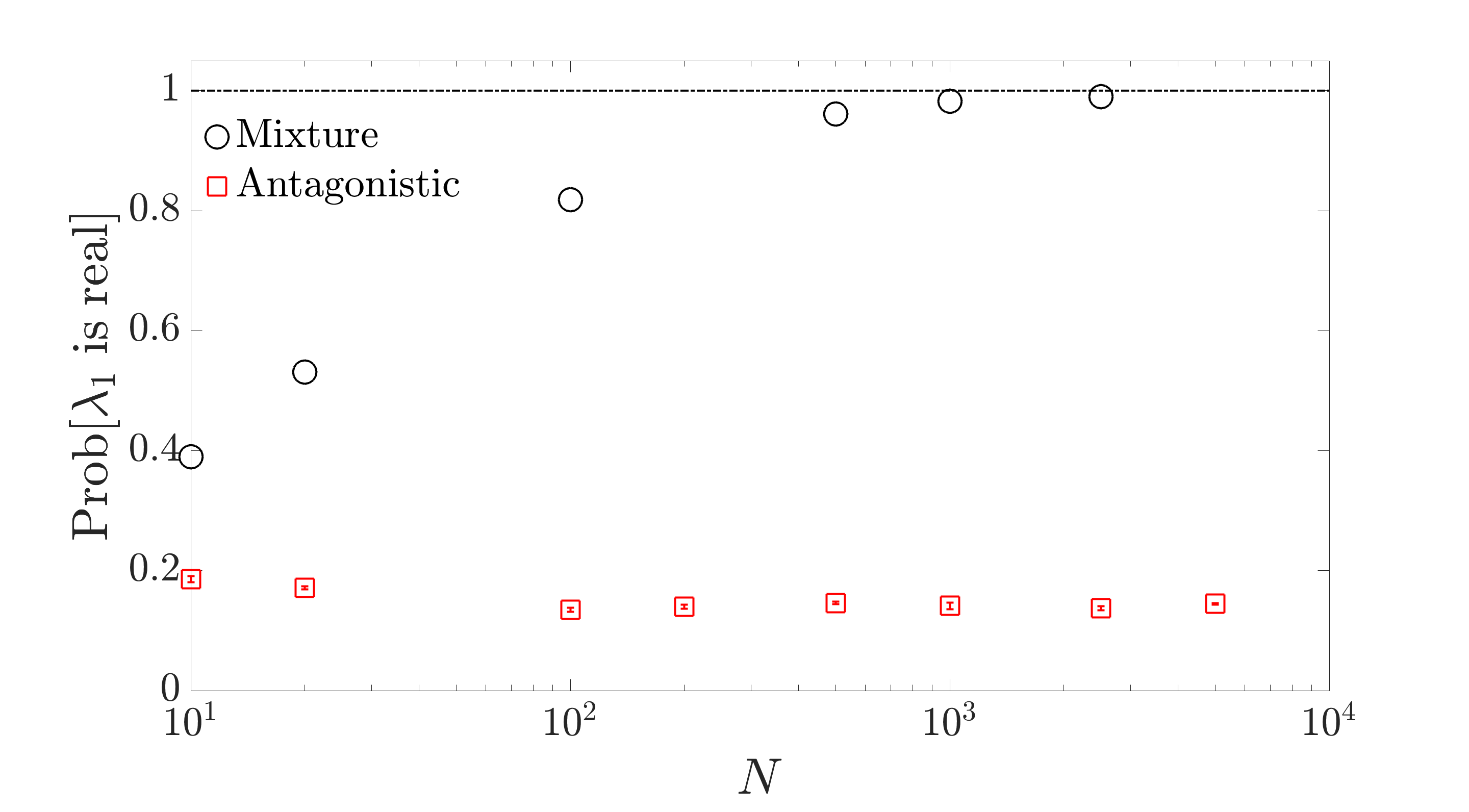}
         \caption{Probability  ${\rm Prob}[\lambda_1\in\mathbb{R}]$    that     $\lambda_1$  is located on the real line as a function of $N$.     Markers are numerical   results obtained from  directly diagonalizing  $10^3$ matrices from the antagonistic ensemble  (Model A, red squares) and the mixture ensemble   (Model B, black circles), both for Erd\H{o}s-R\'{e}nyi  graphs   with a mean degree $c=2$.    
         }
 \label{fig:N_Lambda1RealVSN}
\end{figure}
\subsection{\label{subsec:Oscillations} Large antagonistic systems with a small mean degree exhibit oscillations}
A second main result of this paper is that the dynamics of  infinitely large and sparse systems with  predator-prey interactions   exhibit oscillations in the vicinity of a fixed point.   In particular, we obtain that  infinitely large systems  defined on antagonistic matrices exhibit oscillations    when the mean degree $c$ is small enough.   This follows from an analysis of the imaginary component of the leading eigenvalue.

Figure~\ref{fig:N_Lambda1RealVSN} gives a first evidence that the leading eigenvalue of infinitely large mixture matrices is real, while the leading eigenvalue of antagonistic matrices can have a nonzero imaginary part.      Indeed, the plot shows the probability ${\rm Prob}[\lambda_1(\mathbf{A})\in\mathbb{R}]$  that the leading eigenvalue is real as a function of $N$ for matrices in Model A (antagonistic) and Model B (mixture) with mean degree $c=2$.  
    We observe that for mixture matrices $\lambda_1$ is with probability one a real number when $N$ is large enough, while for antagonistic matrices $\lambda_1$ always has with finite probability a nonzero imaginary component.  We have obtained these results by numerically diagonalizing   $10^3$ matrices and by using the criterion  ${\rm Im}(\lambda_1)<10^{-13}$ to identify an eigenvalue as real,   where $10^{-13}$ is much smaller than the typical distance between two eigenvalues.    
   Remarkably, just as in  Fig.~\ref{fig:ReLambda1VSN}, it is sufficient to have  a small finite fraction of mutualistic and competitive interactions to obtain a real-valued $\lambda_1$.  Hence, almost all interactions must be of the predator-prey type in order to have a nonreal leading eigenvalue~$\lambda_1$.

\begin{figure}[htbp]
         \centering
        \includegraphics[width=0.5\textwidth]{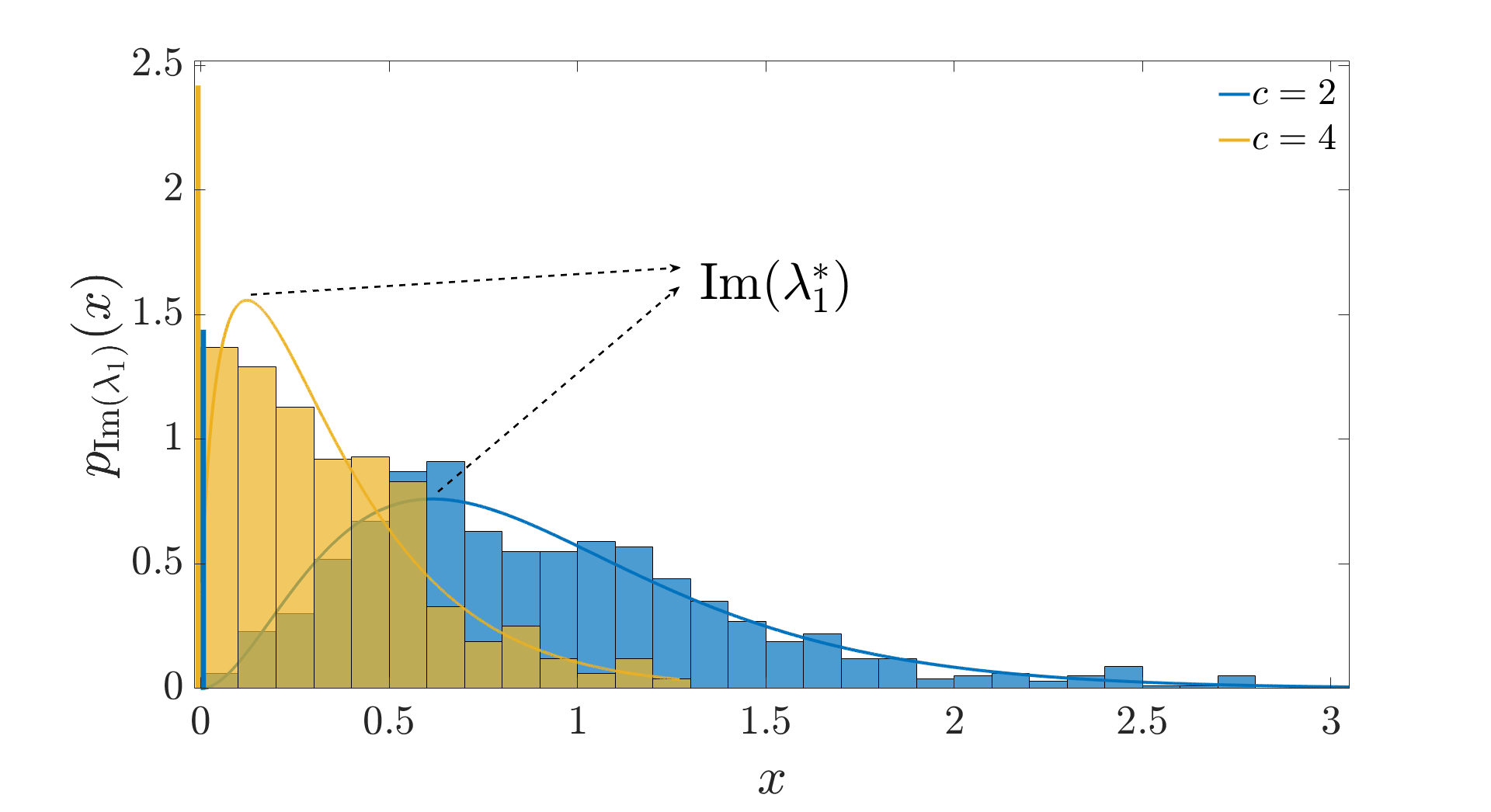}
         \caption{Histograms of the  imaginary part  ${\rm Im}(\lambda_1)$ of the leading eigenvalue $\lambda_1$ in antagonistic matrices  defined on Erd\H{o}s-R\'{e}nyi graphs (Model A) with mean degrees $c=2$ (blue) and $c=4$ (yellow).   Results shown are obtained from diagonalizing $10^3$ matrices of size    $N=5000$.      The thick vertical line at   ${\rm Im}(\lambda_1)=0$   has height   ${\rm Prob}[\lambda_1\in\mathbb{R}]/\delta$, with $\delta = 0.1$ the width of the intervals in the histogram.    Continuous lines are gamma distributions, see Eq.~(\ref{eq:gamma}), fitted to the histograms [fitted parameters are $\alpha=3.04$ and $\beta=3.31$ (blue) and  $\alpha=1.53$ and $\beta=4.35$ (yellow)].
         }
 \label{fig:DistributionImLambda1_c2_4_N5000}
\end{figure}

The leading eigenvalue $\lambda_1$ is not self-averaging %a deterministic variable 
in the limit $N\rightarrow \infty$ as it is  evident from Fig.~\ref{fig:N_Lambda1RealVSN};  since, if $\lambda_1$ were self-averaging, then ${\rm Prob}[\lambda_1(\mathbf{A})\in\mathbb{R}]$ should tend either to one or zero.   %We say that $\lambda_1$ is not self-averaging.   
Since $\lambda_1$ is a random quantity, it is interesting to study its distribution, and in particular in this case the distribution of its imaginary part,
\begin{equation}
p_{{\rm Im}(\lambda_1)}\left(x\right)  = \langle  \delta(x -{\rm Im}(\lambda_1(\mathbf{A})) ) \rangle_{p(\mathbf{A})} \ ,
\end{equation} 
which is plotted in  Fig.~\ref{fig:DistributionImLambda1_c2_4_N5000} for  antagonistic matrices of finite size $N=5000$.     
The distribution takes the form 
 \begin{equation}
p\left({\rm Im}(\lambda_1)\right)  = \pi_{{\rm Re}} \:\delta({\rm Im}(\lambda_1)) + (1-\pi_{{\rm Re}})  p_{\rm c}({\rm Im}(\lambda_1)) \label{eq:pImag}
\end{equation} 
where 
\begin{equation}
\pi_{{\rm Re}} = {\rm Prob}[\lambda_1\in\mathbb{R}]
\end{equation}
is the probability that $\lambda_1$ is real and 
where $p_{\rm c}({\rm Im}(\lambda_1))$ a continuous distribution for nonreal values of ${\rm Im}(\lambda_1)$.  
In turn, from general considerations on sparse graphs  \cite{neri2019spectral} it is expected that in the limit $N\rightarrow \infty$ the continuous component tends to 
\begin{equation}
 p_{\rm c}(x) = a  \delta(x-{\rm Im}(\lambda^\ast_1))+  (1-a) p_{\rm cycle}(x)
\end{equation}
where ${\rm Im}(\lambda^\ast_1)$ is the typical value of ${\rm Im}(\lambda_1)$   and $p_{\rm cycle}(x)$ is the remaining distribution describing the nontypical values of  ${\rm Im}(\lambda_1)$, which are originated due to the presence of small cycles in the graph \cite{ShortestCycleReimer2017}.
Conveniently, theoretical prediction of ${\rm Im}(\lambda^\ast_1)$ in the limit $N\rightarrow \infty$ when the underlying graph has a giant component ($c>1$ for Erd\H{o}s-R\'{e}nyi graphs) is accessible by means of a theory based on the cavity method that we present in detail in Sec.~\ref{sec:Theory}.
Moreover from finite $N$ results, as shown in Fig.~\ref{fig:DistributionImLambda1_c2_4_N5000}, we get an independent estimate of the typical ${\rm Im}(\lambda^\ast_1)$ (see Appendix~\ref{sec:Appendix_NumericsDD} for a  detailed finite size study) by 
identifying it with the mode of a gamma distribution 
\begin{equation}
\gamma(x;\alpha,\beta) =  (1-{\rm Prob}[\lambda_1\in\mathbb{R}])\frac{\beta^\alpha x^{\alpha-1} e^{-\beta x}}{\Gamma(\alpha)} \label{eq:gamma}
\end{equation}
fitted on the histogram of nonreal values ${\rm Im}(\lambda_1)$,  
with $\Gamma(\alpha)$ the gamma function and $\beta,\alpha\in \mathbb{R}_+$ two fitting parameters.

\begin{figure}[htbp]
        \includegraphics[width=0.5\textwidth]{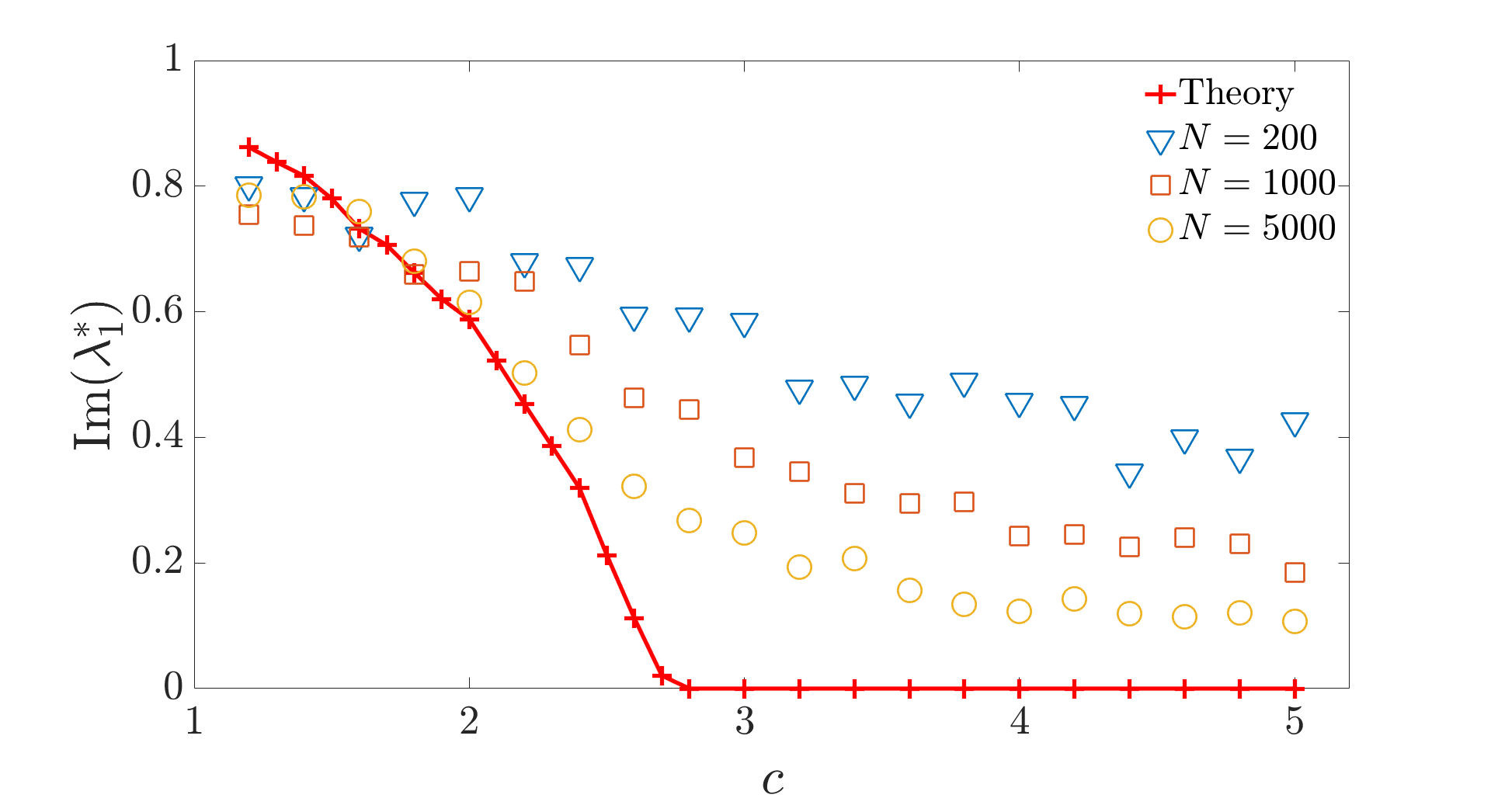} 
         \put(-142,7){\large$\uparrow$}
                  \put(-142,-5){\large$c_{\rm crit}$}
         \caption{Imaginary part  of the typical value  $
         {\rm Im}(\lambda^\ast_1)$ of the leading eigenvalue $\lambda_1$ as a function of the mean degree $c$ for antagonistic matrices defined on Erd\H{o}s-R\'{e}nyi graphs  (Model A).     Red crosses denote $
         {\rm Im}(\lambda^\ast_1)$ in the limit of $N\rightarrow \infty$ computed with the cavity method (see Sec.\ref{sec:Theory}), while the solid line is a guide to the eye.
         Unfilled markers denote   $
         {\rm Im}(\lambda^\ast_1)$  obtained from directly diagonalizaing $10^3$ matrices of a given size $N$ as shown in Fig.~\ref{fig:DistributionImLambda1_c2_4_N5000}.    Results shown are for $c>1$ since Erd\H{o}s-R\'{e}nyi graphs do not have a giant component when  $c<1$.
         }
 \label{fig:ImLambda1VSc}
 \end{figure}

Figure~\ref{fig:ImLambda1VSc} shows that these two independent methods give compatible results on the typical value ${\rm Im}(\lambda^\ast_1)$ of antagonistic random matrices, which interestingly it is seen to exhibit a phase transition as a function of the mean degree $c$.   
In particular, Fig.~\ref{fig:ImLambda1VSc}  shows that there exists a critical $c_{\rm crit}$ such that  ${\rm Im}(\lambda^\ast_1)$ converges as $N\rightarrow \infty$ towards zero for $c>c_{\rm crit}$ and to a nonzero value for $c<c_{\rm crit}$.

\begin{figure}[htbp]
         \centering
        \includegraphics[width=0.5\textwidth]{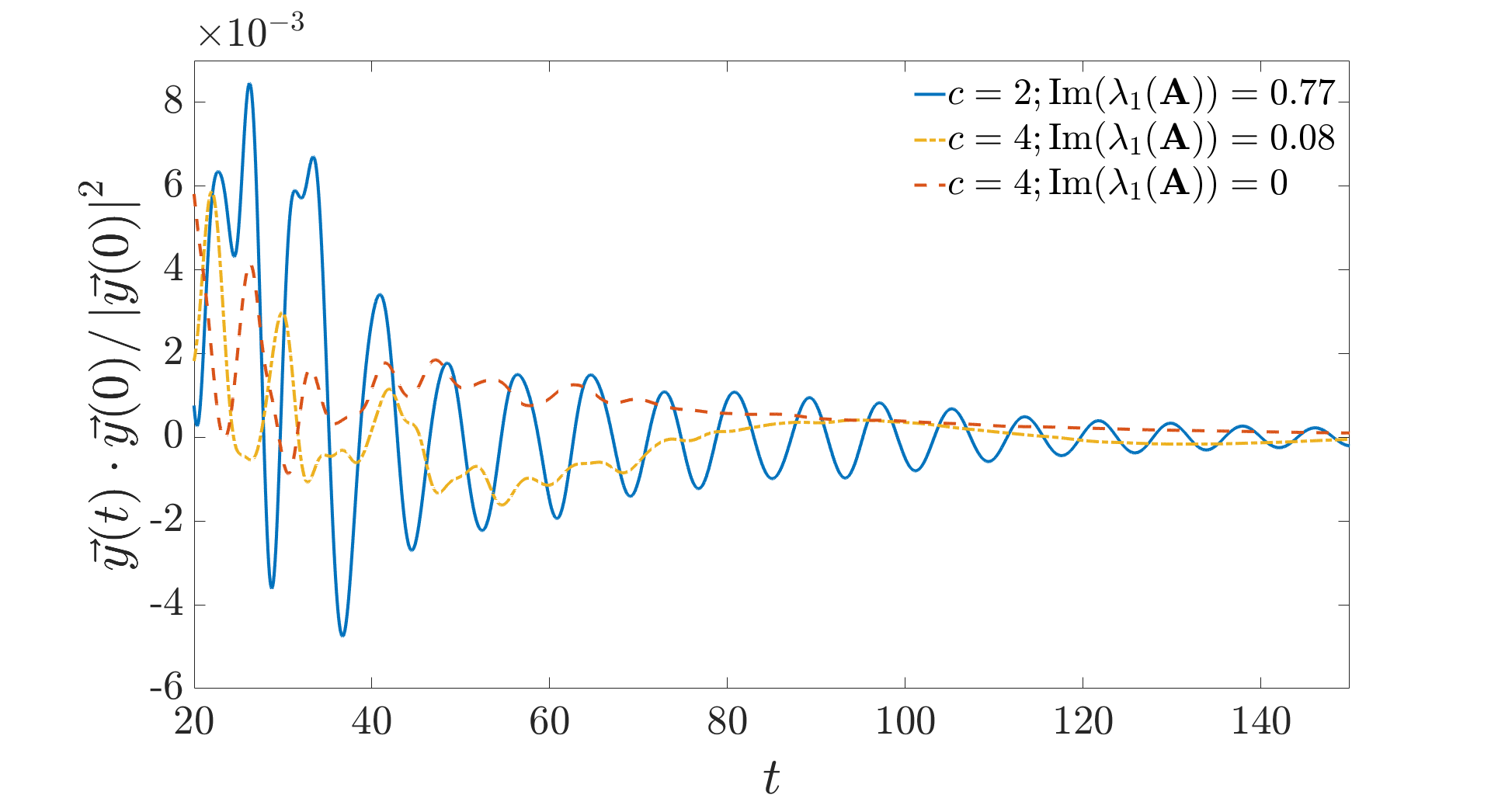}
         \caption{Plot of $\vec{y}(t)\cdot \vec{y}(0)/|\vec{y}(0)|^2$ as a function of $t$  for three different realizations of antagonistic matrices $\mathbf{A}$ drawn from Model A  with $N=10^4$,  and with $c=2$ or $c=4$ as denoted in the legend.   We have chosen matrix realizations such that the ${\rm Im}(\lambda_1(\mathbf{A}))$ (numerical values reported in the figure) correspond to the typical values of ${\rm Im}(\lambda_1)$ in Fig.~\ref{fig:DistributionImLambda1_c2_4_N5000}.
         }
 \label{fig:Dynamics}
\end{figure}
 
The result in Fig.~\ref{fig:ImLambda1VSc}  has deep implications on the dynamics of $\vec{y}(t)$. Indeed, assuming we are in the vicinity of a stable fixed point, the critical mean degree $c_{\rm crit}$ separates a regime at low connectivities, where infinitely large, antagonistic systems oscillate towards the fixed point, from a regime at high connectivities, where infinitely large, antagonistic systems relax monotonously to the fixed point.   Figure~\ref{fig:Dynamics} illustrates the distinction between the dynamics  at $c=2<c_{\rm crit}$
and $ c=4>c_{\rm crit}$ by plotting  trajectories of $\vec{y}$ as a function of $t$ for three different matrix realizations $\mathbf{A}$ drawn from the Model A.    Since $\vec{y}$  is a vector in a high dimensional space, we plot its projection $\vec{y}(t)\cdot\vec{y}(0)/|\vec{y}(0)|^2$ on the initial state, for which we have set all entries equal to one, i.e.,  $\vec{y}(0) = (1,1,\ldots, 1)^T$.     We observe a clear qualitative difference between the $c=2$ and $c=4$ trajectories of $\vec{y}(t)$: for $c=2$ the trajectory  is reminiscent of  a  harmonic oscillator in the underdamped regime while for $c=4$ the trajectory is reminiscent of a harmonic oscillator in the overdamped regime.

More specifically these oscillations emerge most clearly if 
the time scale $\tau_{\rm rel} $ that the system needs to relax to its fixed point 
\begin{equation}
\tau_{\rm rel} = \frac{1}{d - {\rm Re}(\lambda_1)} \ ,
\end{equation}      
is similar or larger than 
the transient time  $\tau_{\rm gap}$ at which the mode associated with the leading eigenvalue starts to dominate the dynamics, {\it i.e.}, $\tau_{\rm rel}\gtrsim \tau_{\rm gap}$. The timescale $\tau_{\rm gap}$ is given by the  inverse of the spectral gap,
\begin{equation}
\tau_{\rm gap} = \frac{1}{{\rm Re}(\lambda_1) - {\rm Re}(\lambda_{M+1})} \ ,
\end{equation}
where $M$ denotes the number of eigenvalues with a real part equal to  ${\rm Re}(\lambda_1)$, see Appendix~\ref{App:Oscil}. Hence ${\rm Re}(\lambda_{M+1})$ denotes the second largest value of the real parts of the eigenvalues of $\mathbf{A}$.
In this situation perturbed systems will show simple exponential recovery of the fixed point if $\lambda_1$ is real or barely visible oscillations if their period
\begin{equation}
\tau_{\rm oscil} = \frac{2\pi}{{\rm Im}(\lambda_1)} 
\end{equation}
is such that $\tau_{\rm oscil} \gg \tau_{\rm rel}$.
Otherwise for $\tau_{\rm oscil} \ll \tau_{\rm rel}$ a typical characteristic oscillatory dynamics of $\vec{y}(t)$ will emerge during relaxation towards equilibrium.
Finally for $\tau_{\rm rel}\ll \tau_{\rm gap}$ the dynamics of $\vec{y}(t)$ appears chaotic.

For large systems, where spectral gap is expected to be small and vanishing in the limit $N\rightarrow\infty$, $\tau_{\rm rel} >\tau_{\rm gap}$ can only be obtained at the verge of instability, {\it i.e.} for $d - {\rm Re}(\lambda_1)$ even smaller than the gap. 
In this setting, systems with $\lambda_1$ typically real, or that rely on finite size fluctuations of ${\rm Im}(\lambda_1)$ to have a large but finite $\tau_{\rm oscil}$, will mostly present a non-oscillatory relaxation dynamics.
This occurs for antagonistic systems characterised by connectivity $c=4>c_{\rm crit}$, an example of which is shown in Fig.~\ref{fig:Dynamics}, where for the instance with $\lambda_1$ real the gap is 0.03 ($\tau_{\rm gap}\sim30$) and it is 0.02 ($\tau_{\rm gap}=50$) in the other case. In this example already a 
$\tau_{\rm rel}=40$ lets the dominating mode emerge at around $t=60$, after a short transient.
Note that some oscillation is still visible in the second case, but they are expected to disappear completely for $N\rightarrow\infty$, as they are originated by finite size fluctuations of the leading eigenvalue.
On the contrary, antagonistic systems with $c=2<c_{\rm crit}$ and typically finite ${\rm Im}(\lambda_1)$ in the large size limit, will necessary be characterised by $\tau_{\rm oscil} \ll \tau_{\rm rel}$ as soon as $\tau_{\rm rel}\gtrsim\tau_{\rm gap}$ and therefore will typically present evident oscillatory dynamics towards equilibrium as in the example of Fig.~\ref{fig:Dynamics} where the gap is $0.02$, $\tau_{\rm gap}=50$ and again we have set $\tau_{\rm rel}=40$.

Interestingly, oscillatory recovery of the fixed point after a perturbation has been observed in experiments on brain functioning during wakefulness and not during sleep \cite{Massimini2228, rogasch2013assessing}.
Several models for different parts of the brain %({\it i.e.} cortex, cerebellum, or even retina tissues)
build on the interconnections of neurons through synapses \cite{brunel2021}, with progressive accent on the non symmetric nature of the interactions and the non trivial topology of the network, including the proposal that a change in the interconnection structure could be responsible of the lost of consciousness during sleep \cite{Massimini2228}. Remarkably some indirect evidence of this phenomenon has emerged in the different response of the brain to controlled external stimuli, showing longer and oscillatory recovery of the pre-stimulus stadium with long range repercussion during wake and a rapid monotonic and localised decay of the stimulus effect during sleep \cite{Massimini2228, rogasch2013assessing}.
The theoretical result discussed in this section in full generality for the schematic antagonistic model proposed provides a general and statistically sound explanation of the typical emergence of similar oscillatory dynamical behaviour for perturbed complex systems characterised by sparse interaction graphs and poised at the verge of instability as well as its typical absence when graph connectivity increases \cite{chaos, Sommers_et_al, marti2018correlations}.

\section{\label{sec:Theory} Theory for the spectra of infinitely large random matrices}
We  present now the theory that  we have used in Figs.~\ref{fig:ReLambda1VSN} and 
 \ref{fig:ImLambda1VSc} to determine the typical value  $\lambda^\ast_1$  of the leading eigenvalue  of infinitely large matrices.       
 
 First, we derive an exact expression for the spectral distribution 
  of infinitely large  random matrices  in the general model defined in Sec.~\ref{sec:generalModel}, which includes antagonistic and mixture matrices.   The spectral distribution of a sequence of square matrices $\mathbf{A}$ of increasing size $N$ is defined by
\begin{eqnarray}
\rho(z)= \lim_{N\rightarrow \infty}\frac{1}{N}\sum^N_{j=1}\delta(z-\lambda_j)
\end{eqnarray} 
for 
\begin{equation}
z = x + {\rm i}y \in \mathbb{C}
\end{equation}
and where 
\begin{equation}
\lambda_1,\lambda_2,\ldots,\lambda_N,
\end{equation}
are the $N$ complex roots of the algebraic equation
\begin{equation}
{\rm det}[\mathbf{A}-z \mathbf{1}_N]=0.   
\end{equation}    
Using the cavity method of Refs.~\cite{cavity_nonH,NeriMetz2012, Izaak_Metz_2019}, we   will derive a closed form expression for the spectral distribution $\rho(z)$ of random matrices in the general model defined in Sec.~\ref{sec:generalModel}.       Note that $\rho$ is in principle a random quantity, as $\mathbf{A}$ is random, but numerical evidence shows that in the limit of large $N$ the spectral distribution converges to a  deterministic limit~\cite{cavity_nonH,NeriMetz2012, Izaak_Metz_2019}.

Since we are mainly interested in the leading eigenvalue, we will focus on the support of $\rho$, which is defined as 
\begin{equation}
\mathcal{S} = \overline{\left\{z\in\mathbb{C}: \rho(z)\neq 0\right\}}
\end{equation}
where $\overline{\Phi}$ denotes the closure of a set $\Phi$.    The typical value of the leading eigenvalue follows from 
\begin{equation}
\lambda^\ast_1 = {\rm argmax}\left\{{\rm Re}(z):z\in\mathcal{S}\right\}, \label{eq:leading}
\end{equation}
where in Eq.~(\ref{eq:leading}) we use the convention, as before, to choose the value with the largest imaginary part in case there are several eigenvalues  with a maximum ${\rm Re}(z)$.
The Eq.\eqref{eq:leading} is only valid in case the leading eigenvalue belongs to the continuous part of the spectrum. This condition holds when the spectrum does not have outlier eigenvalues~\cite{neri2016eigenvalue}.
In particular, a deterministic outlier eigenvalue can be originated when Eq.~\eqref{eq:balanced} is not satisfied.  
Moreover, in the left-hand-side of Eq.~(\ref{eq:leading}), we have set the typical value $\lambda^\ast_1$ of $\lambda_1$.   This is necessary because the leading eigenvalue $\lambda_1$ is, on contrary to the support set $\mathcal{S}$, not a self-averaging quantity.   Indeed,  infinitely large random graphs  contain  cycles of finite length and these may create stochastic outlier eigenvalues in the spectrum that may also be leading eigenvalues,  as discussed in Ref.~\cite{Izaak_Metz_2019,ShortestCycleReimer2017}.   The cavity method    assumes that the ensemble does not contain cycles of finite length in the limit of large $N$.  This assumption  does not influence  the support set $\mathcal{S}$ because $\mathcal{S}$ does not depend on the stochastic outlier eigenvalues in the spectrum.   However, when considering the leading eigenvalue, there is a finite, albeit small, probability that the leading eigenvalue  is contributed by a cycle of finite length \cite{Izaak_Metz_2019} and therefore $\lambda_1$ is not a self-averaging quantity.    Although we cannot compute the contribution to $\lambda_1$ due to cycles, we can use  the cavity method together with Eq.~(\ref{eq:leading}) to  compute the most likely value $\lambda^\ast_1$ of the leading eigenvalue.

\subsection{Spectral distribution for locally tree-like matrices}
We start the theoretical analysis with revisiting the cavity method of Refs.~\cite{cavity_nonH,NeriMetz2012, Izaak_Metz_2019} for the  spectral distribution $\rho$ of random matrices that are locally tree-like.
We say that a random matrix ensemble is locally tree-like if with probability one  in the limit $N\rightarrow \infty$ the finite neighbourhood of a randomly selected node is a tree~\cite{dembo2010gibbs}.       
We note that  an alternative method \cite{Reimer_sparse,EquivalenceCavReplica2011}  based on replica theory could be used to 
 %generalised to asymmetric matrices to 
 obtain  similar results.

As shown in Refs.~\cite{cavity_nonH,NeriMetz2012, Izaak_Metz_2019}, the spectral distribution of a locally tree-like random matrix is given by 
\begin{equation}
\rho(z)=\lim_{\eta\to 0^+}\lim_{N\to \infty}\frac{1}{\pi	N}\sum_{j=1}^N \frac{{\rm d}}{{\rm d}\overline{z}}[\mathsf{G}_j]_{21},\label{eq:densityCavity}
\end{equation}
where 
\begin{equation}
\frac{{\rm d}}{{\rm d}\overline{z}} = \frac{1}{2} \left(\frac{{\rm d}}{{\rm d}x} + {\rm i}\frac{{\rm d}}{{\rm d}y}\right) ,
\end{equation}
and where the $2\times 2$ matrices $\mathsf{G}_j$  satisfy the  relations 
\begin{align}
\mathsf{G}_j&=\left( \mathsf{z}_\eta - \sum_{k\in \partial_j} \mathsf{J}_{jk} \mathsf{G}^{(j)}_k\mathsf{J}_{kj}  \right)^{-1}. \label{eq:CavityEq2Main}
\end{align}  
In Eq.~(\ref{eq:CavityEq2Main}),  we have used the notation 
\begin{equation}
\partial_i = \left\{j: C_{ij}\neq 0 \right\}
    \end{equation}
    for the neighborhood of node $i$,    and we have also used
\begin{equation}\label{eq:J_jk}
 \mathsf{z}_\eta= \begin{pmatrix}
-\mathrm{i}\eta & z \\ \bar{z}& -\mathrm{i}\eta
\end{pmatrix}, \quad {\rm and} \quad \mathsf{J}_{jk}  =\begin{pmatrix}
0 & J_{jk} \\ \bar{J}_{kj}& 0
 \end{pmatrix}. 
 \end{equation}
The matrices $\mathsf{G}_j^{(\ell)}$ on the right-hand side of  Eq.~(\ref{eq:CavityEq2Main}) are $2\times 2$  matrices of complex numbers that satisfy  the recursion relations
\begin{align}
\mathsf{G}_j^{(\ell)}&=\left( \mathsf{z}_\eta  - \sum_{k\in \partial_j\setminus \{\ell\}} \mathsf{J}_{jk} \mathsf{G}^{(j)}_k\mathsf{J}_{kj}  \right)^{-1},\label{eq:CavityEq1Main}
\end{align}   
for each $\ell\in\partial_j$
.

The Eqs.~(\ref{eq:CavityEq2Main})   and (\ref{eq:CavityEq1Main}) are  relations between random variables defined on a  locally tree-like matrix.   In the next section,  we  derive a set of recursive distributional equations for infinitely large matrices  drawn from the general model defined in Sec.~\ref{sec:generalModel}.

\subsection{Spectral distribution for the general model of Sec.~\ref{sec:generalModel}}\label{sec:specDistri}
 Since the general random-matrix model in Sec.~\ref{sec:generalModel} is defined on random graphs with a prescribed degree distribution, it is a locally tree-like ensemble \cite{dembo2010gibbs} and  the cavity method thus applies.

 We use the  Eqs.~(\ref{eq:CavityEq2Main}) and (\ref{eq:CavityEq1Main})  to derive a selfconsistent set of equations in the distributions 
 \begin{equation}
\tilde{q}(\mathsf{g}) := \lim_{N\rightarrow \infty}\frac{1}{N}  \sum^N_{i=1}\delta(\mathsf{g}-\mathsf{G}_i)
 \end{equation}
 and 
\begin{equation}
q(\mathsf{g}) := \lim_{N\rightarrow \infty}\frac{1}{cN}  \sum^N_{i=1}\sum_{\ell\in\partial_i}\delta(\mathsf{g}-\mathsf{G}^{(\ell)}_i).  
 \end{equation} 
Since for the general model defined in  Sec.~\ref{sec:generalModel} the  random variables on the right hand side of  Eqs.~(\ref{eq:CavityEq2Main}) and (\ref{eq:CavityEq1Main})  are independent, we obtain the recursive distributional equations
\begin{align}
& \tilde{q}(\mathsf{g})=\sum_{k=0}^{\infty} p_{\rm deg}(k)\int \prod_{\ell=1}^{k}\mathrm{d} \mathsf{g}_\ell  q(\mathsf{g}_\ell) \nonumber\\ 
&\times \int \prod_{\ell=1}^{k}p(u_\ell,l_\ell)\mathrm{d}u_\ell \mathrm{d}l_\ell   \nonumber\\
& \times\delta\left(\mathsf{g}-\left(\mathsf{z}_\eta- \sum_{\ell=1}^{k}\mathsf{J}_\ell \mathsf{g}_\ell \mathsf{J}_\ell^\dagger  \right)^{-1}    \right)\label{eq:rec1}
\end{align}
and 
\begin{align}
& q(\mathsf{g})=\sum_{k=1}^{\infty}\frac{kp_{\rm deg}(k)}{c}\int \prod_{\ell=1}^{k-1}\mathrm{d} \mathsf{g}_\ell  q(\mathsf{g}_\ell) \nonumber\\ 
&\times \int \prod_{\ell=1}^{k-1}p(u_\ell,l_\ell)\mathrm{d}u_\ell \mathrm{d}l_\ell   \nonumber\\
& \times\delta\left(\mathsf{g}-\left(\mathsf{z}_\eta- \sum_{\ell=1}^{k-1}\mathsf{J}_\ell \mathsf{g}_\ell \mathsf{J}_\ell^\dagger  \right)^{-1}    \right), \label{eq:rec2}
\end{align} 
where
\begin{equation} \label{eq:J_muSec4}
     \mathsf{J}_{\ell}  =\begin{pmatrix}
0 & u_\ell \\ \overline{l}_\ell& 0
 \end{pmatrix}.
\end{equation}
The distribution $\tilde{q}(\mathsf{g})$ provides us with the  spectral distribution, which admits according to Eq.~(\ref{eq:densityCavity}) the expression
\begin{equation}
\rho(z)=\lim_{\eta\to 0^+}\frac{1}{\pi}\frac{{\rm d}}{{\rm d}\overline{z}}\int \mathrm{d}\mathsf{g} \: \tilde{q}(\mathsf{g})\: [\mathsf{g}]_{21}. \label{eq:specrec}
\end{equation}

Since we are mainly interested in the leading eigenvalue $\lambda_1$, we discuss in the next section how to obtain the boundary of the support set $\mathcal{S}$ of $\rho$.

\subsection{Support of the spectral distribution}\label{sec:boundaryTheory}
In this section, we derive an equation for   for the support set $\mathcal{S}$ of the spectral distribution $\rho$.     
We use the fact that the Eqs.~(\ref{eq:rec1}-\ref{eq:specrec})  admit the so-called trivial solution for which
\begin{eqnarray}
 \tilde{q}(\mathsf{g}) =  \int \mathrm{d}g \:  \tilde{q}^{(0)}(g) \delta\left(\mathsf{g} - \left(\begin{array}{cc}0 & -\overline{g} \\ -g &0 \end{array}\right)\right) \label{eq:trivial1}
\end{eqnarray}
and
\begin{eqnarray}
{q}(\mathsf{g}) =  \int \mathrm{d}g \:q^{(0)}(g) \delta\left(\mathsf{g} - \left(\begin{array}{cc}0 & -\overline{g} \\ -g &0 \end{array}\right)\right).\label{eq:trivial2}
\end{eqnarray}
Substituting Eqs.~(\ref{eq:trivial1}) and (\ref{eq:trivial2}) into the  Eqs.~(\ref{eq:rec1}) and (\ref{eq:rec2}), we obtain that 
the $\tilde{q}^{(0)}$ and $q^{(0)}$ solve the equations 
\begin{align}
& \tilde{q}^{(0)}(g)=\sum_{k=0}^{\infty} p_{\rm deg}(k)\int \prod_{l=1}^{k}\mathrm{d}g_\ell  q^{(0)}(g_\ell) \nonumber\\ 
&\times \int \prod_{l=1}^{k}p(u_\ell,l_\ell)\mathrm{d}u_\ell \mathrm{d}l_\ell  \nonumber\\
& \times\delta\left(g+\left(z+\displaystyle \sum_{\ell=1}^{k}u_\ell g_\ell {l}_\ell\right)^{-1} \right), \label{eq:rec1_zero}
\end{align}
and 
\begin{align}
& q^{(0)}(g)=\sum_{k=1}^{\infty} \frac{k p_{\rm deg}(k)}{c}\int \prod_{\ell=1}^{k-1}\mathrm{d}g_\ell  q^{(0)}(g_\ell) \nonumber\\ 
&\times \int \prod_{\ell=1}^{k-1}p(u_\ell,l_\ell)\mathrm{d} u_\ell \mathrm{d} l_\ell   \nonumber\\
& \times\delta\left(g+\left(z+\displaystyle \sum_{\ell=1}^{k-1}u_\ell g_\ell {l}_\ell\right)^{-1} \right). \label{eq:rec1_zero}
\end{align}
In addition, substituting the trivial solution into Eq.~\eqref{eq:specrec} for the spectral distribution, we obtain that
$\rho(z) = 0$, and therefore the trivial solution holds for values of $z\notin \mathcal{S}$.

In order to obtain the boundary of the support $\mathcal{S}$ of the spectral distribution, we perform a linear stability analysis of the  Eqs.~(\ref{eq:rec1}-\ref{eq:rec2}) around the trivial solution given by Eq.~(\ref{eq:trivial2}).   We consider a  perturbation 
\begin{equation}
q(\mathsf{g}) =  \int {\rm d}g  \int {\rm d}h \int {\rm d}h'\: Q(g,h,h') \: \delta\left(\mathsf{g} - \left(\begin{array}{cc}h & -\overline{g} \\ -g &h' \end{array}\right)\right) \label{eq:perturb} 
\end{equation} 
around the trivial solution Eq.~(\ref{eq:trivial2}),
where  the
\begin{eqnarray}
\int {\rm d}g  \int {\rm d}h \int {\rm d}h'  \:  h^n (h')^{m}   Q(g,h, h') \in \mathcal{O}\left(\epsilon^{n+m}\right) \label{eq:expansion}\end{eqnarray}
are assumed to be  of order  $\mathcal{O}\left(\epsilon^{n+m}\right)$  and where $\epsilon\ll 1$ is a small number that quantifies the strength of the perturbation around the trivial solution.  
 
 In Appendix~\ref{sec:Appendix_StabAnal_Boundary}, we show that 
Eqs.~(\ref{eq:rec2}), (\ref{eq:perturb}) and (\ref{eq:expansion})  imply up to order $O(\epsilon)$ that 
\begin{align}
& Q(g,h) =\sum_{k=1}^{\infty}\frac{kp_{\rm deg}(k)}{c}\int \prod_{\ell=1}^{k-1} \mathrm{d}g_\ell \mathrm{d} h_\ell Q(g_{\ell},h_{\ell})  \nonumber\\ 
&\times \int \prod_{\ell=1}^{k-1}p(u_\ell,l_\ell)\mathrm{d}u_\ell \mathrm{d}l_\ell   \nonumber\\
&\times \delta\left(h - |g|^2 \sum_{\ell=1}^{k-1} h_\ell {l}_\ell^2  \right) \nonumber\\  
& \times\delta\left(g+\left[z+ \sum_{\ell=1}^{k-1}u_\ell g_\ell {l}_\ell\right]^{-1}  \right), \label{eq:StabBoundary}
\end{align}
where 
\begin{equation}
Q(g,h) = \int {\rm d} h' Q(g,h, h') .
\end{equation}

Since the Eq.~(\ref{eq:StabBoundary}) is obtained through a linear stability analysis of the recursive distributional equations (\ref{eq:rec2}) at the trivial solution Eq.~(\ref{eq:trivial2}), we obtain that   
$z\notin \mathcal{S}$
if 
\begin{eqnarray}
Q(g,h) = q^{(0)}(g)\delta(h), \label{eq:q0Delta}
\end{eqnarray}
is a stable solution of the Eq.~(\ref{eq:StabBoundary}), where $q^{(0)}$ solves the relations in Eq.~(\ref{eq:rec1_zero}).   On the other hand, if  
$z\in \mathcal{S}$
then the  Eq.~(\ref{eq:StabBoundary}) will not be stable at the trivial solution given by Eq.~(\ref{eq:q0Delta}).    Hence, the boundary of the support set $\mathcal{S}$ is given by the edge of stability of  Eq.~(\ref{eq:StabBoundary}) at   the fixed point solution (\ref{eq:q0Delta}).

There exist two limiting case for which we can obtain the edge of stability analytically.  First, there is the case of a highly connected graph with $c\rightarrow \infty$, which we will discuss later.    Second, there is the case of matrices with oriented interactions for which $p$ is of the form given by Eq.~(\ref{eq:oriented}).  In this case,  we recover that 
\begin{equation}
 \mathcal{S} = \left\{z\in \mathbb{C}: 
\left|z \right|^2\leq \frac{\left\langle k (k-1) \right\rangle_{p_{\rm deg}}  }{2c}\left\langle u^2 \right\rangle_{\tilde{p}}.
  \right\} ,
 \end{equation}
  which is consistent with the results in
 Ref.~\cite{neri2019spectral}, as we show in Appendix~\ref{sec:App_OrientedBoundary}.

For antagonistic and mixture random matrices, it is difficult to make analytical progress.  However, we can determine the edge of stability of Eq.~(\ref{eq:StabBoundary}) at   the fixed point solution (\ref{eq:q0Delta}), and thus also the support set $\mathcal{S}$, with a population dynamics algorithm that we describe in detail in  Appendices~\ref{sec:Appendix_PopDyn} and \ref{sec:App_SpectralDensityResults}.       We have used this algorithm to determine the real part and imaginary part of $\lambda^\ast_1$ in Figs.~\ref{fig:ReLambda1VSN} and 
 \ref{fig:ImLambda1VSc}.  In the following section, we use this algorithm to determine the support set $\mathcal{S}$ of both antagonistic and mixture random matrices.

\section{Spectra of antagonistic and mixture matrices}\label{sec:BoundaryResults}
%BELOW THE PROPOSAL OF PUTTING c=2 AND c=4 IN THE SAME FIGURE
\begin{figure*}[htbp]
\centering
\subfigure[$N=10$]{\includegraphics[width=0.17\textwidth]{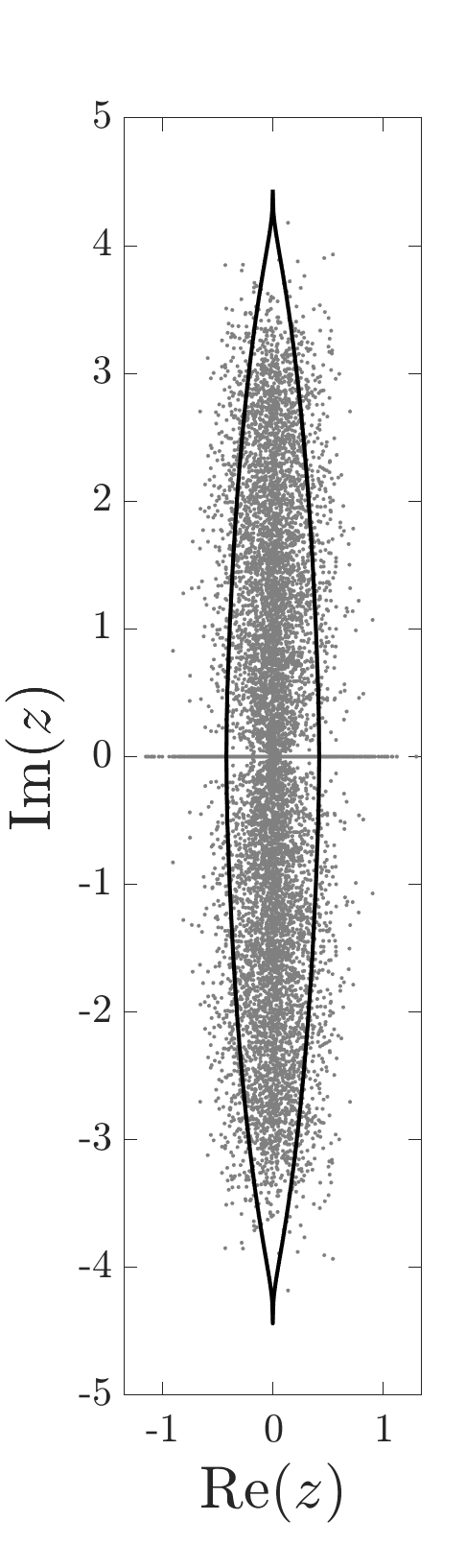}
\label{subfig:AntPoissN10S1000c4}}
\subfigure[$N=10^2$]{\includegraphics[width=0.17\textwidth]{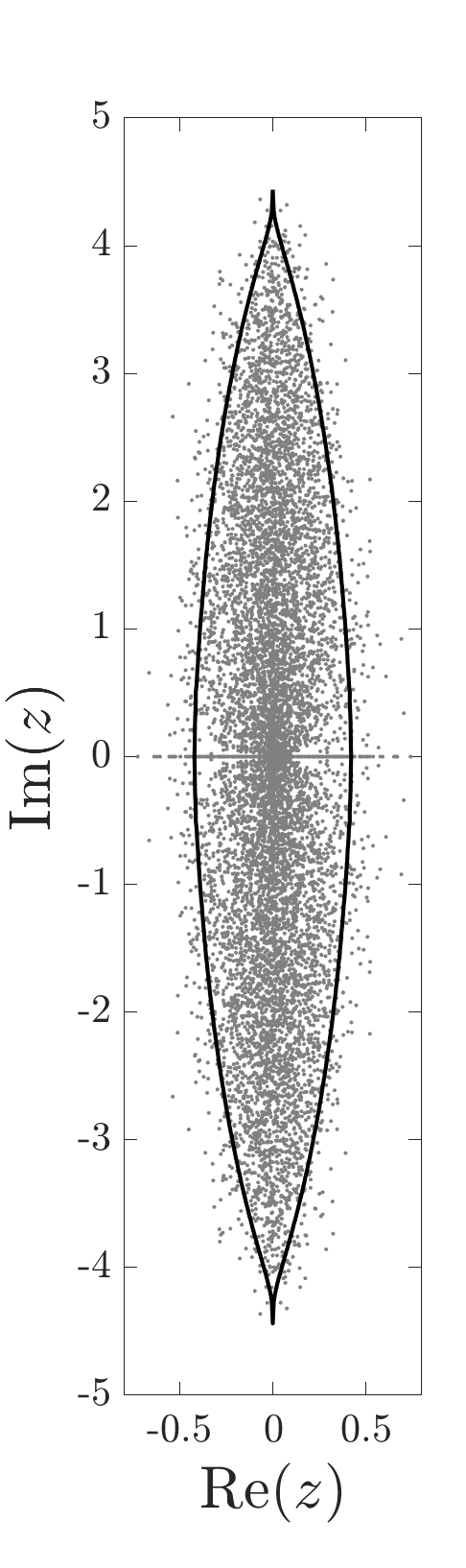}
\label{subfig:AntPoissN100S100c4}}
\subfigure[$N=10^3$]{\includegraphics[width=0.17\textwidth]{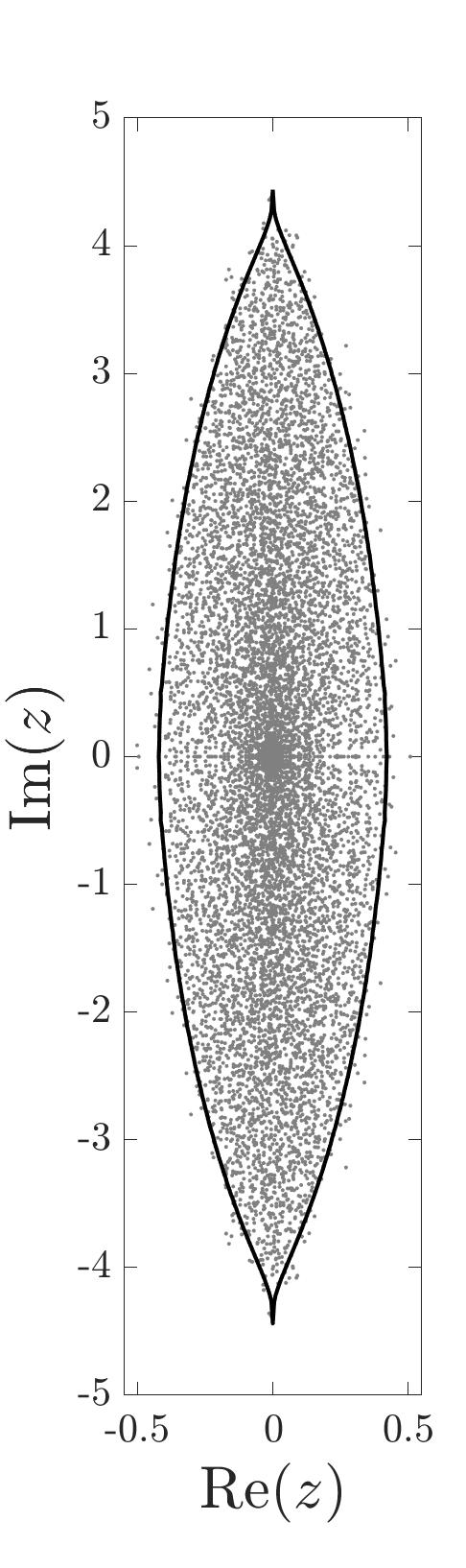}
\label{subfig:AntPoissN1000S10c4}}
\subfigure[$N=10^4$]{\includegraphics[width=0.17\textwidth]{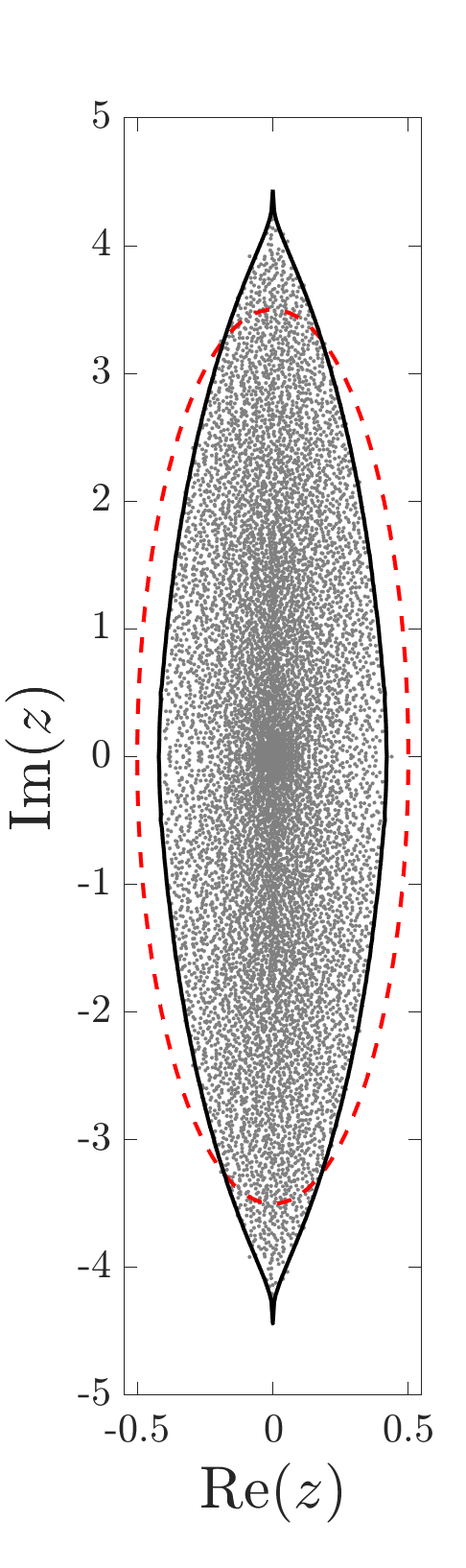}\label{subfig:AntPoissN10000S1c4Im0}}
\\
\subfigure[$N=10$]{\includegraphics[width=0.17\textwidth]{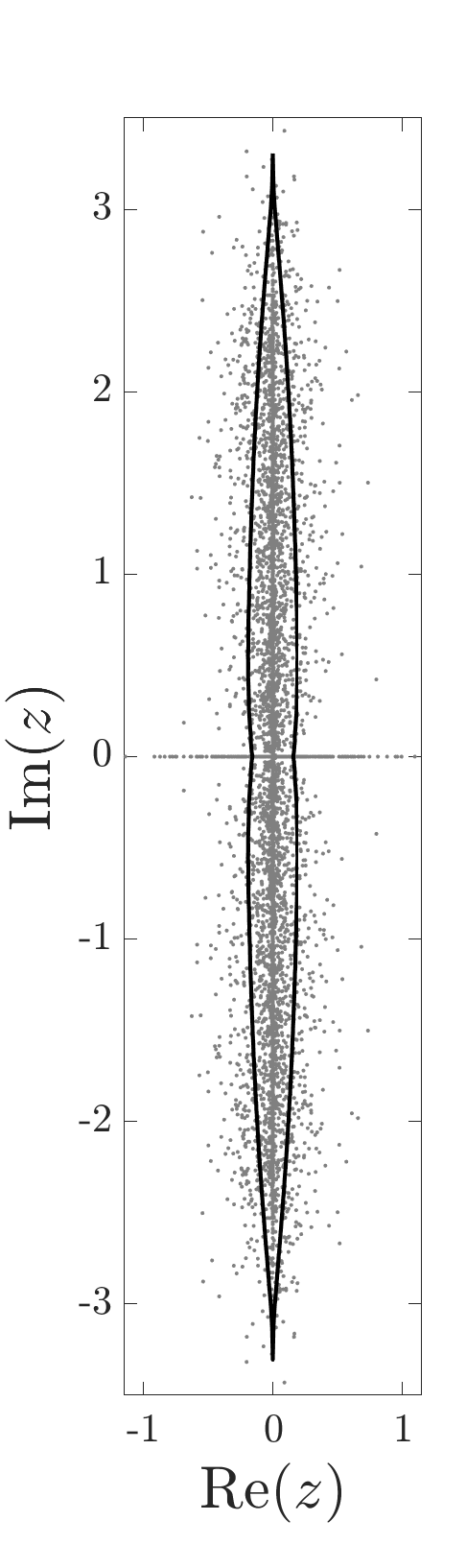}
\label{subfig:AntPoissN10S1000c2}}
\subfigure[$N=10^2$]{\includegraphics[width=0.17
\textwidth]{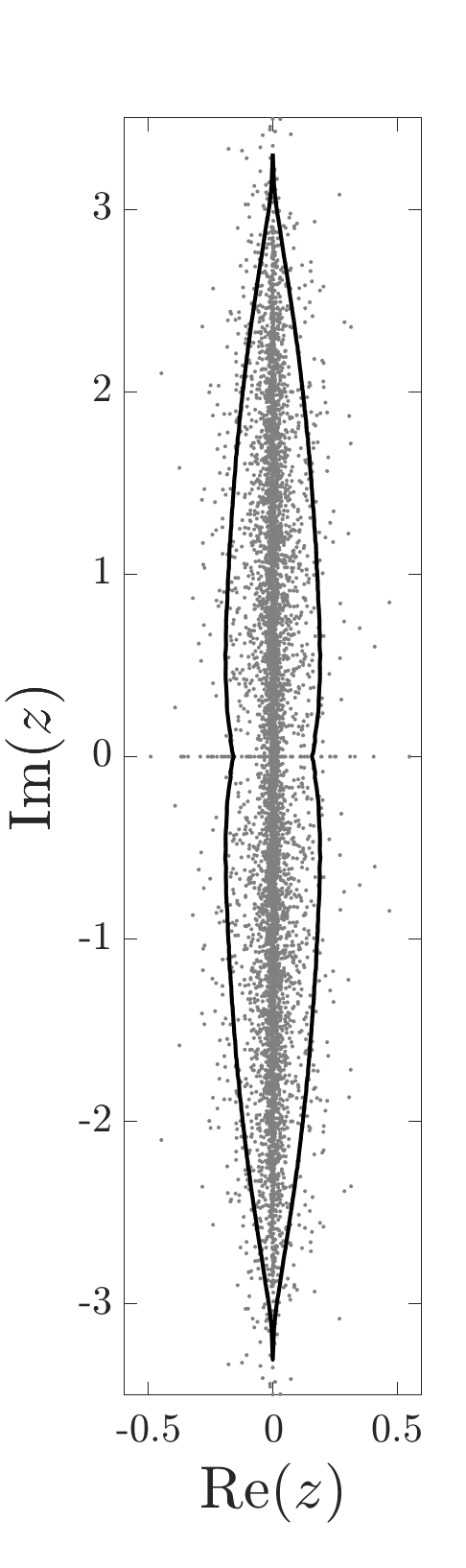}
\label{subfig:AntPoissN100S100c2}}
\subfigure[$N=10^3$]{\includegraphics[width=0.17\textwidth]{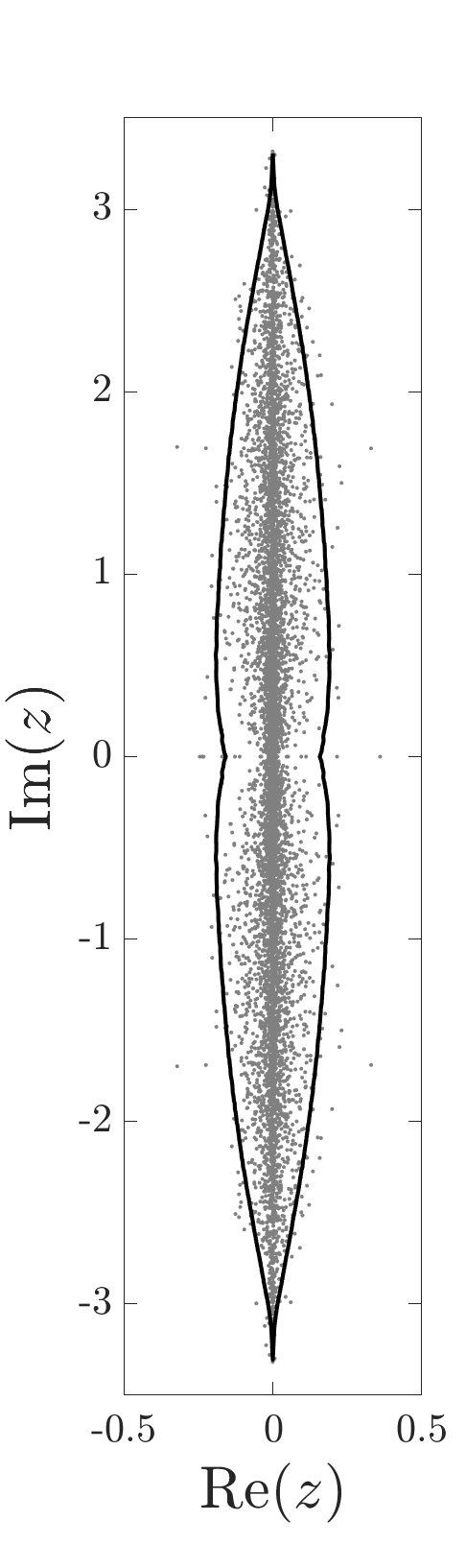}
\label{subfig:AntPoissN1000S10c2}}
\subfigure[$N=10^4$]{\includegraphics[width=0.17\textwidth]{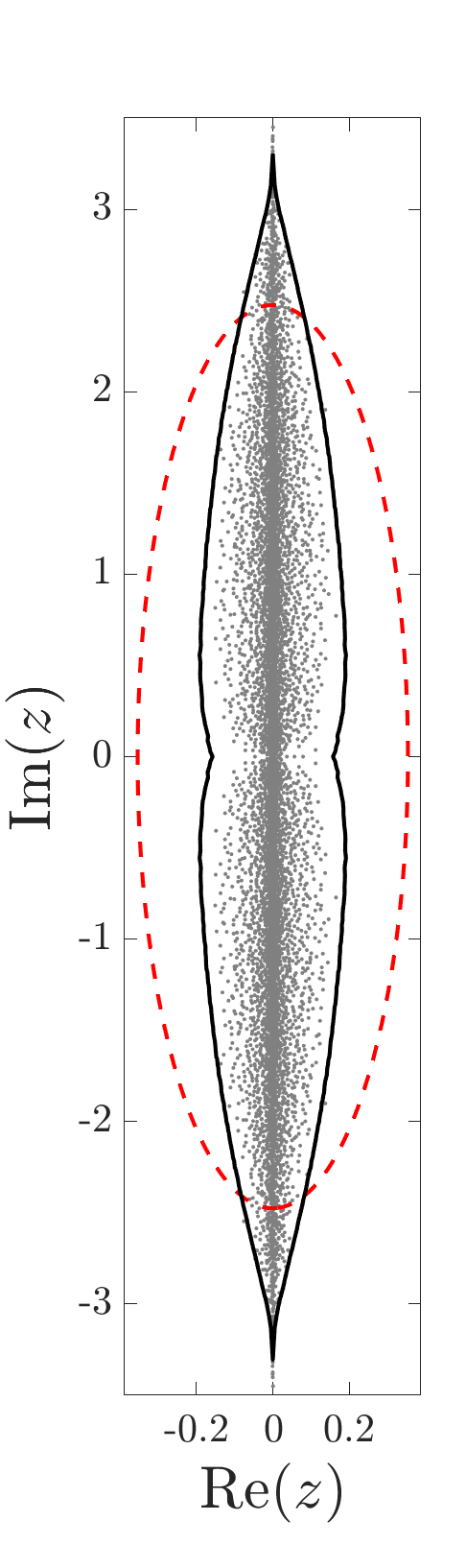}\label{subfig:AntPoissN10000S1c2Im0p77}}
 \caption{Spectra of antagonistic  random matrices on Erd\H{o}s-R\'{e}nyi graphs (Model A  in Sec.~\ref{sec:relEx})  with mean degree $c=4$ (panels from (a) to (d)) and $c=2$ (panels from (e) to (h)). Grey markers are the eigenvalues of  $m_{\rm s}$ matrices of size $N$, with $m_{\rm s}=10^4/N$, that are randomly drawn from this ensemble and are obtained through direct diagonalization routines.    Continuous black lines are theoretical results for $N\rightarrow \infty$ obtained with the cavity theory of  Sec.~\ref{sec:Theory} solved by population dynamics algorithm with population size $N_p=25000$, note that the boundary would not change using larger $N_p$. Red dashed lines shown in  panels (d) and (h) represent the boundary of the elliptic law given by Eqs.~(\ref{eq:elliptic1}-\ref{eq:tau_elliptic_main})  with $\sigma^2= c$ and $\tau  = -3c/4$. 
 }
\label{fig:BoundaryFiniteNc2_4Poiss}
\end{figure*}    
In Sec.~\ref{sec:MainResults}, we have  found  that the   leading eigenvalues of antagonistic and mixture matrices behave in qualitatively different ways.    As a consequence, we expect  that also the spectra of these ensembles  are qualitatively different.      Therefore,  in this section we analyse the spectra of antagonistic and mixture ensembles and provide a holistic view on the results for the leading eigenvalue in  Sec.~\ref{sec:MainResults}.
\subsection{Antagonistic  matrices}   
We first consider the spectra of antagonistic matrices given by Model A, defined in Sec.~\ref{sec:relEx} and whose leading eigenvalue results are shown in Figs.~\ref{fig:ReLambda1VSN}-\ref{fig:ImLambda1VSc}. 
Figure~\ref{fig:BoundaryFiniteNc2_4Poiss}  presents the spectra of antagonistic  matrices   on   graphs with mean degrees $c=4$ and $c=2$.   Each panel  shows $10^4$ eigenvalues obtained from diagonalizing      $m_{\rm s} = 10^4/N$ matrices.      In addition to these results from numerical experiments, the plot also shows  the boundary of the spectrum for $N$ infinitely large, which is obtained with  the cavity theory  in Sec.~\ref{sec:boundaryTheory}.  

We observe a very good correspondence between  theory and  numerical results  when $N\approx 10^4$,  while for smaller $N$ deviation appears, due to 
fluctuations in the spectral properties for matrices of finite size.   For example, from  Fig.~\ref{fig:BoundaryFiniteNc2_4Poiss} it appears that the leading eigenvalue at  low values of $N=10$  is larger than  the leading eigenvalue at  $N=10^4$.  However, as shown in Fig.~\ref{fig:ReLambda1VSN},  the average of the leading eigenvalue is independent of $N$, and therefore what we observe in  Fig.~\ref{fig:BoundaryFiniteNc2_4Poiss}  are sample-to-sample fluctuations, which are significant for small system sizes $N$.     
In Appendix~\ref{sec:App_SpectralDensityResults}, we compare theoretical results for the   spectral distribution $\rho(z)$ with histograms of eigenvalues at  finite $N$, and we obtain again an excellent agreement between theory and numerical experiments, which further corroborates the theory.

The most striking feature observed in Fig.~\ref{fig:BoundaryFiniteNc2_4Poiss} is the qualitative difference between $c=4$ and $c=2$ in the boundaries of the spectra of antagonistic matrices.    For $c=4$, the boundary of the  spectrum  has a  shape similar to the elliptic law given by Eqs.~(\ref{eq:elliptic1}-\ref{eq:tau_elliptic_main}), while for  $c=2$ a qualitatively new feature appears in the profile of the boundary: at the intersection with the real axis it shows negative curvature accompanied by a rarefaction of eigenvalues in correspondence of the real axis (excepts from some noteworthy finite size effects).
This phenomenon, hereafter called {\it reentrance} behaviour or reentrance effect, corresponds to the fact that typically, for large systems, the leading eigenvalue of the spectrum is not real but it is a pair of complex conjugate numbers with finite imaginary parts. 
The qualitative change observed when comparing the boundary of the spectra at $c=4$ and $c=2$ provides a holistic view on the phase transition at $c_{\rm crit} \approx 2.75$, depicted in Fig.~\ref{fig:ImLambda1VSc} for the imaginary part of the leading eigenvalue, which is real for $c=4>c_{\rm crit}$, while for  $c=2<c_{\rm crit}$ it has a finite nonzero imaginary part because of the reentrance  effect in the spectrum boundary. Hence, the phase transition in Fig.~\ref{fig:ImLambda1VSc} reflects a qualitative change in the spectrum of antagonistic matrices at low $c$.

\begin{figure*}[htbp]
\centering
\subfigure[$N=10$]{\includegraphics[width=0.17\textwidth]{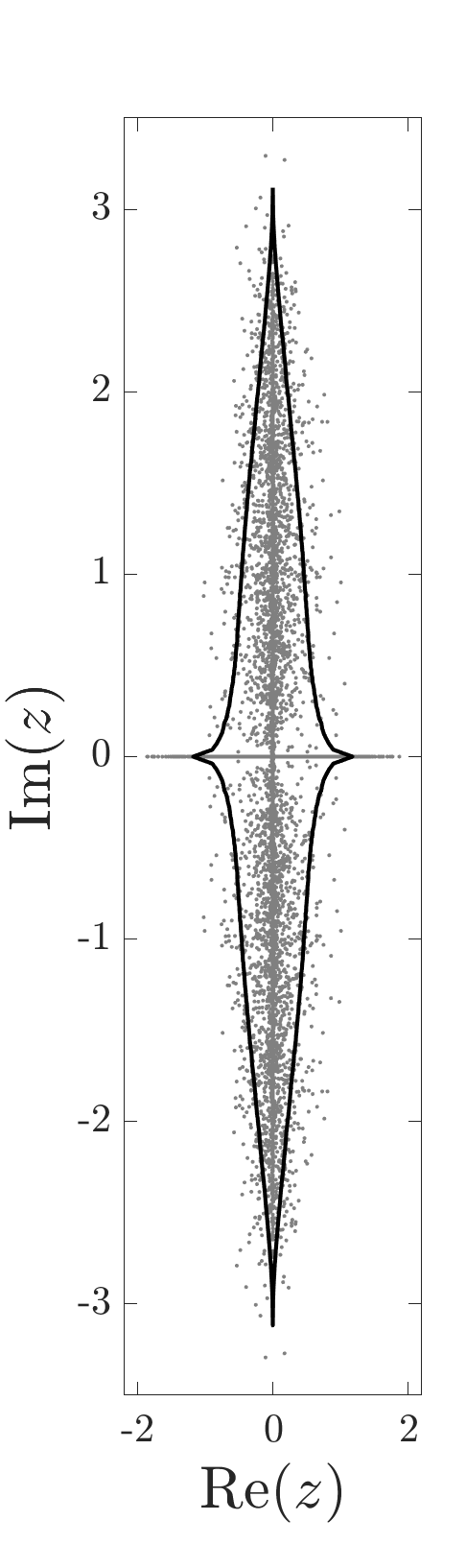}
\label{subfig:MixtPoissN10S1000c2}}
\subfigure[$N=10^2$]{\includegraphics[width=0.17\textwidth]{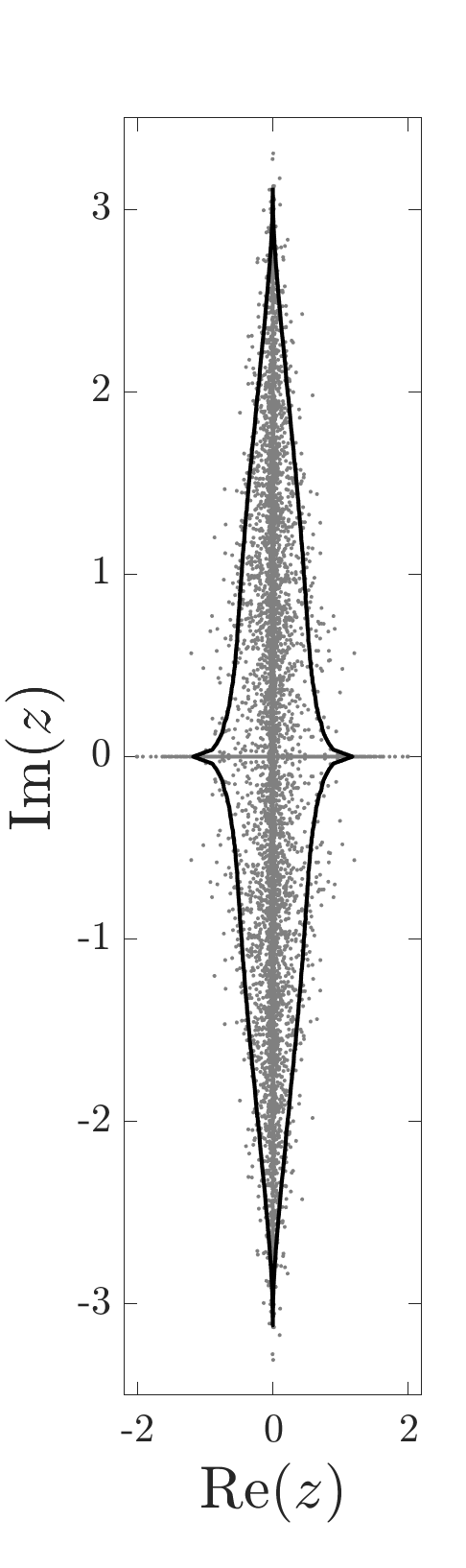}
\label{subfig:MixtPoissN100S100c2}}
\subfigure[$N=10^3$]{\includegraphics[width=0.17\textwidth]{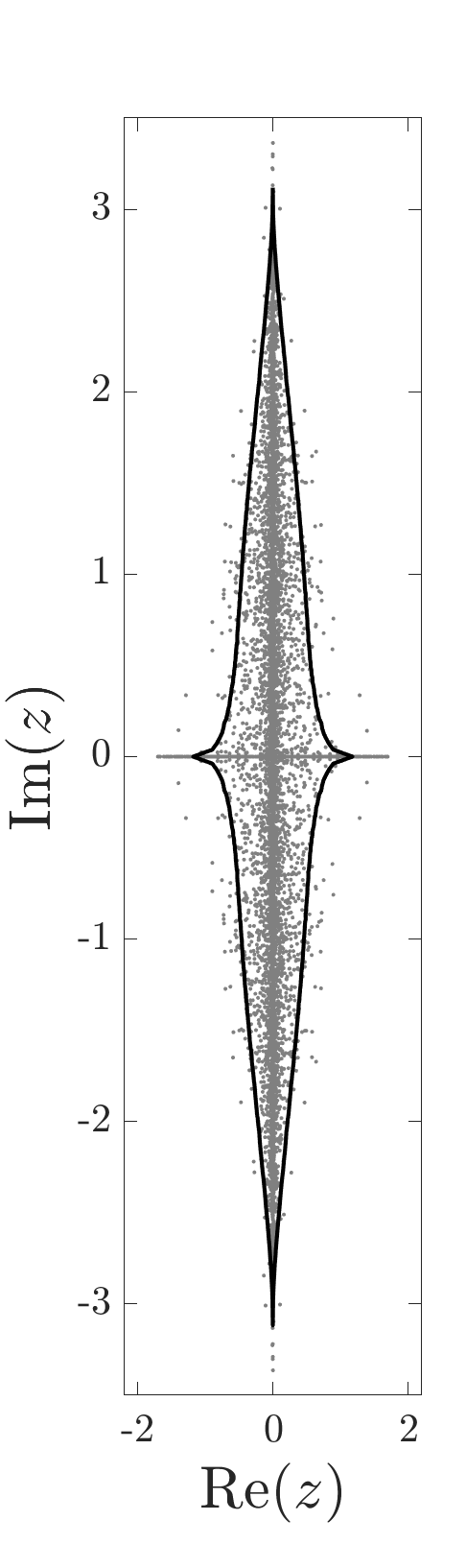}
\label{subfig:MixtPoissN1000S10c2}}
\subfigure[$N=10^4$]{\includegraphics[width=0.17\textwidth]{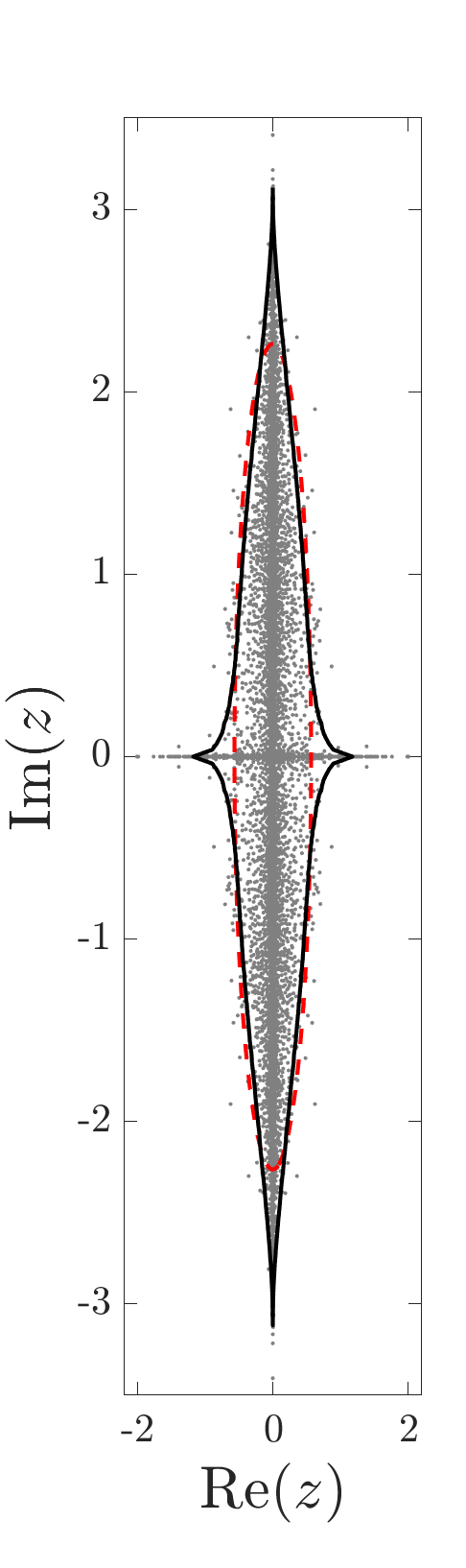}\label{subfig:MixtPoissN10000S1c2}}
 \caption{ Spectra of mixture  random matrices on Erd\H{o}s-R\'{e}nyi graphs with mean degree $c=2$ (Model B in Sec.~\ref{sec:relEx}).
Grey markers are the eigenvalues of  $m_{\rm s}$ matrices of size $N$, with $m_{\rm s}=10^4/N$,  randomly drawn from this ensemble and obtained through direct diagonalization routines.   Continuous black lines are theoretical results for $N\rightarrow \infty$ obtained with the cavity theory of  Sec.~\ref{sec:Theory} solved by population dynamics algorithm with population size $N_p=25000$, note that the boundary will change using larger $N_p$, see Fig.~\ref{fig:ComparisonAntMixture_LogEpsVSReLambda} in Appendix~\ref{sec:Appendix_PopDyn}.    Red dashed lines shown in  panel (d)  represents the boundary of the elliptic law given by Eqs.~(\ref{eq:elliptic1}-\ref{eq:tau_elliptic_main})  with $\sigma^2= c$ and $\tau  = -3c/5$. }
\label{fig:BoundaryFiniteNc2Poiss}
\end{figure*}

\begin{figure}[htbp]
    \centering
    \subfigure[Antagonistic]{\includegraphics[width=0.5\textwidth]{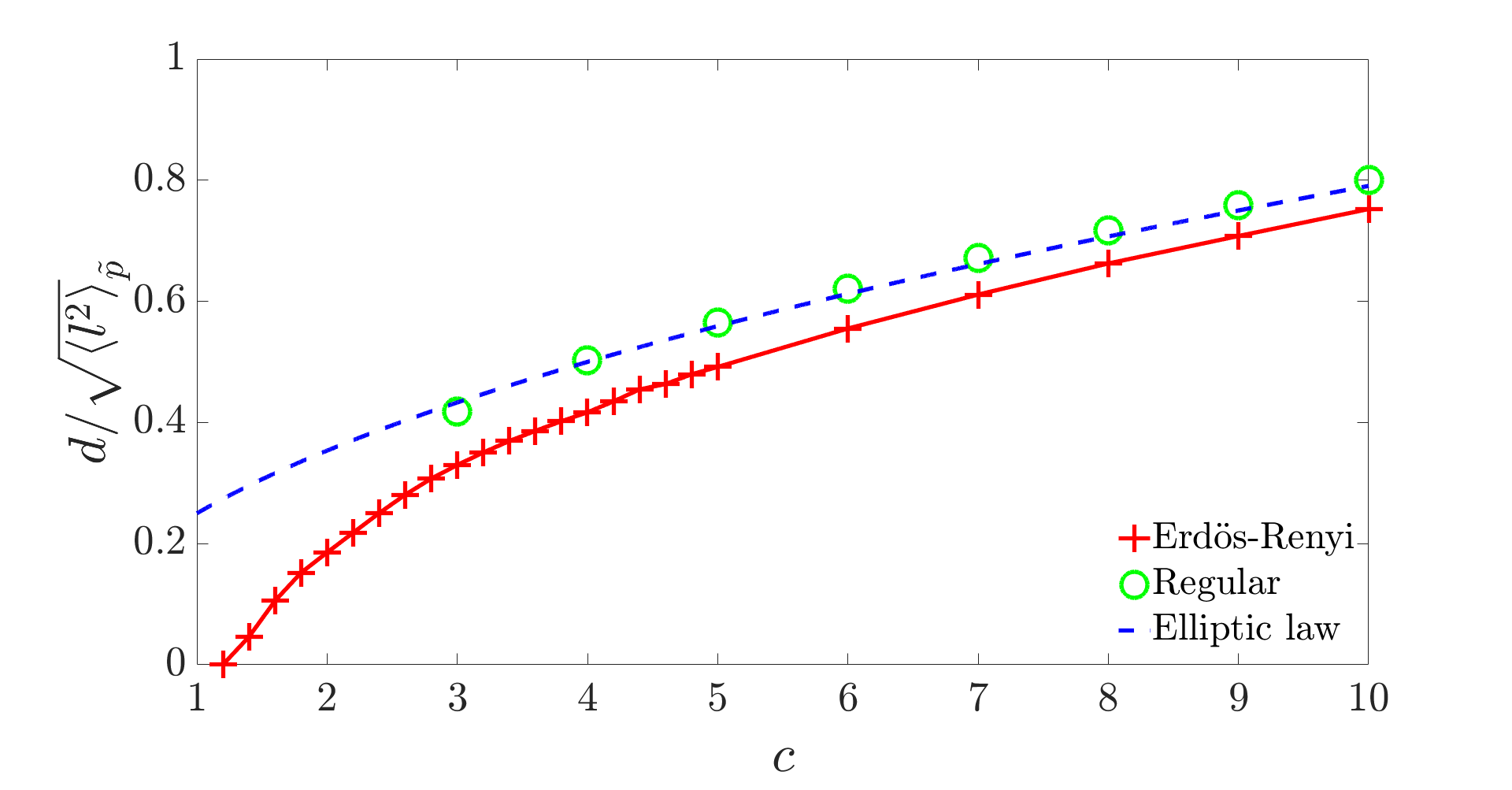}}
    \put(-150,50){{\bf Unstable}}
    \put(-200,100){{\bf Stable}} \\
    \subfigure[Oriented]{\includegraphics[width=0.5\textwidth]{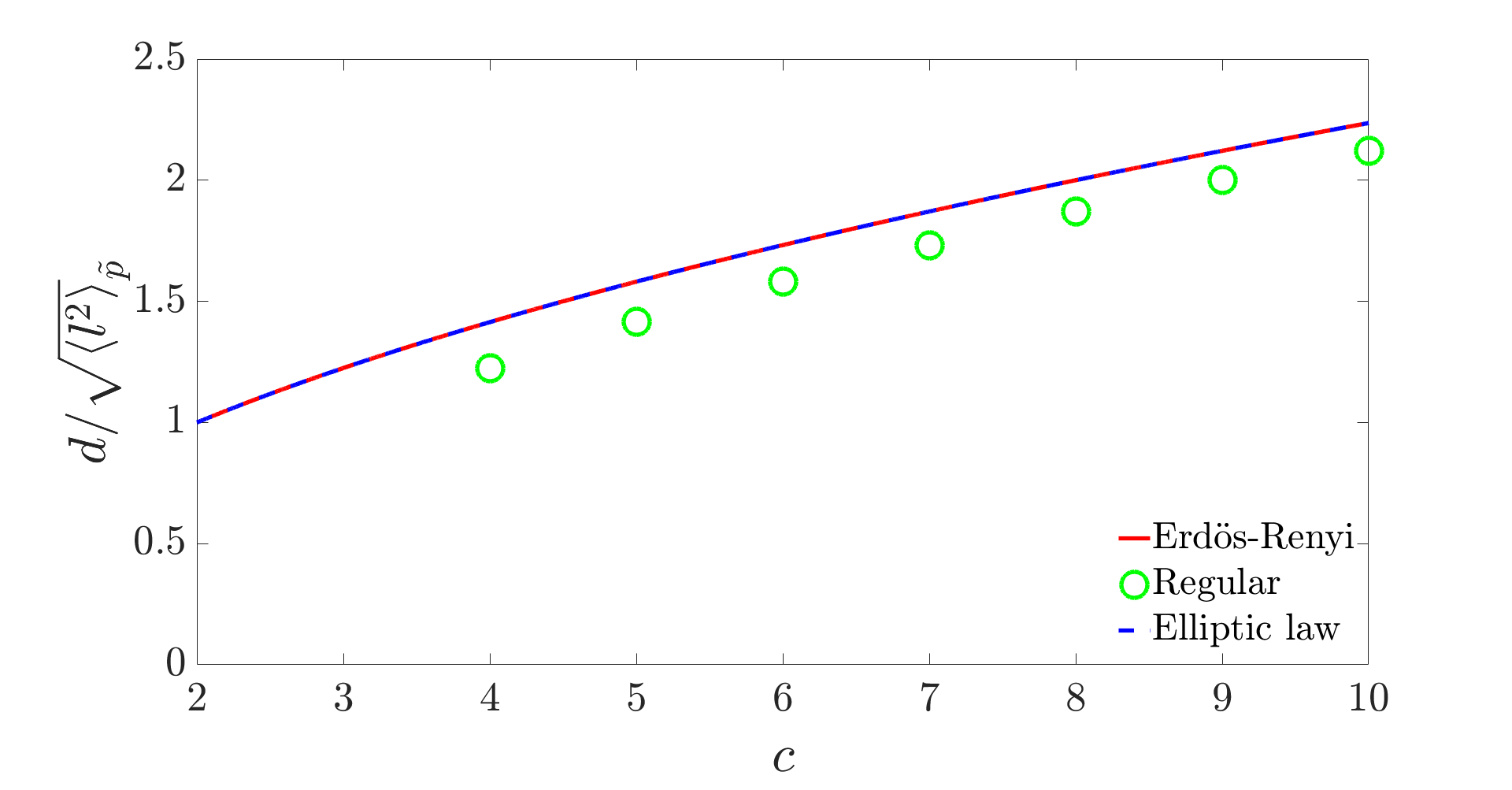} \label{subfig:ReLambda1VSc_Oriented}}
    \put(-150,50){{\bf Unstable}}
    \put(-200,100){{\bf Stable}}
    \caption{
  Phase diagrams for the linear stability of fixed points for antagonistic  matrices   [Panel (a)]  and oriented matrices [Panel (b)], both defined on either  Erd\H{o}s-R\'{e}nyi  or  regular graphs.  The lines denote the edge of stability given by the values of the system parameters for which the leading eigenvalue of $\textbf{M}$ has null real part, which is obtained where $d={\rm Re}(\lambda^\ast_1)$, as ${\rm Re}(\lambda^\ast_1)$ is the real part of the typical leading eigenvalue of the coupling matrix {\textbf{A}}.
      For antagonistic matrices $\tilde{p}$     is given by  Eq.~(\ref{eq:uniform}) and for oriented matrices $\tilde{p}$ is an arbitrary distribution with unit variance and zero mean. 
    Panel (a): predictions  from the theory in Sec.~\ref{sec:Theory}  (markers) are compared with   the elliptic law   given by Eqs.~(\ref{eq:elliptic1}-\ref{eq:tau_elliptic_main})  (dashed blue line).  The red line connecting the red crosses is a guide to the eye.   Panel (b): analytical predictions from Eq.~(\ref{eq:leadOriented}) are compared with the elliptic law given by Eqs.~(\ref{eq:elliptic1}-\ref{eq:tau_elliptic_main}).
}\label{fig:meandegree}
\end{figure}

\subsection{Mixture matrices} 
We consider now the spectra of mixture matrices (Model B),
for which we have seen  that  the leading eigenvalue diverges as a function of $N$ (Fig.~\ref{fig:ReLambda1VSN}), and is real (Fig.~\ref{fig:N_Lambda1RealVSN}).            Figure~\ref{fig:BoundaryFiniteNc2Poiss}, which is the equivalent  for mixture matrices of the  Fig.~\ref{fig:BoundaryFiniteNc2_4Poiss} for antagonistic matrices,  shows a good agreement between numerical results and the boundary of the spectrum, a part from important sample fluctuations for small sizes.
Remarkably the big difference with the antagonistic case is visible in proximity of the real axis, where eigenvalues seem this time to accumulate and every remnant of the reentrant behaviour previously emerging at small $c$ has disappeared. 
Moreover we observe  that mixture  matrices develop long tails of eigenvalues on the real axis, which are absent in the spectra of antagonistic matrices.  These tails are directly responsible of the divergence as a function of $N$ of the leading eigenvalue, which turns out to be real with probability one.  The observed  tails are reminiscent of the Lifshitz tails in nondirected graphs  \cite{krivelevich2003largest, bauer2001random, khorunzhiy2006lifshitz, EquivalenceCavReplica2011, bapst2011lifshitz, slanina2012localization}, which also appear in nonHermitian random matrices.    
Further details about the tails of the spectrum are discussed in Appendix~\ref{sec:Appendix_PopDyn}, where we focus on the boundary of the spectrum in proximity of the real axis as obtained by the cavity method (see in particular  Fig.~\ref{fig:ComparisonAntMixture_LogEpsVSReLambda}). Interestingly we find that strong finite size effects affect these  theoretical results obtained with population dynamics similarly to what happens for results from direct diagonalisation giving evidence that the support set $\mathcal{S}$ of the spectrum contains the entire real line. 
Note that in particular the result shown in Fig.~\ref{fig:BoundaryFiniteNc2Poiss} has been obtained for population size $N_p=25000$. Larger population size would push the boundary on the real line further away from the origin.  On the other hand we have observed that the boundary far from the real line is not affected by population size.
We anticipate here that this result is particularly relevant for its implications for structural stability as we will discuss in Sec.~\ref{sec:disc}. 

\subsection{Comparing the spectra of sparse matrices with the elliptic law} 
 In the  limiting case of sparse   matrices of high connectivity, i.e.~$c\rightarrow \infty$,  the Eqs.~(\ref{eq:rec1}-\ref{eq:specrec})  imply    the elliptic law   for the spectral distribution \cite{Sommers_et_al, Nguyen,gotzeelliptic, allesina2012}, as we show in Appendix~\ref{sec:App_Large_cLimit}, namely, 
 \begin{equation}
    \rho(z) = \left\{ \begin{array}{ccc}  \displaystyle\frac{\sigma^2}{\pi (\sigma^4-\tau^2) } \ &{\rm if}& \quad  \displaystyle \frac{{\rm Re}(z)^2}{\left(\sigma^2+\tau\right)^2}+\frac{{\rm Im}(z)^2}{\left(\sigma^2-\tau\right)^2}\leq \frac{1}{\sigma^2}, \\  \displaystyle 0 &{\rm if}& \quad  \displaystyle \frac{{\rm Re}(z)^2}{\left(\sigma^2+\tau\right)^2}+\frac{{\rm Im}(z)^2}{\left(\sigma^2-\tau\right)^2}> \frac{1}{\sigma^2}, \end{array}\right. \label{eq:elliptic1}
 \end{equation} 
where 
\begin{equation}\label{eq:tau_elliptic_main}
 \tau = \lim_{c\rightarrow \infty}  c \left\langle ul \right\rangle_p, \quad  {\rm and} \quad 
\sigma^2 = \lim_{c\rightarrow \infty}c \left\langle u^2 \right\rangle_p ,
\end{equation}
and where  $p$ is the distribution of $u$ and $l$ in the general model of Sec.~\ref{sec:generalModel}.
Hence, in this limit it holds  
that \begin{equation}
 \mathcal{S} = \left\{z\in \mathbb{C}: \frac{{\rm Re}(z)^2}{\left(\sigma^2+\tau\right)^2}+\frac{{\rm Im}(z)^2}{\left(\sigma^2-\tau\right)^2}\leq \frac{1}{\sigma^2}  \right\}, \label{eq:SEllipt}
 \end{equation}
 where $\sigma$ and $\tau$ are defined as in Eq.~(\ref{eq:tau_elliptic_main}).  Equation~(\ref{eq:SEllipt}) 
   is consistent with the elliptic law Eq.~(\ref{eq:elliptic1}).

The elliptic law derived to describe the boundary of the spectrum of dense matrices is by definition insensitive of the local network topology.
To highlight how important is the influence of network topology on the spectral results presented so far 
we adapt the elliptic law \cite{allesina2012} of Eqs.~(\ref{eq:elliptic1}-\ref{eq:tau_elliptic_main}) to matrices from Model A, and to matrices of Model B as they were dense matrices, and thus ignoring the network structure.    
To do so, we consider the elliptic law for an  i.i.d.~matrix ensemble whose entries $(A_{ij},A_{ji})$  have the same  variance and  correlations as in the  sparse ensembles represented by Model A, $\sigma^2 = c$ and $\tau = -3c/4$, and Model B, $\sigma^2 = c$ and $\tau = -3c/5$.
The first consequence of this adaptation is that the elliptic law has finite support, at variance with what happens for $N \times N$ dense matrices with finite elements.
In other words the elliptic law adapted to the sparse case predicts that systems whose coupling strengths are described by sparse matrices can be stable in all cases as the leading eigenvalue does not diverge with $N$. 

In Fig.~\ref{fig:BoundaryFiniteNc2_4Poiss}, we observe that the elliptic law thus obtained gives an acceptable quantitative prediction  for the boundary of the spectrum and the  leading eigenvalue of an antagonistic matrix with $c=4$.    On the other hand,  for $c=2$  the spectrum and the leading eigenvalue  deviate considerably between the elliptic and sparse ensembles.  The difference is in this case mainly due to the reentrance effect, which is absent in the dense ensemble. 

Discrepancies emerge also for mixture matrices, as shown in Fig.~\ref{fig:BoundaryFiniteNc2Poiss}, and this time are due to the emergence of tails on the real line, which cannot be described by the elliptic law. Such tails reinstate the divergence of the leading eigenvalue of sparse large matrices which is not simply originated by an extensive mean connectivity as for dense matrices, but it is due to unbounded maximum degree $k_{\rm max}$ similarly to what occurs for symmetric matrices as discussed in the introduction.
Therefore, especially for mixture matrices, the predictions on the leading eigenvalue that can be obtained by neglecting the local graph topology and using an adaptation of the elliptic law to the sparse case are completely unreliable.

Finally, although for antagonistic matrices with $c=4$ the elliptic law predicts  well   the boundary of  the support set $\mathcal{S}$, this is not the case for the spectral distribution $\rho$.  
In Appendix~\ref{sec:App_SpectralDensityResults}, we plot  $\rho$ for antagonistic random  matrices  in  Model A and find that  their spectral distribution  deviates significantly from the uniform elliptic law.    In fact, for sparse ensembles, there is even a divergence for $z\rightarrow 0$.   
We expect this discrepancy to hold more generally for $c\gtrsim 4$ and an eventual recovery of the elliptic law at much larger $c$.
 
\section{Influence of network topology on system stability}\label{sec:networkTop} 
So far, we have studied how  the properties of the interactions $(J_{ij},J_{ji})$, which may be of the predator-prey, competitive or mutualistic type, affect the stability of large dynamical systems.   However, we have focused on only one random graph ensemble, namely, the Erd\H{o}s-R\'{e}nyi ensemble with a Poisson degree distribution.   Here, on the other hand, we will study  how graph structure affects system stability.    We discuss how the leading eigenvalue $\lambda_1$ depends on mean degree $c$  and  the  variance 
\begin{equation}
{\rm var}(k) = \langle k^2\rangle_{p_{\rm deg}} - c^2.
\end{equation} 

 We  determine the parameter regimes for which fixed points in antagonistic systems, which have exclusively predator-prey interactions, are stable and compare the resultant phase diagrams with those of models with unidirectional interactions.     We do not consider models with a mixture of interactions as for such systems the stability of fixed points is dependent on system size and therefore we would obtain a trivial phase diagram in the limit of infinitely large $N$, as in the infinite size limit the system is unstable for all parameter values.   
 
 For the antagonistic ensemble, we consider the distribution $\tilde{p}$  given by Eq.~(\ref{eq:uniform}) and for the oriented ensemble $\tilde{p}$ is an arbitrary distribution with unit variance and zero mean; for oriented matrices the precise form of $\tilde{p}$ does not matter as ${\rm Re}(\lambda_1)$ only depends on its variance and mean value \cite{neri2019spectral}.

\begin{figure*}[htbp]
\centering
    \subfigure[Antagonistic matrices]{\includegraphics[width=0.44\textwidth]{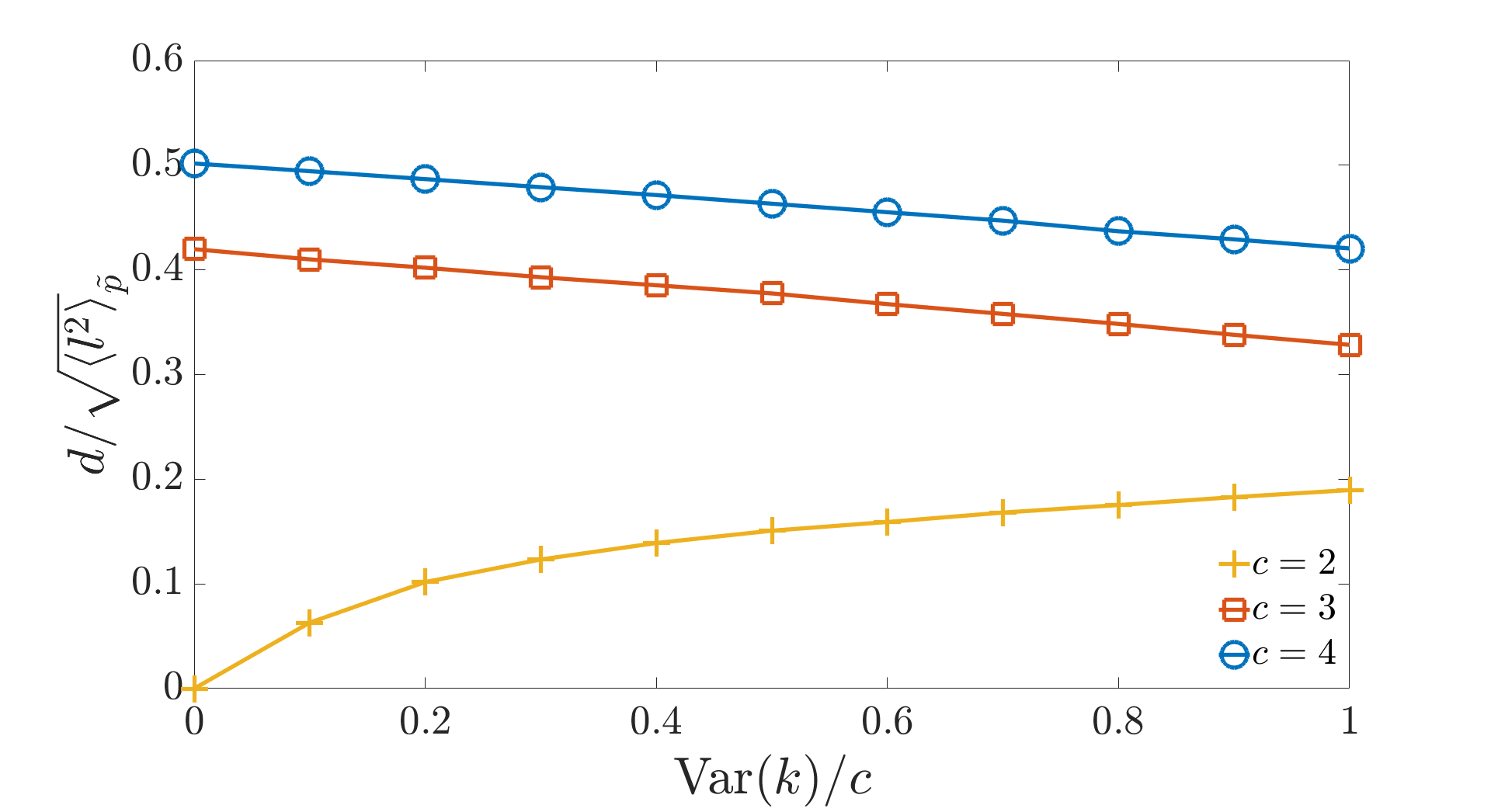}
    \label{subfig:ReLambda1VSVar_Antagonistic}}
         \subfigure[Oriented matrices]{\includegraphics[width=0.44\textwidth]{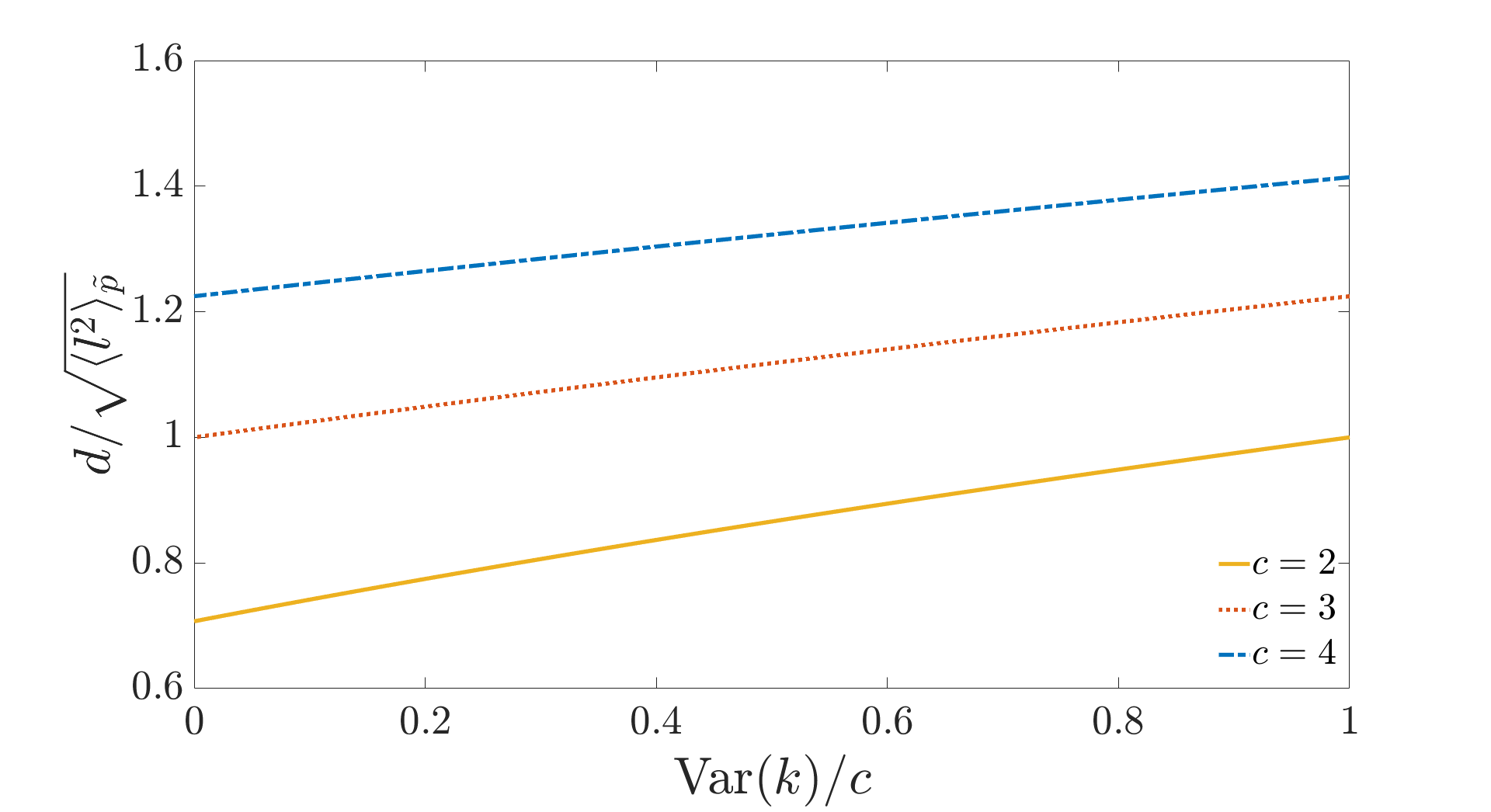}
    \label{subfig:ReLambda1VSVar_Oriented}}
      \put(-328,27){{\bf Unstable}}
      \put(-328,100){{\bf Stable}}
      \put(-100,32){{\bf Unstable}}
      \put(-180,105){{\bf Stable}}
    \caption{Comparison between the phase diagrams for the linear stability of fixed points in systems with predator-prey interactions  [Panel (a)]
     and those with unidirectional interactions  [Panel (b)].    The lines denote the edge of stability, this is given by the values of the system parameters ${\rm Var}[k]/c$ and $d/\sqrt{\langle l^2\rangle_{\tilde{p}}}$  (note that for antagonistic matrices $\langle l^2\rangle_{\tilde{p}} = \langle l^2\rangle_{p}$ while for oriented matrices $\langle l^2\rangle_{\tilde{p}} = \langle l^2\rangle_{p}/2$) %${\rm Re}(\lambda^\ast_1)$, equivalent to  
for which the leading eigenvalue of $\textbf{M}$ has null real part, which is obtained where $d={\rm Re}(\lambda^\ast_1)$.  Results are for random graphs with the prescribed degree distribution  Eq.~(\ref{eq:p_k_linearcomb}).   For antagonistic matrices  [Panel (a)] $\tilde{p}$     is given by  Eq.~(\ref{eq:uniform}) and for oriented matrices [Panel (b)] $\tilde{p}$ is an arbitrary distribution with unit variance and zero mean.  The  markers in Panel (a) are obtained with the  theory in Sec.\ref{sec:Theory} and lines are guides to the eye. The lines in Panel (b) are the theoretical results given  by Eq.~(\ref{eq:leadOriented}). 
    }\label{fig:leading}
\end{figure*}

 \subsection{Large mean degree has for all interaction types a negative effect on system stability}  
Figure~\ref{fig:meandegree}(a) shows  the edge of stability, given by the values of the parameters for which ${\rm Re}(\lambda^\ast_1)=d$, for infinitely large, antagonistic, random  matrices on Erd\H{o}s-R\'{e}nyi graphs (Model A) as a function of~$c$, with $c>1$ so that there  exists a giant component (see  Appendix~\ref{App:LCC}).    We find that the value ${\rm Re}(\lambda^\ast_1)$, and therefore the edge of the stability,  increases as a function of $c$ implying that  interactions destabilise fixed points in dynamical systems.      Note that on the y-axis we have $d/\sqrt{\langle l^2\rangle}$ as the result is invariant under a rescaling of both $d$ and $\sqrt{\langle l^2\rangle} = \sqrt{\langle u^2\rangle}$ by the same factor.

In order to better understand the effect of network topology on system stability, we  compare the results for  Erd\H{o}s-R\'{e}nyi graphs with those for    regular graphs, for which 
\begin{equation}
p_{\rm deg}(k) = \delta_{k,c}, 
\end{equation}
with $c\in \left\{3,4,\ldots\right\}$ so that there exists a giant component (see  Appendix~\ref{App:LCC})).  Figure~\ref{fig:meandegree}(a) shows  that if interactions are of the predator-prey type, then Erd\H{o}s-R\'{e}nyi graphs are more stable than regular graphs.

Comparing the  results for sparse matrices  obtained with the cavity method with predictions from the elliptic law, which ignores the presence of an underlying network, we find that the elliptic law provides a reasonable quantitative prediction of the leading eigenvalue of antagonistic matrices for values $c  \gtrsim4$.

Figure~\ref{fig:meandegree}(b)  presents similar results as in Figure~\ref{fig:meandegree}(a), but this time for oriented matrices with $(J_{ij},J_{ji})$ random variables drawn from the distribution given by Eq.~(\ref{eq:oriented}) with $\tilde{p}$ an arbitrary distribution with zero mean and unit variance.   In this case, the boundary of the support set $\mathcal{S}$ is given by \cite{neri2019spectral}
\begin{equation}\label{eq:boundary_oriented}
\left|z \right|^2=\frac{\left\langle k (k-1) \right\rangle_{p_{\rm deg}}  }{2c}\left\langle l^2 \right\rangle_{\tilde{p}}, 
\end{equation}      
as shown in Appendix~\ref{sec:App_OrientedBoundary}, and therefore
\begin{equation}
{\rm Re}(\lambda^\ast_1) =  \sqrt{\langle l^2 \rangle_{\tilde{p}} \frac{{\rm var}(k) + c^2-c}{2c}}. \label{eq:leadOriented}
\end{equation}   
The Eq.~(\ref{eq:leadOriented}) applies as long as the graph has a giant strongly connected component (otherwise, the leading eigenvalue is determined by small cycles, see Ref.~\cite{neri2019spectral}).  As shown in  Appendix~\ref{App:LCC}, for Erd\H{o}s-R\'{e}nyi graphs the condition for the existence of a giant strongly connected component corresponds with $c>2$, while for regular graphs $c>3$.     

Comparing Figs.~\ref{fig:meandegree}(a) and (b), we observe that dynamical systems  with predator-prey interactions are more stable than those with unidirectional interactions.   Moreover, we find that for systems with predator-prey  interactions   Erd\H{o}s-R\'{e}nyi graphs are more stable than  regular graphs, while for systems with unidirectional interactions it is the other way around.    Hence, how network topology affects system stability depends on the nature of the interactions.

  To focus more specifically on the role played by network topology in the next subsection we directly study how degree fluctuations affect system stability interpolating between the two extreme cases of Erd\H{o}s-R\'{e}nyi (maximum degree fluctuations) and random regular graphs (no degree fluctuations).

 \subsection{The effect of degree fluctuations on system stability depends on the nature of the interactions} 
 One of the biggest advantage of the cavity approach is to be able to include in the computation and show in the results the role of local network topology on spectral properties, which in this case turn out to be far from trivial.
 From Figs.~\ref{fig:meandegree}(a) and (b) we gathered first evidences that  degree fluctuations can have both a stabilizing and a destabilizing effect on system stability, depending on the nature of the interactions. 
   Here, we aim at systematically studying the  effect of  degree fluctuations on  system stability by  analysing the dependency of the   leading eigenvalue  on the variance of the degree distribution at a fixed value of the mean degree $c$ interpolating between the random regular and Erd\H{o}s-R\'{e}nyi graphs discussed in the previous subsection.   We therefore consider 
 random graphs with the degree distribution 
\begin{align}
p_{\rm deg}&(k)=a\delta_{c,k}+(1-a)e^{-c}\frac{c^k}{k!},\label{eq:p_k_linearcomb}
\end{align}
where $a\in[0,1]$.    By varying the  parameter $a$, we  modulate   the variance of the degree distribution,  which is given by 
\begin{align}
{\rm Var}(k)&=c(1-a), \label{Eq:Var_kLinearComb}
\end{align}  
while keeping the mean degree $c$ fixed.

In Fig.~\ref{fig:leading}(a), we plot the edge of stability, i.e. the line where leading eigenvalue of $\textbf{M}$ has null real part, which is obtained where $d={\rm Re}(\lambda^\ast_1)$, for antagonistic matrices with $\tilde{p}$  given by Eq.~(\ref{eq:uniform}) as a function of the  ratio between the variance of the degree distribution $p_{\rm deg}$ in Eq.~(\ref{eq:p_k_linearcomb}).     We observe that degree fluctuations generally tend to stabilize antagonistic dynamical systems as the area where linear stability holds increases 
as a function of ${\rm Var}(k)$. Therefore a smaller $d$ will suffice to stabilise the system. 
A notable exception is when the mean degree is $c=2$, in which case 
the regions of the phase diagram where stability holds shrinks when degree fluctuations get larger.        

In Fig.~\ref{fig:leading}(b), we plot the edge of stability, for oriented matrices with $\tilde{p}$ a arbitrary distribution with zero mean and unit variance.   Remarkably, in this case we obtain the opposite result, namely, that degree fluctuations always destabilize systems with unidirectional interactions.  In fact, this result  follows readily  from Eq.~(\ref{eq:leadOriented}).    

It is surprising that for antagonistic systems variability in the degrees can enhance the stability of fixed points. 
The fact that this behaviour emerges only for $c>2$, while $c=2$ shows a more standard destabilising effect due to degree variability, suggests that it is connected with the absence or presence of the reentrance effect in the boundary of the spectral distribution. A more accurate study of the effect of degree variability of linear stability and its relation to the shape of the boundary of the spectral distribution is left to future inspection.

Taken together, we have found that in certain situations systems with degree fluctuations have the tendency to be more stable than systems without degree fluctuations  and this is an unexpected example of how complexity of a disordered system can  increase its stability.

\section{\label{sec:disc} Discussion and outlook}    
We have analysed the stability of linear dynamical systems defined on sparse random graphs that  contain interactions of the predator-prey, competitive, and mutualistic type, extending previous studies that considered the stability of     systems defined on dense graphs, see Refs.~\cite{Allesina2008, allesina2012,  mougi2012diversity, mougi2014stability, allesina2015predicting}.    For random graphs with a prescribed degree distribution that has unbounded support, we  have shown  that system stability is strongly dependent on the  interaction type.    Indeed,  we have shown that the stability of  general systems that contain a mixture of interactions is size-dependent  as ${\rm Re}(\lambda_{1})$ diverges as a function of $N$.   For such systems there exists for each finite value of $d$ a critical size $N^\ast$ above which the system is unstable implying a tradeoff between stability and diversity.     On the other hand, when the interactions are exclusively of the predator-prey type, then there exists values of $d$ for which the  system exhibits   absolute stability as  ${\rm Re}(\lambda_1)$ converges to a finite value as a function of $N$.     These results are unexpected and noteworthy as previous studies  have shown that for dynamical systems on dense graphs  the scaling of the leading eigenvalue  with system size does not depend on the interaction type, see Refs.~\cite{Allesina2008, allesina2012,  mougi2012diversity, mougi2014stability, allesina2015predicting}.

We can provide an intuitive interpretation   for the  enhanced stability of  dynamical systems   with predator-prey interactions that  is based on an analysis of the  local neighbourhood of a randomly selected node. 
Although random graphs contain a large number of cycles of length $
\log N$~\cite{bianconi2005loops}, the local neighbourhood of a node in  a large random graph is with probability one a tree graph~\cite{dembo2010gibbs}.   
As we discuss in Appendix~\ref{App:last}, numerical evidences suggest that the eigenvalues of antagonistic, tree matrices are imaginary. 
Therefore,  the stability of antagonistic dynamical systems defined on tree graphs is granted for any  $-d<0$.
For antagonistic matrices, this implies that the local neighbourhood of a randomly selected node represents a stable dynamical system enhancing global stability.   
In contrast, the  leading eigenvalues of tree graphs with competitive and mutualistic interactions have unbounded real part, which typically increases with the maximum degree and the strength of the interactions involved.
This implies that, for any $d$, large mixture matrices with unbounded degree distribution or unbounded interaction distribution always contain local neighbourhoods that are unstable.   
Taken together, stability of dynamical systems defined on random graphs is strongly related to their local structure and this     is captured by the cavity method.    
Note that the above reasoning  suggests that antagonistic systems exhibit absolute stability as long as the local neighbourhood of a randomly selected node is with probability one an  antagonistic tree graph.   Hence, more general antagonistic ensembles than the one described in Eq.~(\ref{eq:Antagonisticp}), also  including asymmetries between predators and preys, could be studied and are expected to give qualitatively similar results.
Moreover, it is not required that the interaction strengths are i.i.d. random variables drawn from a certain distribution $p(u,l)$ as considered in Eq.~(\ref{eq:puv}).   Any random coupling matrix is expected to lead to absolute stability for antagonistic systems as long as it is locally tree like and antagonistic.

%Even though the linear  stability  of fixed points in large systems defined on complex networks is a problem of broad interest    (see  introduction for several applications of linear stability analysis), we  discuss here in more depth the relevance of the results for theoretical ecology as this has been a subject of intensive study in the literature. For ecosystems described by randomly coupled Lotka-Volterra  equations,  the Refs.~\cite{rossberg2013food, rossberg2017structural, dougoud2018feasibility, o2019metacommunity} suggest that structural stability is more relevant to describe ecosystem stability.  
Linear  stability  of fixed points in large systems defined on complex networks is a problem of broad interest (see  introduction for several applications of linear stability analysis). In theoretical ecology it has also been the subject of intensive studies, including the still open discussion about how well the Jacobian can be approximated by a random matrix~\cite{stone2018feasibility, biroli2018marginally, barbier2021fingerprints}.
On the other hand, it has been argued~\cite{rossberg2013food, arnoldi2016resilience, rossberg2017structural, dougoud2018feasibility, o2019metacommunity} that another kind of stability, called structural stability, is more relevant to describe ecosystem stability.
Importantly, the  main result of this paper, namely that  dynamical systems on  random graphs with  predator-prey interactions   exhibit  absolute stability  while those with a  mixture of interactions  exhibit  size-dependent stability,  remains valid when one uses  structural stability to characterise the stability of an ecosystem as discussed in the following.

Consider an ecosystem described by a set of generalised Lotka-Volterra equations
\begin{equation}
\partial_t x_i =  x_i  \left(r_i + \sum^N_{j=1;j\neq i}B_{ij} x_j  - \delta x_i\right)\ , \label{eq:LV}
\end{equation} 
where $x_i$ denotes the population abundance of the $i$-th species, $r_i$ is the intrinsic growth rate of the $i$-th species, $\textbf{B}$ is the {\it interaction matrix} that describes the effect species $j$ has on the population abundance of species $i$, and $\delta$ is a parameter that determines the finite carrying capacity of species $i$.  
Note that within the linear stability analysis the Jacobian at a fixed point $\vec{x}^\ast$ for such model has elements $x^\ast_i B_{ij}$ for $i\neq j$, which preserve the antagonistic, mutualistic or competitive structure of interactions contained in $\textbf{B}$, as $x^\ast_i >0 \ \ \forall  \ \ i$.
More generally a fixed point solution  $\vec{x}^\ast$ must solve 
\begin{equation}
\vec{x}^\ast = (-\mathbf{B}+\delta\mathbf{1}_N)^{-1}\vec{r} \ .
\end{equation}
Therefore, a necessary condition for the mere existence of a fixed point  is that the matrix  $-\mathbf{B}+\delta\mathbf{1}_N$ is invertible, {\it i.e.} has no zero eigenvalues, and it is said that a system is structurally stable if this condition is satisfied~~\cite{rossberg2013food, rossberg2017structural, dougoud2018feasibility, o2019metacommunity}.
If the community matrix $\mathbf{B}$ is modelled as a random matrix describing either mixture or antagonistic interactions on a sparse random graph, the results obtained in this work apply.
In particular, it is impossible for a system to exhibit structural stability in the limit $N\rightarrow \infty$ if the spectrum of $\mathbf{B}$ covers the whole real axis, as it is the case for mixture systems. Indeed, for any finite $\delta$ the spectrum of $\mathbf{B}$ will be shifted by a finite amount but will always contain zero eigenvalues (note that for large but finite $N$ the matrix $-\mathbf{B}+\delta\mathbf{1}_N$ will have many  near zero eigenvalues leading to strong sensitivity from $\vec{r}$ fluctuations). 
Moreover, for a finite $\delta$, structural stability will depend on the size of the system giving rise to a tradeoff between diversity and stability.
On the other hand, 
the spectra of antagonistic community matrices $\mathbf{B}$ defined on random graphs form a  compact set for any system size and therefore there exists a finite value of $\delta$ for which the ecosystem exhibits absolute structural  stability (where the attribute "absolute" is intended as discussed in the Introduction).   The values of $\delta$ that render the system structurally stable are in general different  than the values of $d$ that render the system linearly stable, but the most important point is that such a finite value of $\delta$ exists.  

We have also analyzed how network topology affects system stability in dynamical systems described by antagonistic and oriented matrices.   The mean degree of a graph has a clear destabilizing effect on system stability, as it was first shown in other contexts~\cite{may1972will}.  However, the impact of degree fluctuations  on system stability is more subtle: it depends on the nature of the interactions and on the mean degree of the graph.   For example,  for antagonistic systems degree fluctuations can have a stabilizing and a destabilizing effect, depending on whether the mean degree $c$ is large or small.     On the other hand, for oriented matrices, degree fluctuations always destabilize large systems. 
All these results hold both for linear stability and structural stability.

The second main result of this paper instead is entirely relying on linear stability approach. 
We have found that the dynamics in the vicinity of a fixed point is oscillatory only if almost all interactions are of the predator-prey type and the mean degree of the graph $c$ is small enough.     In particular we have shown that the typical value of the imaginary part of the leading eigenvalue of antagonistic sparse matrices exhibits a phase transition as a function of $c$, as shown in Fig.~\ref{fig:ImLambda1VSc}.     This phase transition is due to a reentrance behaviour in the spectra of antagonistic matrices at low values of $c$, as shown in Fig.~\ref{fig:BoundaryFiniteNc2_4Poiss}.    
Conversely, for mixture matrices, the leading eigenvalue, not only diverges as a function of $N$, but is also real with probability one. This is because the spectra of mixture matrices are characterized by long tails  on the real line, as shown in Fig.~\ref{fig:BoundaryFiniteNc2Poiss}.   These tails are reminiscent of Lifshitz tails in symmetric random matrices \cite{krivelevich2003largest, bauer2001random, khorunzhiy2006lifshitz, EquivalenceCavReplica2011, bapst2011lifshitz, slanina2012localization}. 
Remarkably, it is sufficient to have a small, finite fraction of competitive and mutualistic interactions in order for Lifshitz tails to develop. 
We have thus found that antagonistic sparse systems can oscillate in response to an external perturbation because the spectra of corresponding matrices do not develop Lifshitz tails and undergo a transition in the shape of the spectral boundary.

It is important to stress again that, within the linear stability approach, the present paper relies on the assumption that the Jacobian of the nonlinear dynamics describing an ecosystem  in the vicinity of a fixed point is well described by a random matrix ensemble~\cite{Allesina2008, allesina2012, mougi2012diversity, mougi2014stability, allesina2015predicting}.   If and how generally this assumption is valid is still under debate~\cite{stone2018feasibility, biroli2018marginally, barbier2021fingerprints}. 

We end the paper with a few open questions.  

It will be interesting to relate the  derived results for  the spectra of antagonistic and mixture matrices to the properties of the corresponding right and left eigenvectors for two reasons. First,   since the right eigenvectors of random directed graphs exhibit power-law  localization at small values of the mean degree $c$, see  Ref.~\cite{metz2020localization}, one expects the same to happen for antagonistic matrices.   As a consequence, the  phase transition in Fig.~\ref{fig:ImLambda1VSc} from a nonoscillatory to an oscillatory phase could be related to a localization-delocalization phase transition.        Second, for sparse symmetric matrices the Lifshitz tails correspond with exponentially localized modes in the spectrum~\cite{krivelevich2003largest, bauer2001random, khorunzhiy2006lifshitz, EquivalenceCavReplica2011, bapst2011lifshitz, slanina2012localization}. It would be interesting to investigate whether this is also the case for Lifshitz tails developing in mixture matrices. 

So far we, have considered a situation for which $d_j=d$ for all $j$ and we have assumed that the spectrum does not contain outlier eigenvalues.   Note that  the latter assumption   is valid as we have considered a balanced scenario for which $\langle u\rangle = \langle l\rangle = 0$.    These two conditions can however  be relaxed.     It is straightforward to incorporate variable diagonal elements in the cavity method  analysis of the present paper.     A study of the outlier eigenvalues is more challenging, but it can also be dealt with a cavity method approach, see Refs.~\cite{neri2016eigenvalue,  neri2019spectral}.   

Finally, it  will be interesting to compare predictions of sparse random graph theory  with spectra of  real ecosystems.  In this regard, Ref.~\cite{james2015constructing} shows that the spectra of  foodwebs depend strongly on the type of interactions that cover the foodweb, and Ref.~\cite{dunne2002food} shows that foodwebs contain a small number of cycles.  These two  empirical observations are   consistent with the theoretical results found in this paper.   It will also be interesting to 
 relate the  phase transition in the imaginary component of the leading eigenvalue of antagonstic random matrices, reported in Fig.~\ref{fig:ImLambda1VSc}, to the oscillatory response observed in real-world systems~\cite{Massimini2228, rogasch2013assessing, hermann2012heterogeneous, del2013synchronization, bimbard2016instability, arnoldi2018ecosystems}.

\acknowledgements

We acknowledge P.~Vivo and R.~K\"uhn who contributed in the initial stages of the project and for insightful discussions on the replica method.  We thank J-P. Bouchaud, F.L.~Metz, T. Galla, G.~Torrisi, and V.A.R.~Susca  for interesting discussions.

\onecolumngrid
\appendix

\section{Leading eigenvalue: stability criterion and frequency of oscillations}\label{App:Oscil} 
We show that the leading eigenvalue $\lambda_1$ of $\mathbf{A}$ governs the  dynamics of  $\vec{y}$  in the limit $t\gg 1$.   In particular, we derive the conditions given by Eqs.~(\ref{eq:crit1}) and (\ref{eq:crit2}) in the main text for the stability of linear  systems and the conditions  given by Eqs.~(\ref{eq:crit3}) and (\ref{eq:crit4}) for oscillations in the dynamical response.

We order the eigenvalues of  the $N\times N$  matrix $\mathbf{A}$ such that  
\begin{equation}
{\rm Re}(\lambda_1) \geq {\rm Re}(\lambda_2) \geq \ldots \geq {\rm Re}(\lambda_N) .
\end{equation}
If  two consequent eigenvalues, say $\lambda_j$ and $\lambda_{j+1}$, have the same real part, then we use the convention that ${\rm Im}[\lambda_j] \geq {\rm Im}(\lambda_{j+1})$.   

We assume that the matrix $\mathbf{A}$ is diagonalizable so that it can be decomposed as 
\begin{equation}
\mathbf{A} =  \sum^N_{j=1}\lambda_j  \vec{R}_j\vec{L}^\dagger_j, \label{eq:decomp}
\end{equation}
where $\vec{R}_j$ is a right eigenvector associated with $\lambda_j$, $\vec{L}_j$ is a left eigenvector associated $\lambda_j$, and $\vec{L}^\dagger_j$ denotes the complex conjugate of $\vec{L}_j$.   We  normalize left and right eigenvectors, such that, 
\begin{equation}
\vec{R}_j\cdot \vec{L}_k= \delta_{j,k},
\end{equation} 
for all $j,k\in \left\{1,2,\ldots,N\right\}$.

Substitution of  Eq.~(\ref{eq:decomp}) in Eq.~(\ref{eq:ODEs_linearised}) yields for $d_j=d$,
\begin{equation}
\vec{y}(t) =  e^{-dt} \sum^N_{j=1} e^{\lambda_j t} [\vec{L}_j\cdot\vec{y}(0) ] \vec{R}_j .
\end{equation} 
Hence, in the limit $t\rightarrow \infty$, we obtain 
\begin{equation}
\vec{y}(t) =  e^{({\rm Re}(\lambda_1)-d) t} \left[ \sum^{M}_{j=1} e^{{\rm i} {\rm Im}(\lambda_j) t} [\vec{L}_j\cdot\vec{y}(0) ] \vec{R}_j  + O\left( e^{({\rm Re}(\lambda_{M+1})-{\rm Re}(\lambda_1))t}\right) \right], \label{eq:dynamics}
\end{equation}
where $M$ denotes the number of eigenvalues  for which ${\rm Re}(\lambda_1) = {\rm Re}(\lambda_2) =\ldots = {\rm Re}(\lambda_M)$.    From Eq.~(\ref{eq:dynamics}) both the stability conditions, given by Eqs.~(\ref{eq:crit1}) and (\ref{eq:crit2}),  and the conditions for oscillations in $\vec{y}(t)$, given by Eqs.~(\ref{eq:crit3}) and (\ref{eq:crit4}), readily follow.

 The conditions  Eqs.~(\ref{eq:crit1})-(\ref{eq:crit4})   also apply when $\mathbf{A}$ is nondiagonalizable.  However, in this case we cannot employ the eigendecomposition Eq.~(\ref{eq:decomp}) and we should instead rely on a Jordan decomposition, see  Ref.~\cite{neri2019spectral}.
 
\section{Finite size study of leading eigenvalue}\label{sec:Appendix_NumericsDD}%\textbf{Gamma Fit}    
Figure~\ref{fig:ImLambda1VSc} in the main text shows  that finite size effects are significant in sparse random matrices.     Therefore, we  analyze here how  the  distribution $p({\rm Im}(\lambda_1))$, plotted in Fig.~\ref{fig:DistributionImLambda1_c2_4_N5000}, depends on $N$.

\begin{figure*}[htbp]
\centering
\subfigure[$c=2$]{\includegraphics[width=0.49\textwidth]{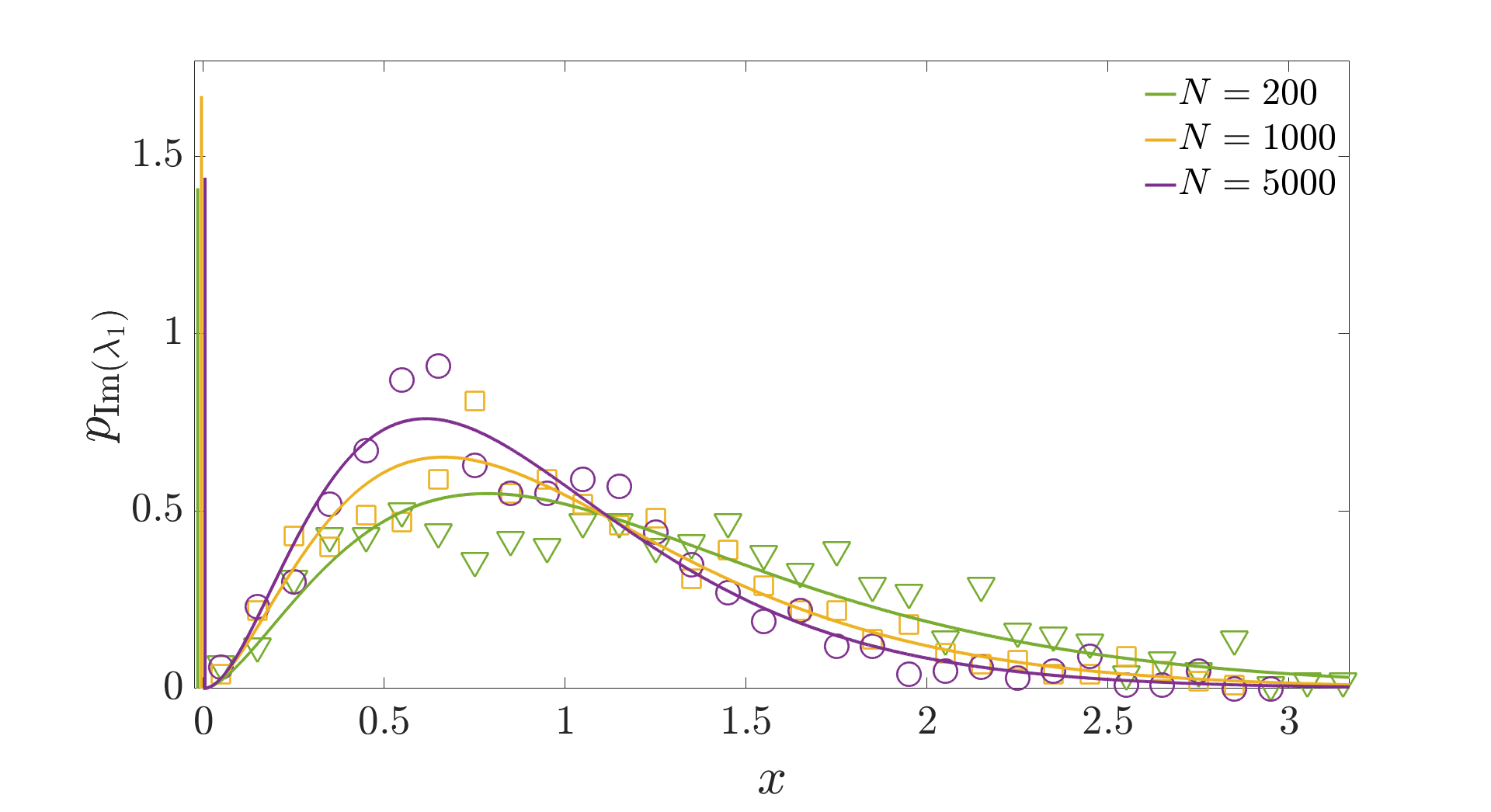}\label{subfig:DistributionImLambdaMax_c2_NStudy}}
\subfigure[$c=4$]{\includegraphics[width=0.49\textwidth]{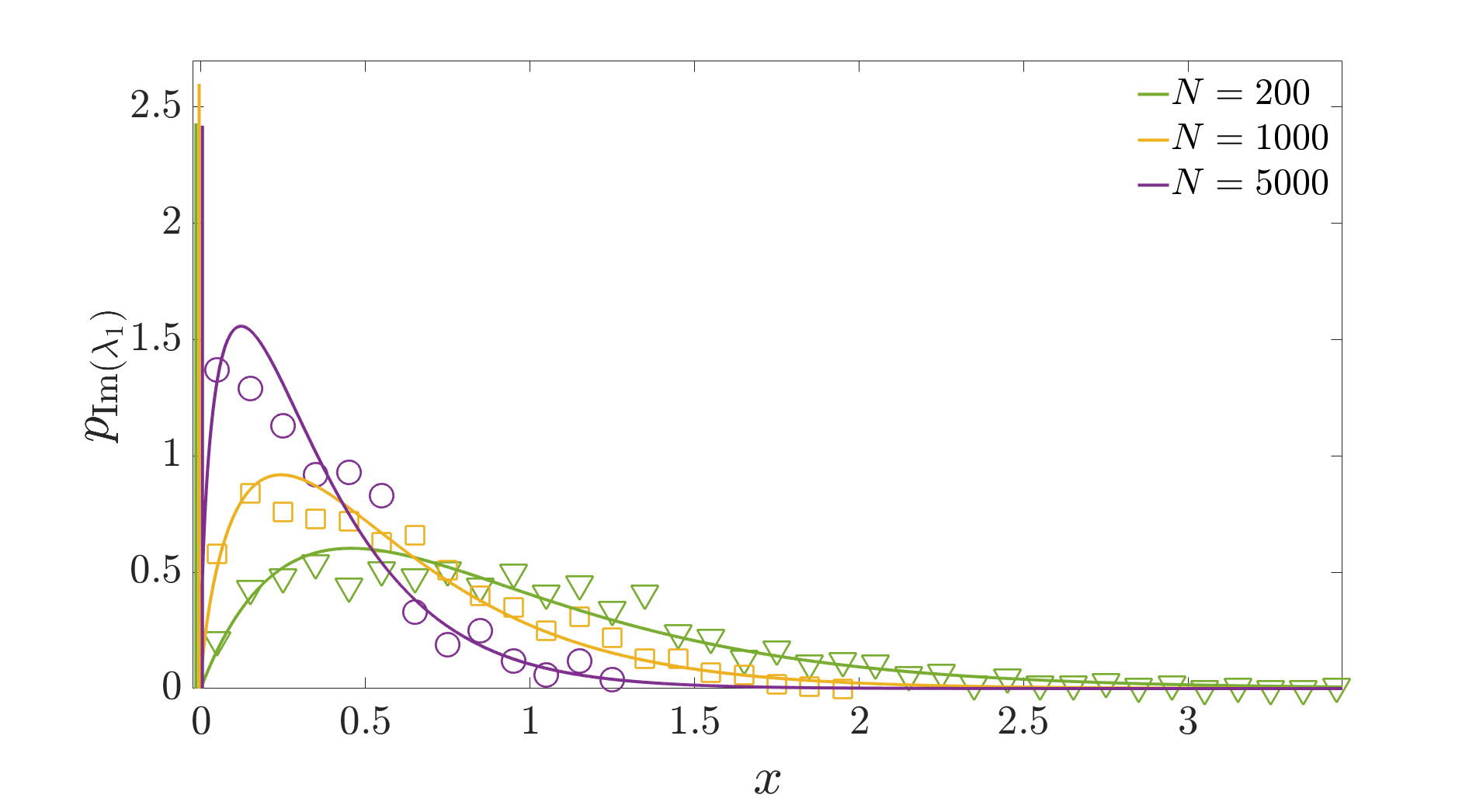}\label{subfig:DistributionImLambdaMax_c4_NStudy}}
  \caption{Distributions of the imaginary part of the leading eigenvalue for  antagonistic matrices (Model A) with $c=2$ [Panel (a)] and $c=4$ [Panel (b)].  The thick vertical line at   ${\rm Im}(\lambda_1)=0$   has height   ${\rm Prob}[\lambda_1\in\mathbb{R}]/\delta$, with $\delta = 0.1$ the width of the intervals in the histogram.    Markers are histograms of imaginary part of the leading eigenvalues obtained through direct diagonalisation of $m_{\rm s} = 1000$ antagonistic matrices for  different values of $N$. Continuous lines are obtained by fitting the Gamma distribution on these data.}
 \label{fig:DistributionImLambdaMax_c2&4_NStudy}
\end{figure*}

Figure~\ref{fig:DistributionImLambdaMax_c2&4_NStudy}  presents empirical data  for  the distribution of ${\rm Im}(\lambda_1)$ in antagonistic matrices with parameters that are the same as in Fig.~\ref{fig:DistributionImLambda1_c2_4_N5000}, except  for the system size $N$, which now takes three values $N=200$, $N=1000$ and $N=5000$.  Just as in Fig.~\ref{fig:DistributionImLambda1_c2_4_N5000}, we observe that the distribution of  ${\rm Im}(\lambda_1)$ consists of two parts and is of the form  given by Eq.~(\ref{eq:pImag}) in the main text.  We make a couple of interesting observations  from  Fig.~\ref{fig:DistributionImLambdaMax_c2&4_NStudy}.     First, we observe that the probability ${\rm Prob}[\lambda_1\in \mathbb{R}]$ that the leading eigenvalue is real is independent of $N$, consistent with the results obtained in Fig.~\ref{fig:N_Lambda1RealVSN}.   A possible explanation for  the  observed $N$-independence of ${\rm Prob}[\lambda_1\in \mathbb{R}]$ is  that the leading eigenvalue  is real when the matrix  $\mathbf{A}$  contains a  cycle that induces  a strong   enough feedback loop.   Since for sparse random graphs the number of cycles of a given fixed length is independent of $N$, and  since cycles of finite length are not accounted for by the  cavity method, this explanation is consistent with  the numerical diagonalization results and the theoretical results obtained in this paper.      Second, we observe that the mode $ {\rm Im}(\lambda^\ast_1)$ of the continuous part of the distribution  decreases as a function of $N$.  For $c=4>c_{\rm crit}$ the distribution moves swiftly towards zero while for $c=2<c_{\rm crit}$ the mode appears to converge to  a finite nonzero value, which is consistent with the phase transition at $N\rightarrow \infty$ shown in Fig.~\ref{fig:ImLambda1VSc} and the conjecture that the cavity method provides an estimate for the mode  of the continuous part of the distribution of $|{\rm Im}(\lambda_1)|$.  

\begin{figure*}[htbp]
\centering
\subfigure[$c=2$]{\includegraphics[width=0.49\textwidth]{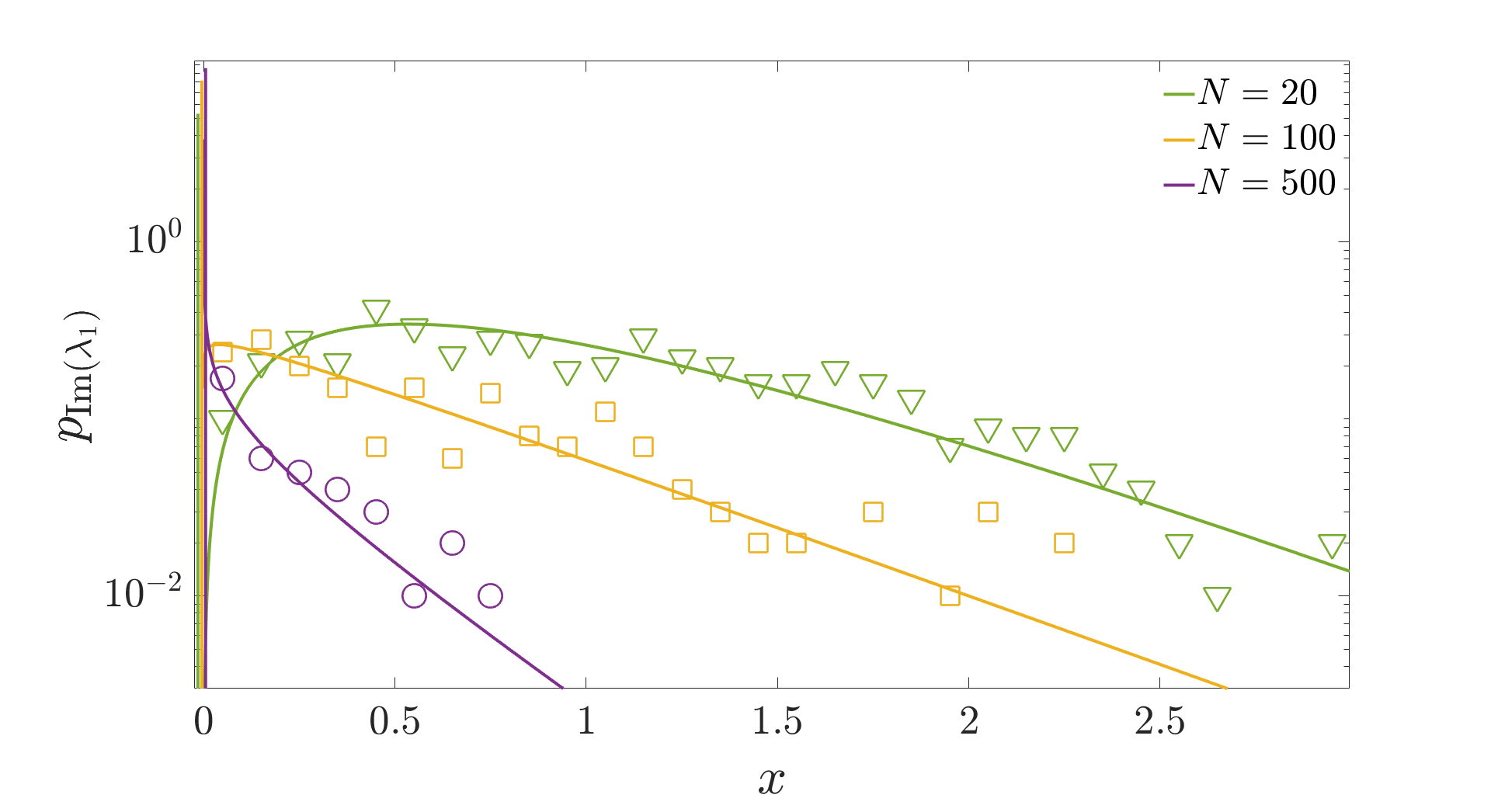}\label{subfig:MixtureDistributionImLambdaMax_c2_NStudy}}
\subfigure[$c=4$]{\includegraphics[width=0.49\textwidth]{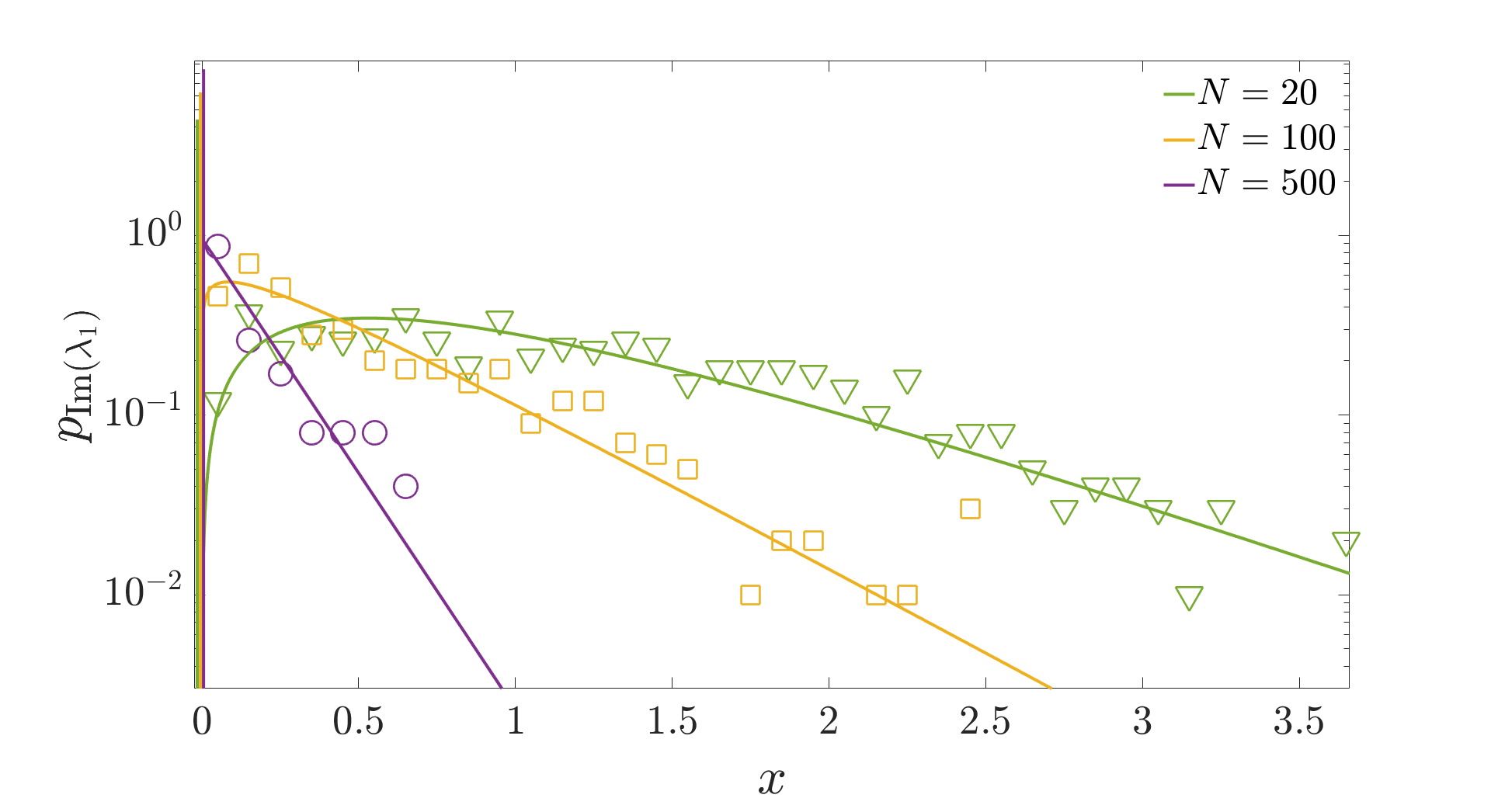}\label{subfig:MixtureDistributionImLambdaMax_c4_NStudy}}
  \caption{Distributions of the imaginary part of the leading eigenvalue  mixture matrices (Model B) with $c=2$ and $c=4$.  The thick vertical line at   ${\rm Im}(\lambda_1)=0$   has height   ${\rm Prob}[\lambda_1\in\mathbb{R}]/\delta$, with $\delta = 0.1$ the width of the intervals in the histogram.     Markers are histograms of imaginary part of the leading eigenvalues obtained through direct diagonalisation of $m_{\rm s} = 1000$ mixture matrices  and for  different values of $N$. Continuous lines are obtained by fitting the Gamma distribution on these data. Vertical axes are in log-scale to make visible the continuous part of the distributions.}
 \label{fig:MixtureDistributionImLambdaMax_c2&4_NStudy}
\end{figure*}
 
Figure~\ref{fig:MixtureDistributionImLambdaMax_c2&4_NStudy} plots $p({\rm Im}(\lambda_1))$ for mixture matrices, which is the equivalent of Fig.~\ref{fig:DistributionImLambdaMax_c2&4_NStudy} for antagonistic matrices.    Comparing the distribution in  Figs.~\ref{fig:DistributionImLambdaMax_c2&4_NStudy} and \ref{fig:MixtureDistributionImLambdaMax_c2&4_NStudy}, we see that the main difference is the  behaviour of ${\rm Prob}[\lambda_1\in \mathbb{R}]$, which rapidly converges to $1$ for mixture matrices, as also shown in Fig.~\ref{fig:N_Lambda1RealVSN}.   As a consequence, for mixture matrices the  continuous part of the distribution   $p({\rm Im}(\lambda_1))$  disappears for large enough $N$.

\section{Derivation of the iteration Equation (\ref{eq:StabBoundary}), in the main text, for the linear stability of the trivial solution}\label{sec:Appendix_StabAnal_Boundary}
We perform a linear stability analysis  of Eq.~(\ref{eq:rec2}) around the trivial solution given by Eq.~(\ref{eq:trivial2}).    To this aim, we consider a perturbation around the trivial solution given by Eqs.~(\ref{eq:perturb})-(\ref{eq:expansion}).

After substitution of Eqs.~(\ref{eq:perturb})-(\ref{eq:expansion}) into  Eq.~(\ref{eq:rec2}), we obtain in the argument of the delta distribution on the right-hand-side of Eq.~(\ref{eq:rec2}) the following  
\begin{eqnarray}
\lefteqn{\left( \begin{pmatrix}
0 & z \\ \bar{z}& 0
\end{pmatrix} - \sum^{k-1}_{\ell=1} \begin{pmatrix}
0 & u_{\ell} \\l_{\ell}& 0
\end{pmatrix}
\begin{pmatrix}
{h}_\ell & -\bar{g}_\ell \\ -{g}_\ell&h'_\ell
\end{pmatrix}
 \begin{pmatrix}
0 & l_\ell \\ u_{\ell}& 0
\end{pmatrix}\right)^{-1} }&&  \nonumber\\ 
&=&\begin{pmatrix}
-\displaystyle\sum^{k-1}_{\ell=1}h'_\ell u_{\ell}^2 &z+ \displaystyle\sum^{k-1}_{\ell=1}g_{\ell}u_{\ell}l_{\ell} \\ \bar{z}+ \displaystyle\sum^{k-1}_{\ell=1}\bar{g}_{\ell}u_{\ell}l_{\ell}  &-\displaystyle\sum^{k-1}_{\ell=1}h_\ell l_{\ell}^2
\end{pmatrix}^{-1}   \nonumber \\
&=&\frac{\begin{pmatrix}
-\displaystyle\sum^{k-1}_{\ell=1}h_\ell l_{\ell}^2
    &-z- \displaystyle\sum^{k-1}_{\ell=1}g_{\ell}u_{\ell}l_{\ell}\\ -\overline{z}- \displaystyle\sum^{k-1}_{\ell=1}\overline{g}_{\ell}u_{\ell}l_{\ell}  & -\displaystyle\sum^{k-1}_{\ell=1}h'_\ell u_{\ell}^2
\end{pmatrix}}
{\sum^{k-1}_{\ell=1}h'_\ell u_{\ell}^2 \: \sum^{k-1}_{\ell=1}h_\ell l_{\ell}^2 -\left( \overline{z}+ \displaystyle\sum^{k-1}_{\ell=1}\overline{g}_{\ell}u_{\ell}l_{\ell}   \right)  \left(z+ \displaystyle\sum^{k-1}_{\ell=1}g_{\ell}u_{\ell}l_{\ell}  \right)}. \label{step:inverse}
\end{eqnarray}

Expanding up to linear order in $h$ and $h'$, we obtain 
\begin{eqnarray}
 &&Q(g,h, h') = \sum_{k=1}^{\infty}\frac{k\:p_{\rm deg}(k)}{c} \int \prod_{\ell=1}^{k-1} \mathrm{d}g_\ell \mathrm{d} h_\ell \mathrm{d} h'_\ell  \:Q(g_{\ell},h_{\ell},h'_{\ell})  \int \prod_{\ell=1}^{k-1}\mathrm{d}u_\ell \mathrm{d}l_\ell\: p(u_\ell,l_\ell) \nonumber\\ 
&&  \times \delta\left(g+\left[z+ \sum_{\ell=1}^{k-1}u_\ell g_\ell {l}_\ell\right]^{-1}  \right)\delta\left(h - \frac{\sum^{k-1}_{\ell=1}h_{\ell}l^2_{\ell}}{|z +\sum^{k-1}_{\ell=1}g_{\ell}u_{\ell}l_{\ell}  |^2} \right)  \delta\left(h' - \frac{\sum^{k-1}_{\ell=1}h'_{\ell}u^2_{\ell}}{|z +\sum^{k-1}_{\ell=1}g_{\ell}u_{\ell}l_{\ell}  |^2}\right)  + \mathcal{O}\left(\epsilon^2\right) ,\nonumber\\ \label{eq:Q}
\end{eqnarray}
where $\epsilon$ is defined in Eq.~(\ref{eq:expansion}).

In Eq.~(\ref{eq:Q}), the recursions for $h$ and $h'$ are decoupled.   Hence, we can integrate out one of these variables.  Defining 
\begin{eqnarray}
Q(g,h) = \int {\rm d}h'\: Q(g,h,h'),
\end{eqnarray}
we obtain from Eq.~(\ref{eq:Q}) that $Q(g,h)$ obeys  Eq.~(\ref{eq:StabBoundary}).

\section{Computing the boundary of the support set for infinitely large matrices}\label{sec:Appendix_PopDyn}
We detail the numerical algorithm we use to obtain the boundary of the support set  $\mathcal{S}$  in the Figs.~\ref{fig:BoundaryFiniteNc2_4Poiss} and \ref{fig:BoundaryFiniteNc2Poiss} and  the typical leading eigenvalue $\lambda^\ast_1$ in Figs.~\ref{fig:ReLambda1VSN}, \ref{fig:DistributionImLambda1_c2_4_N5000}, \ref{fig:meandegree}(a) and \ref{fig:leading}(a).      We first present in  Subsec.~\ref{sec:App_Boundary_NumericalImplementation} the population dynamics algorithm we use  to obtain the boundary of $\mathcal{S}$, and  in Subsec.~\ref{sec:leadingEigv} we show the method we use to obtain $\lambda^\ast_1$ from the population dynamics results.
\subsection{Population dynamics algorithm}\label{sec:App_Boundary_NumericalImplementation}
Recursive distributional equations of the form Eq.~(\ref{eq:StabBoundary}) can be solved numerically with a population dynamics algorithm, see Refs.~\cite{Abou_Chacra_1973, Mezard2001, Reimer_sparse, metz2010localization}.

The population dynamics algorithm represents the distribution $Q(g,h)$ with a population of  $N_{\rm p}$ realizations of the  random variables $(g,h)$.   The population is initialized and updated as follows:  
\begin{enumerate}
\item Initialise  the population by drawing  $N_{\rm p}$   independent realizations of random variables $(g^{(i)},h^{(i)})$  from a certain distribution $p_{\rm init}(g,h)$;
\item Generate a degree $k$ from the distribution $\frac{k}{c}p_{\rm deg}(k)$;
\item Uniformly and  randomly select $k-1$ elements  $(g_\ell,h_\ell)$ from the population and draw $k-1$ random variables $(u_\ell,l_\ell)$ from the distribution $p(u_\ell,l_\ell)$, with $\ell=1,2,\ldots, k-1$;
\item Compute 
\begin{equation}\label{eq:UpdateRuleBoundaryStabilityPopDyn}
g=-\left[z+ \sum_{\ell=1}^{k-1} u_\ell g_\ell l_\ell\right]^{-1}, \  {\rm and} \quad h = |g|^2 \sum_{\ell=1}^{k-1} h_\ell \left|l_\ell\right|^2 ;
\end{equation}
\item Uniformly  and randomly select an index   $i\in \{1,\ldots,N_p\}$ and replace $(g^{(i)},h^{(i)})$  by $(g,h)$.
\end{enumerate}  
Steps (2-5) are  repeated for a certain number $N_{\rm s} = n_{\rm s}N_{\rm p}$  of iterations, where $n_{\rm s}$ is the number of sweeps for which the whole population is updated.   After $n_{\rm s}$ sweeps, the steps (2-5) are repeated for another $N_{\rm r} = n_{\rm r}N_{\rm p}$ of iterations after which  the population is  estimated as 
\begin{equation}
\hat{Q}(g,h) = \frac{1}{n_{\rm r}N_{\rm p} } \sum^{n_{\rm r}N_{\rm p}}_{j=1}\delta(g-g^{(j)})\delta(h-h^{(j)}).
\end{equation}  
If   $N_{\rm p}\rightarrow \infty$, then the  pairs $(g^{(j)},h^{(j)})$  in the population are independent realizations drawn from the distribution  $Q(g,h)$  and the algorithm is exact.    

Since the Eq.~(\ref{eq:StabBoundary}) follows from a stability analysis, it holds that if the initial population is of the form given by Eq.~(\ref{eq:q0Delta}), then
\begin{eqnarray}
\lim_{n_{\rm s}\rightarrow\infty}\lim_{N_{\rm p}\rightarrow\infty}\left\langle |h|\right\rangle_{\hat{Q}}  = \left\{\begin{array}{ccc} 0, &&z \notin \mathcal{S}, \\ \infty,&& z\in \mathcal{S} .\end{array}\right. \label{eq:absH}
\end{eqnarray}
  Hence, we obtain the boundary of $\mathcal{S}$ by determining the value of $z$ that separates the region where     
$\left\langle |h|\right\rangle_{\hat{Q}} $  diverges  from the region where $\left\langle |h|\right\rangle_{\hat{Q}} $  converges to zero.

\subsection{Numerical results for finite population size of the boundary}\label{AppD2}
\begin{figure*}[htbp]
\centering
\subfigure[Antagonistic]{\includegraphics[width=0.49\textwidth]{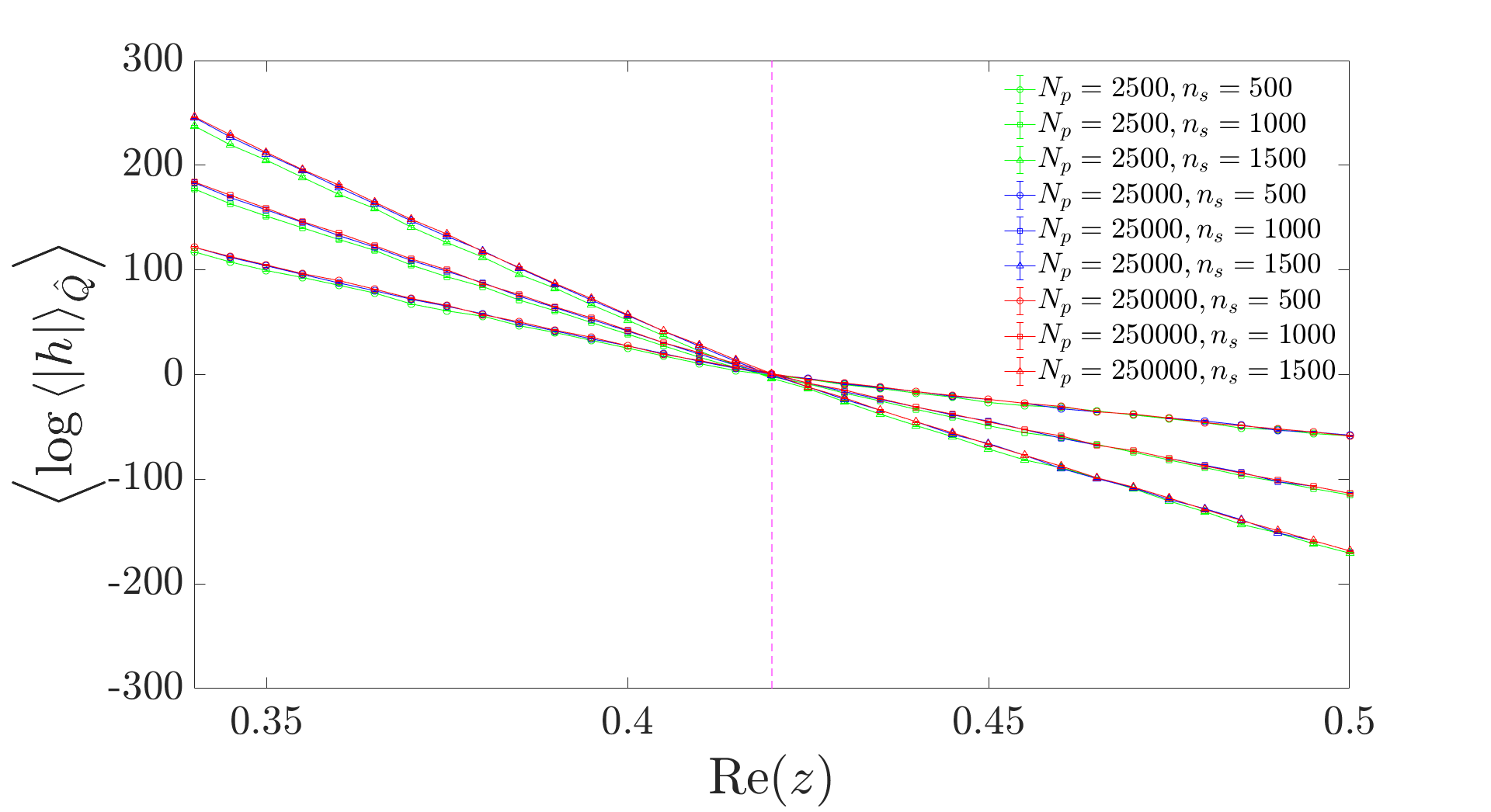}\label{subfig:1}}
\subfigure[Mixture]{\includegraphics[width=0.49\textwidth]{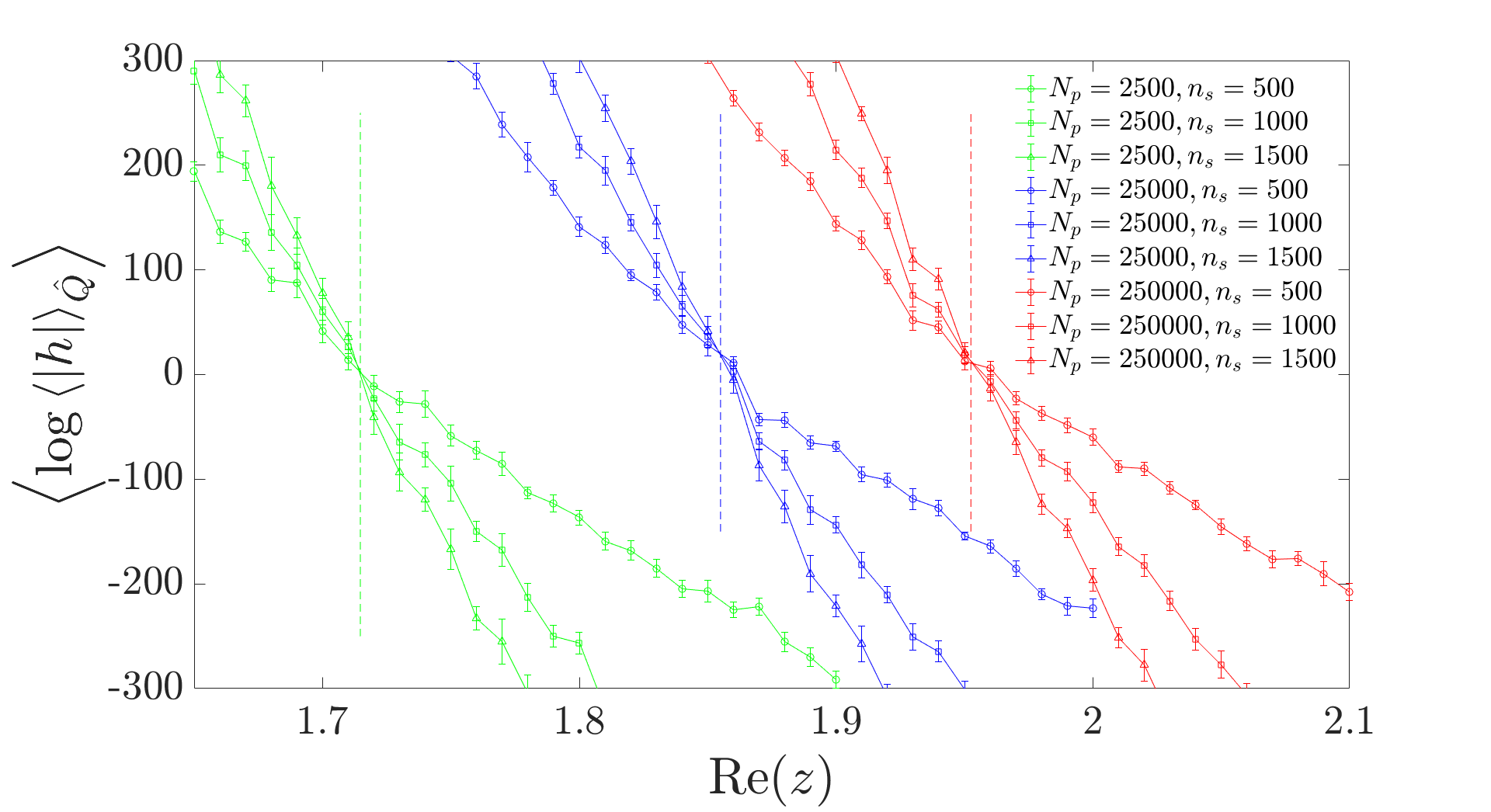}\label{subfig:2}}
  \caption{Plots of  $\log \left\langle |h|\right\rangle_{\hat {Q}}$  as a function of ${\rm Re}(z)$  for  antagonistic matrices  (Model A in Sec.~IID, panel (a)) and mixture matrices (Model B in  Sec.~IID, panel (b)).   The mean degree $c=4$ and  ${\rm Im}(z)=0$.   The markers are obtained with the  population dynamics algorithm described in Sec.~\ref{sec:App_Boundary_NumericalImplementation} with the population size $N_{\rm p}$  and the number of sweeps  $n_{\rm s}$  as given in the legend, and the error bars denote the estimated error obtained with repeated realizations of the population dynamics algorithm.   }
 \label{fig:ComparisonAntMixture_LogEpsVSReLambda}
\end{figure*}

Figure~\ref{fig:ComparisonAntMixture_LogEpsVSReLambda} shows the   mean of  $\log \left\langle |h|\right\rangle_{\hat{Q}}$ taken over the $\mathcal{N}$ realizations of the population dynamics algorithm  as a function of  ${\rm Re}(z)$ for antagonistic matrices  (Model A in Sec.~IID) and mixture matrices (Model B in  Sec.~IID) with   mean degree $c=4$ and ${\rm Im}(z)=0$.    Practically, we evaluate $\left\langle |h|\right\rangle_{\hat{Q}} $ as follows.  
We initialize the  $(g^{(i)},h^{(i)})$ with  the uniform distribution
\begin{equation}
p_{\rm init}(g,h) =  \frac{1}{\Delta^2}, \quad g\in [-\Delta,\Delta], h \in  [-\Delta,\Delta], 
\end{equation}   
for which we have set $\Delta = 10$, but the precise value of $\Delta$ does not matter much.   Subsequently, we compute
\begin{equation}
\left\langle |h|\right\rangle_{\hat{Q}}  = \frac{1}{n_{\rm r}N_{\rm p} }  \sum^{n_{\rm r}N_{\rm p}}_{j=1} |h^{(j)}|
\end{equation} 
with $n_{\rm r} = 500$.    In addition, in order to obtain an estimate of the fluctuations in $\left\langle |h|\right\rangle_{\hat{Q}}$ between different realizations of the population dynamics algorithm, we repeat this procedure a $\mathcal{N}=10$ times, i.e., we compute $\left\langle |h|\right\rangle_{\hat{Q}}$ for $\mathcal{N}$ runs of the population dynamics algorithm with different initial realizations of $(g^{(i)},h^{(i)})$.    Plots show $\log \left\langle |h|\right\rangle_{\hat{Q}}$  for various values of the population size $N_{\rm p}$ and the number of sweeps $n_{\rm s}$; the error on the mean value of   $\log \left\langle |h|\right\rangle_{\hat{Q}}$ is obtained from  the standard deviation of $\log \left\langle |h|\right\rangle_{\hat{Q}}$ on the sample of   $\mathcal{N}=10$ realizations.     

In the case of antagonistic matrices, all lines intersect in a common point, which provides the estimate of the boundary of the support set~$\mathcal{S}$.     On the other hand, in the case of mixture matrices,  the intersection point for different $n_{\rm s}$ increases as a function of the population size $N_{\rm p}$.  This implies that the intersection point diverges as a function of $N_{\rm p}$  and  the real axis belongs to the support set $\mathcal{S}$.   This corroborates the result of Fig.~\ref{fig:ReLambda1VSN}  that show that the leading eigenvalue of mixture matrices diverges as a function of $N$, while the leading eigenvalue of antagonstic matrices converges to a finite value as a function of $N$.

\subsection{Determination of  the leading eigenvalue}\label{sec:leadingEigv}
We discuss how in Figs.~\ref{fig:ReLambda1VSN}, \ref{fig:DistributionImLambda1_c2_4_N5000}, \ref{fig:meandegree}(a) and \ref{fig:leading}(a),  we have   implemented  Eq.~(\ref{eq:leading}) to obtain $\lambda^\ast_1$, the typical value of the leading eigenvalue, from the population dynamics results.

\begin{figure*}[htbp]
\centering
\subfigure[$c=4$]{\includegraphics[width=0.2\textwidth]{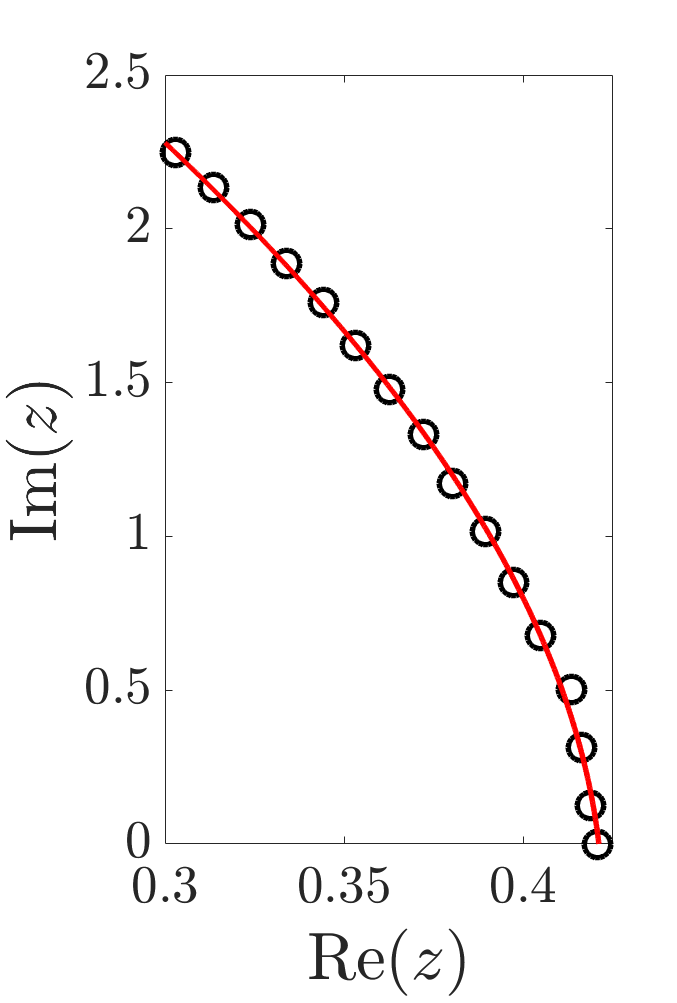}\label{subfig:1}}
\subfigure[$c=2$]{\includegraphics[width=0.2\textwidth]{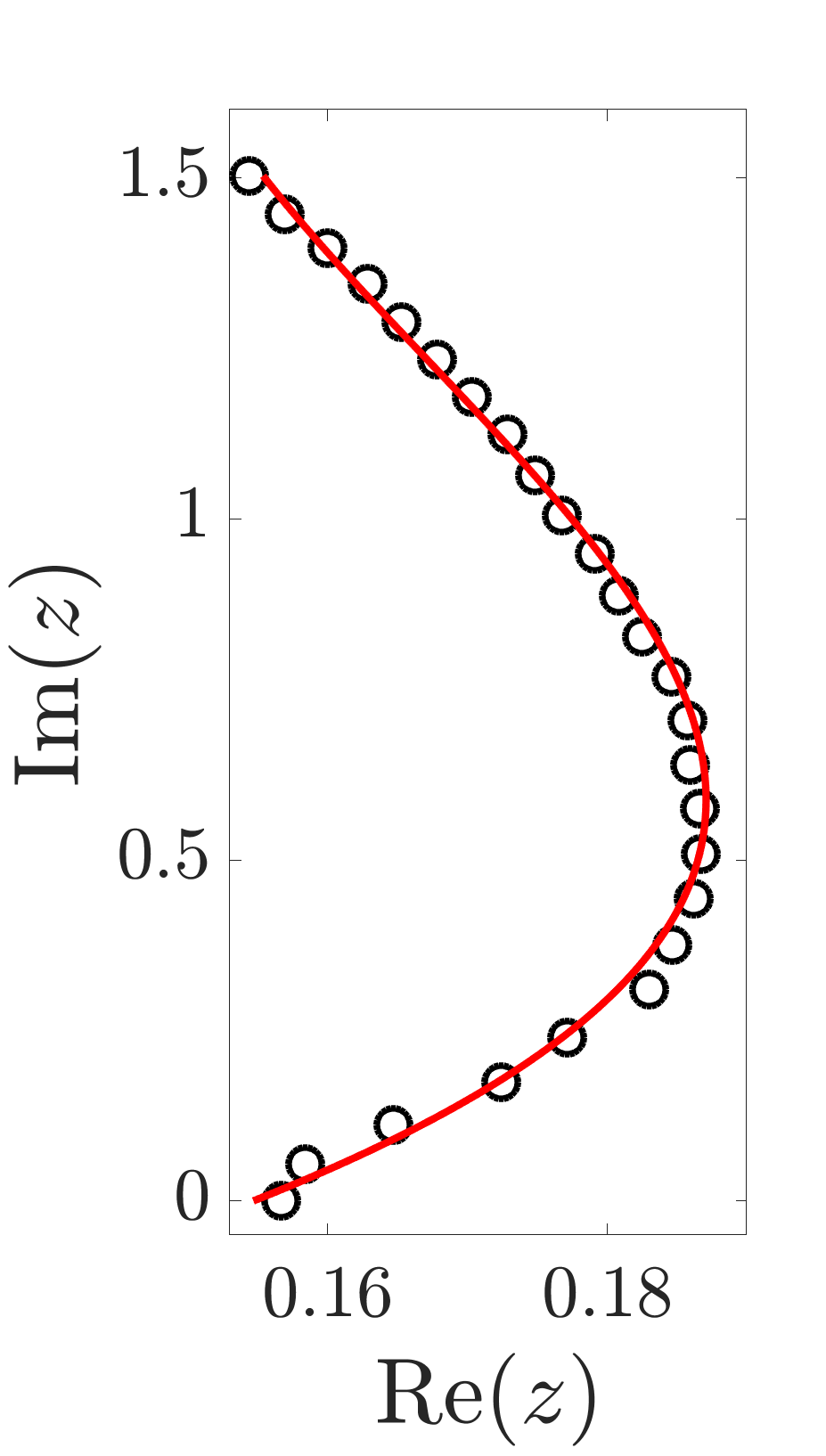}\label{subfig:2}}
\subfigure[$c=1.3$]{\includegraphics[width=0.2\textwidth]{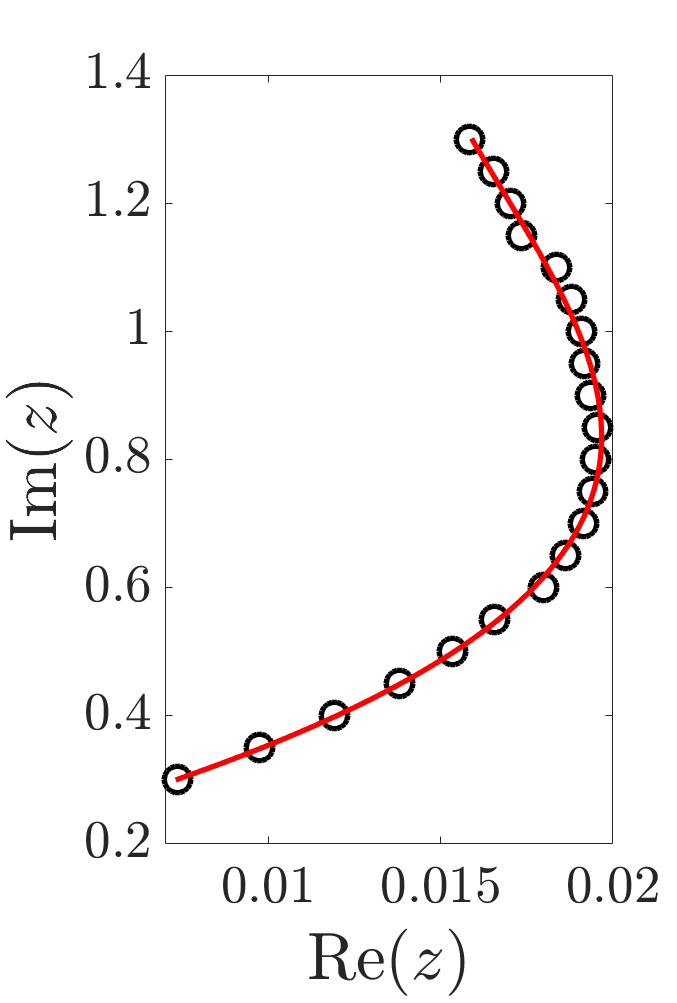}\label{subfig:2}}
  \caption{Cubic fits to the data points, obtained with the population dynamics algorithm described in Appendix~\ref{sec:App_Boundary_NumericalImplementation}, for the boundary of the support set in the vicinity of $\lambda^\ast_1$.   Results shown are for     Model A with mean degrees $c=4$, $c=2$ and $c=1.3$.    Panels (a) and (b) are a zoom of the spectra shown in Fig.~\ref{fig:BoundaryFiniteNc2_4Poiss}. } 
 \label{fig:lead}
\end{figure*}

Since for antagonistic matrices the slope of the boundary of $\mathcal{S}$ is vertical, as shown in Fig.~\ref{fig:BoundaryFiniteNc2_4Poiss},  one needs to control  the fluctuations in the population dynamics algorithm to obtain an accurate value of $\lambda^\ast_1$.   
To this aim, we  use a cubic fit on the values for the boundary of $\mathcal{S}$ obtained with the population dynamics algorithm.  This procedure is shown in Fig.~\ref{fig:lead}, which  shows data points for the boundary of $\mathcal{S}$ in the vicinity of  $\lambda^\ast_1$ for three values of $c$ and also shows a cubic fit through these data points.    We obtain an estimate of $\lambda^\ast_1$ by computing the maximum value of the fitted cubic polynomial.   It this   estimate for  $\lambda^\ast_1$ that we plotted in Figs.~\ref{fig:ReLambda1VSN}, \ref{fig:DistributionImLambda1_c2_4_N5000}, \ref{fig:meandegree}(a) and \ref{fig:leading}(a).   

\section{Spectral distribution}\label{sec:App_SpectralDensityResults}
We compute the spectral distribution $\rho(z)$ for random antagonistic matrices defined on sparse graphs.   In principle, we can solve Eqs.~(\ref{eq:rec1})-(\ref{eq:specrec}) with a population dynamics algorithm to obtain the spectral distribution $\rho$.   However,  this requires one to take a numerical derivative, which leads to  a large numerical error when the population size is small.      One can avoid this numerical error by considering a joint distribution for $\mathsf{G}$ and its derivative  $\frac{{\rm d}}{{\rm d}\overline{z}}\mathsf{G}$,  as discussed in  Ref.~\cite{cavity_nonH}.  We apply this approach to the general model of Sec.~II A and then present numerical results for the spectral distribution of antagonistic, random matrices.
\subsection{Alternative expression  for the spectral distribution}
We  first take the derivative of the Eqs.~(\ref{eq:CavityEq2Main}) and (\ref{eq:CavityEq1Main}).   
Using the chain rule 
\begin{equation}
\frac{{\rm d}}{{\rm d}\overline{z}}\mathsf{A}^{-1}    = -\mathsf{A}^{-1}  \left(\frac{{\rm d}}{{\rm d}\overline{z}}\mathsf{A}\right) \mathsf{A}^{-1} \label{eq:CavityEq1Der}
\end{equation}
on Eqs.~(\ref{eq:CavityEq2Main}), we obtain 
\begin{equation}
\frac{{\rm d}}{{\rm d}\overline{z}}\mathsf{G}_j=-\mathsf{G}_j\left[\sigma_-  - \sum_{k\in \partial_j} \mathsf{J}_{jk} \left(\frac{{\rm d}}{{\rm d}\overline{z}}\mathsf{G}_k^{(j)} \right) \mathsf{J}_{kj}  \right]\mathsf{G}_j, \label{eq:CavityEq2Der}
\end{equation} 
where 
\begin{equation}
\sigma_- = \frac{{\rm d}}{{\rm d}\overline{z}}\mathsf{z}_\eta= \begin{pmatrix}
0 & 0 \\ 1& 0
\end{pmatrix}.
\end{equation} 
Analogously, we obtain from Eqs.~(\ref{eq:CavityEq1Main}), 
\begin{equation}
\frac{{\rm d}}{{\rm d}\overline{z}}\mathsf{G}^{(\ell)}_j=-\mathsf{G}^{(\ell)}_j\left[ \sigma_- - \sum_{k\in \partial_j\setminus \{l\}} \mathsf{J}_{jk} \left(\frac{{\rm d}}{{\rm d}\overline{z}}\mathsf{G}^{(j)}_k \right) \mathsf{J}_{kj}  \right]\mathsf{G}^{(\ell)}_j. \label{eq:CavityEq1Der}
\end{equation} 
The Eqs.~(\ref{eq:CavityEq2Main}), (\ref{eq:CavityEq1Main}),  (\ref{eq:CavityEq2Der}), and (\ref{eq:CavityEq1Der}), together with 
\begin{eqnarray}
\rho(z) =  \frac{1}{\pi N} \lim_{\eta\rightarrow 0^+} \sum^N_{j=1} \frac{{\rm d}}{{\rm d}\overline{z}}\mathsf{G}_j \label{eq:specH}, 
\end{eqnarray} 
 provides the spectral distribution of a locally tree-like random matrix model.

For the general model defined in Sec.~II A, we derive now a set of recursion relations in the distributions  
 \begin{equation}
\tilde{q}(\mathsf{g},\mathsf{g}') := \lim_{N\rightarrow \infty}\frac{1}{N}  \sum^N_{j=1}\delta(\mathsf{g}-\mathsf{G}_j)\delta(\mathsf{g}'- \frac{{\rm d}}{{\rm d}\overline{z}}\mathsf{G}_j)
 \end{equation}
 and 
\begin{equation}
q(\mathsf{g}, \mathsf{g}') := \lim_{N\rightarrow \infty}\frac{1}{cN}  \sum^N_{j=1}\sum_{\ell\in\partial_i}\delta(\mathsf{g}-\mathsf{G}^{(\ell)}_j)\delta(\mathsf{g}'-\frac{{\rm d}}{{\rm d}\overline{z}}\mathsf{G}^{(\ell)}_j).  
 \end{equation} 
Taking an ensemble average of the  Eqs.~(\ref{eq:CavityEq2Main}), (\ref{eq:CavityEq1Main}),  (\ref{eq:CavityEq2Der}), and (\ref{eq:CavityEq1Der}), we obtain the recursive distributional equations
\begin{eqnarray}
\lefteqn{q(\mathsf{g},\mathsf{g}')=\sum_{k=1}^{\infty}\frac{kp_{\rm deg}(k)}{c}\int \prod_{\ell=1}^{k-1}\mathrm{d} \mathsf{g}_\ell \mathrm{d}\mathsf{g}'_\ell  \: q(\mathsf{g}_\ell,\mathsf{g}'_{\ell})\int \prod_{\ell=1}^{k-1}\mathrm{d}u_\ell \mathrm{d}l_\ell \: p(u_\ell,l_\ell) } && \nonumber \\
&\times&  \delta\left[\mathsf{g}-\left(\mathsf{z}_\eta- \sum_{\ell=1}^{k-1}\mathsf{J}_\ell \mathsf{g}_\ell \mathsf{J}_\ell^\dagger  \right)^{-1}    \right] \delta\left[\mathsf{g}'+\mathsf{g}\left( \begin{pmatrix}
0 & 0 \\ 1& 0
\end{pmatrix}-  \sum_{\ell=1}^{k-1}\mathsf{J}_\ell \mathsf{g}'_\ell \mathsf{J}_\ell^\dagger  \right)  \mathsf{g}    \right],\label{eq:SpDensityQTilxx}
\end{eqnarray}
and
\begin{eqnarray}
\lefteqn{\tilde{q}(\mathsf{g},\mathsf{g}')=\sum_{k=0}^{\infty}p_{\rm deg}(k)\int \prod_{\ell=1}^{k}\mathrm{d}\mathsf{g}_\ell \mathrm{d}\mathsf{g}'_\ell q(\mathsf{g}_\ell,\mathsf{g}'_{\ell})\int \prod_{\ell=1}^{k}\mathrm{d}u_\ell \mathrm{d}l_\ell \: p(u_\ell,l_\ell)}&& \nonumber \\
&\times&  \delta\left[\mathsf{g}-\left(\mathsf{z}_\eta- \sum_{\ell=1}^{k}\mathsf{J}_\ell \mathsf{g}_\ell \mathsf{J}_\ell^\dagger  \right)^{-1}    \right]  \delta\left[\mathsf{g}'+\mathsf{g}\left( \begin{pmatrix}
0 & 0 \\ 1& 0
\end{pmatrix}-  \sum_{\ell=1}^{k}\mathsf{J}_\ell \mathsf{g}'_\ell \mathsf{J}_\ell^\dagger  \right)  \mathsf{g}    \right]. \label{eq:rho2}
\end{eqnarray}

The spectral distribution follows from the ensemble averaged version of Eq.~(\ref{eq:specH}), which is given by 
\begin{equation}\label{eq:SpDensityQTil}
\rho(z)=\lim_{\eta\to 0^+}\frac{1}{\pi}\int \mathrm{d}\mathsf{g}\mathrm{d}\mathsf{g}'\:\tilde{q}(\mathsf{g},\mathsf{g}')[\mathsf{g}']_{21}. 
\end{equation}

In the next section, we solve the Eqs.~(\ref{eq:SpDensityQTilxx})-(\ref{eq:SpDensityQTil}) with a population dynamics algorithm.  Note that the 
Eqs.~(\ref{eq:SpDensityQTilxx})-(\ref{eq:SpDensityQTil}) do not involve a numerical derivative ${\rm d}/{\rm d}\overline{z}$.
\subsection{Population dynamics algorithm for the spectral distribution}
We use a population dynamics algorithm similar as described in Sec.~\ref{sec:App_Boundary_NumericalImplementation} to solve the Eq.~(\ref{eq:SpDensityQTilxx}).  We represent the 
distributions $q(\mathsf{g},\mathsf{g}')$ with a population of pairs  $(\mathsf{g}^{(j)},\mathsf{g}'^{(j)})$ of $2\times2$ matrices $\mathsf{g}^{(j)}$ and $\mathsf{g}'^{(j)}$ with complex entries.      

The population is initialized and updated as follows:  
\begin{enumerate}
\item Initialise  the population by drawing  $N_{\rm p}$   independent realizations of random variables $(\mathsf{g}^{(i)},\mathsf{g}'^{(i)})$  from a certain distribution $p_{\rm init}(\mathsf{g},\mathsf{g}')$;
\item Generate a degree $k$ from the distribution $\frac{k}{c}p_{\rm deg}(k)$;
\item Uniformly and  randomly select $k-1$ elements  $(\mathsf{g}_\ell,\mathsf{g}'_\ell)$ from the population and draw $k-1$ random variables $(u_\ell,l_\ell)$ from the distribution $p(u_\ell,l_\ell)$, with $\ell=1,2,\ldots, k-1$;
\item Compute 
\begin{equation}\label{eq:UpdateRuleBoundaryStabilityPopDyn}
\mathsf{g} = \left(\mathsf{z}_\eta- \sum_{\ell=1}^{k-1}\mathsf{J}_\ell \mathsf{g}_\ell \mathsf{J}_\ell^\dagger  \right)^{-1}, \  {\rm and} \quad \mathsf{g}'  = -\mathsf{g}\left( \begin{pmatrix}
0 & 0 \\ 1& 0
\end{pmatrix}-  \sum_{\ell=1}^{k-1}\mathsf{J}_\ell \mathsf{g}'_\ell \mathsf{J}_\ell^\dagger  \right)  \mathsf{g}  ;
\end{equation}
\item Uniformly  and randomly select an index   $i\in \{1,\ldots,N_p\}$ and replace $(\mathsf{g}^{(i)},\mathsf{g}'^{(i)})$  by $(\mathsf{g},\mathsf{g}')$.
\end{enumerate}

In this case, the precise form of the distribution $p_{\rm init}$ does not matter.  The steps (2-5) are repeated a number $N_{\rm eq}$ of times until the estimated distribution 
\begin{equation}
\hat{q}(\mathsf{g}, \mathsf{g}') = \frac{1}{N_{\rm p}} \sum^{N_{\rm p}}_{j=1}\delta(\mathsf{g} - \mathsf{g}^{(j)})\delta(\mathsf{g}' - \mathsf{g}'^{(j)})
\end{equation}
has converged to its stationary value.

After the distribution $\hat{q}(\mathsf{g}, \mathsf{g}')$ has converged, we  compute a first estimate $\hat{\rho}_1$ of $\rho$ from Eqs.~(\ref{eq:rho2}) and (\ref{eq:SpDensityQTil}) with a Monte-Carlo integration algorithm.  We then repeat the steps (2-5) a number $N_{\rm p}$ of times and then compute a second estimate of $\hat{\rho}_2$ of $\rho$.  We repeat this procedure a number $n_{\rho}$ of times to obtain the final estimate 
  \begin{equation}
\label{eq:rho_s}
\hat{\rho} =\frac{1}{n_\rho}\sum_{i=1}^{n_\rho} \hat{\rho}_i.
\end{equation} 
The error on $\hat{\rho}$ is computed based on the standard deviation of the set of  sampled $\hat{\rho}_i$.

\subsection{Numerical results for antagonistic matrices on Erd\H{o}s-R\'{e}nyi graphs}

\begin{figure}[htbp]
     %\begin{subfigure}[t]{0.4\textwidth}
         %\centering
        \includegraphics[width=0.4\textwidth]{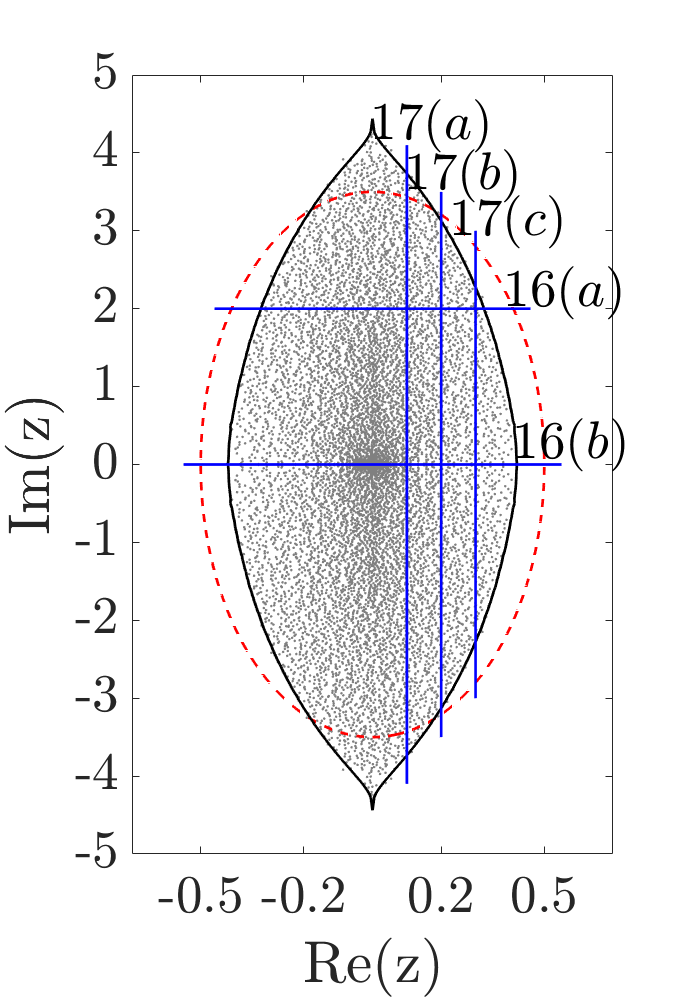}
         \caption{ Eigenvalues of one matrix sampled from Model A  with $c=4$, as in Figs.~\ref{fig:BoundaryFiniteNc2_4Poiss}(a)-(d),  but now with $N=5000$.   The blue lines denote the cuts along which we compute the spectral distribution $\rho$ in Figs.~\ref{fig:SPDEnsityCutImFixed}  and \ref{fig:SPDEnsityCutReFixed}.  The red dashed line is the elliptic law given by  Eqs.~(66-68)  with $\sigma^2= c$ and $\tau  = -3c/4$, and the black solid line denotes the boundary of the spectrum in the limit of infinitely large $N$ obtained with the cavity theory of  Sec.~\ref{sec:Theory}.}
         %\label{fig:y equals x}
     %\end{subfigure}
 \label{fig:AntagonisticBoundaryc4WithCuts}
\end{figure}

 We show the spectral distribution of antagonistic matrices on Erd\H{o}s-R\'{e}nyi graphs (Model A) with $c=4$ along cuts in the complex plane that are parallel to the real or imaginary axis, as indicated in Fig.~\ref{fig:AntagonisticBoundaryc4WithCuts}. 

\begin{figure*}[htbp]
\centering
%\subfigure[${\rm Im}(z)=0$]{\includegraphics[width=0.7\textwidth]{FinalLatexHardAnt_SpectralDensityVSDirectDiagNormalized_CutHistogramIm0_20000Populationsize4000Samples_rho4c300eta.png}\label{subfig:SPDEnsityCutIm0}}
\subfigure[${\rm Im}(z)=2$]{\includegraphics[width=0.45\textwidth]{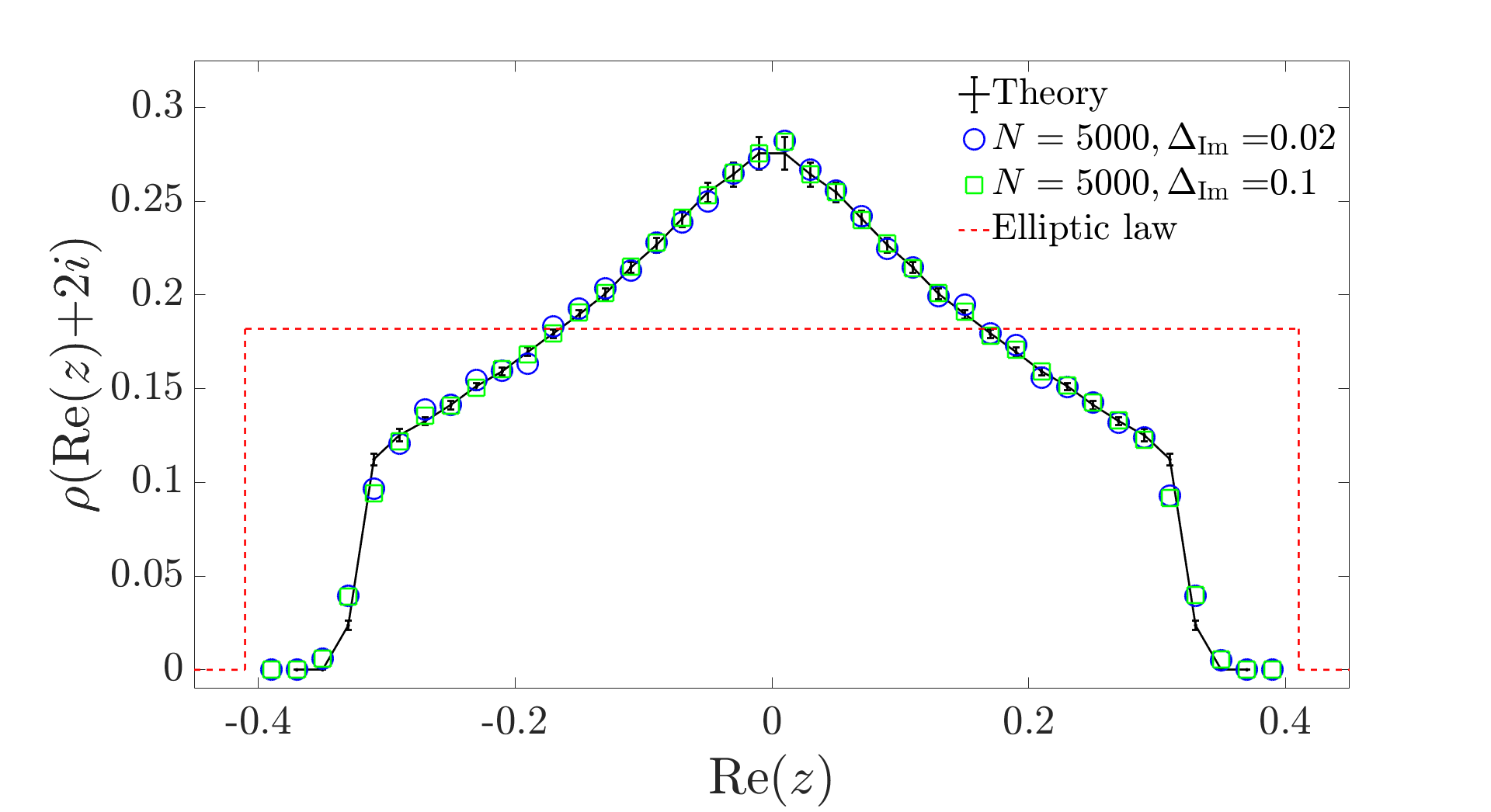}\label{subfig:SPDEnsityCutIm2}}
\subfigure[${\rm Im}(z)=0$]{\includegraphics[width=0.45\textwidth]{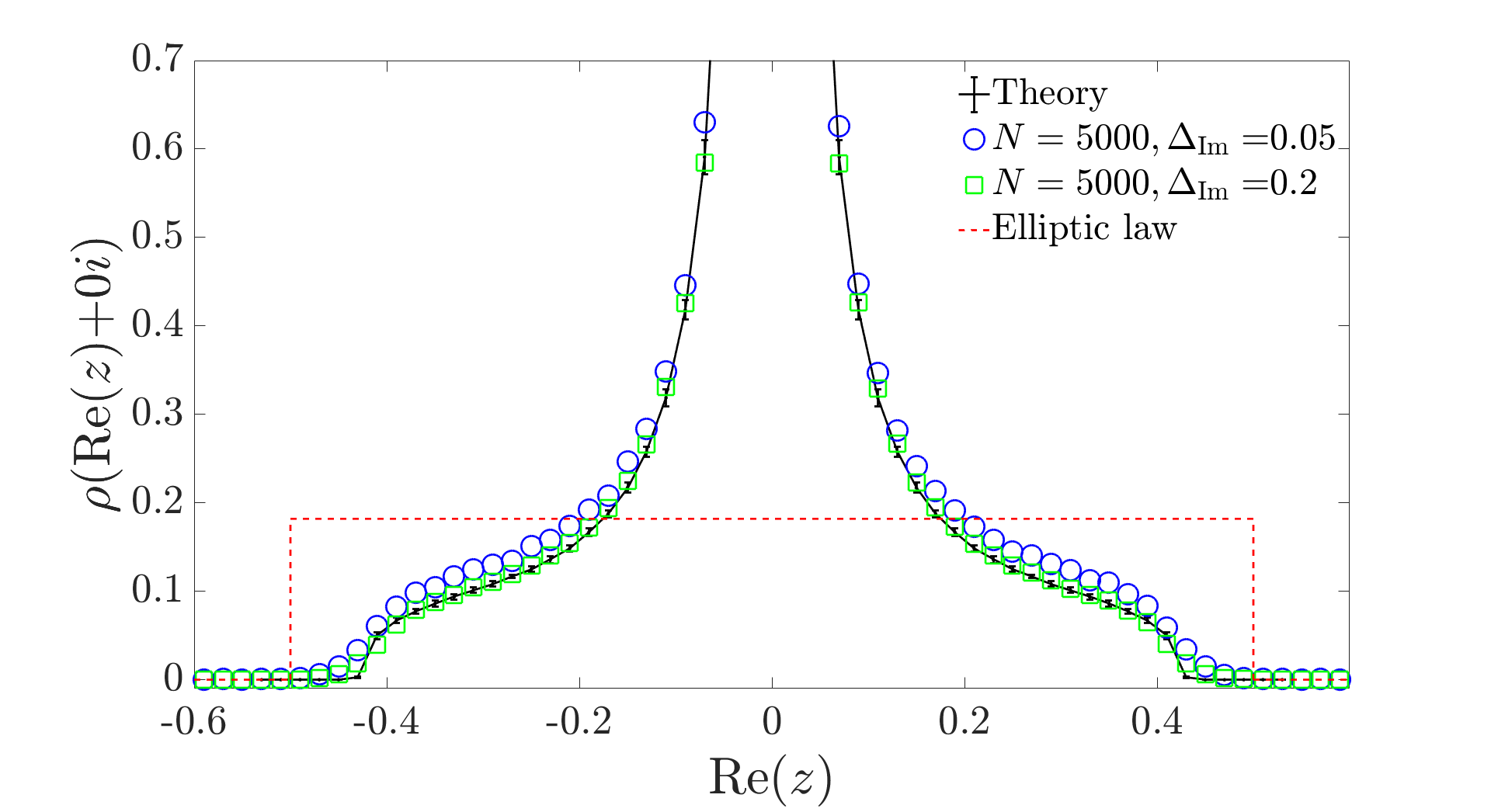}\label{subfig:SPDEnsityCutIm0Zoom}}
  \caption{The spectral distribution $\rho$ along   cuts  parallel to the real axis for random matrices of Model A with $c=4$.   The values  of ${\rm Im}(z)$ are indicated in the captions and the cuts are shown in Fig.~\ref{fig:AntagonisticBoundaryc4WithCuts}.   Theoretical results from the cavity method (solid black line) are compared with histograms obtained   by numerically diagonalizing $10^4$ matrices of size $N =5000$ and collecting all eigenvalues in a strip of width $\Delta_{\rm Im}$ (markers). The spectral distribution is also compared with the elliptic law given by Eqs.~(66-68)   with $\sigma^2= c$ and $\tau  = -3c/4$ (red dashed line).   Error bars denote the numerical error on the $\rho$  value computed with population dynamics,  as explained in the main text.}
 \label{fig:SPDEnsityCutImFixed}
\end{figure*}

\begin{figure*}[htbp]
\centering
\subfigure[${\rm Re}(z)=0.1$]{\includegraphics[width=0.45\textwidth]{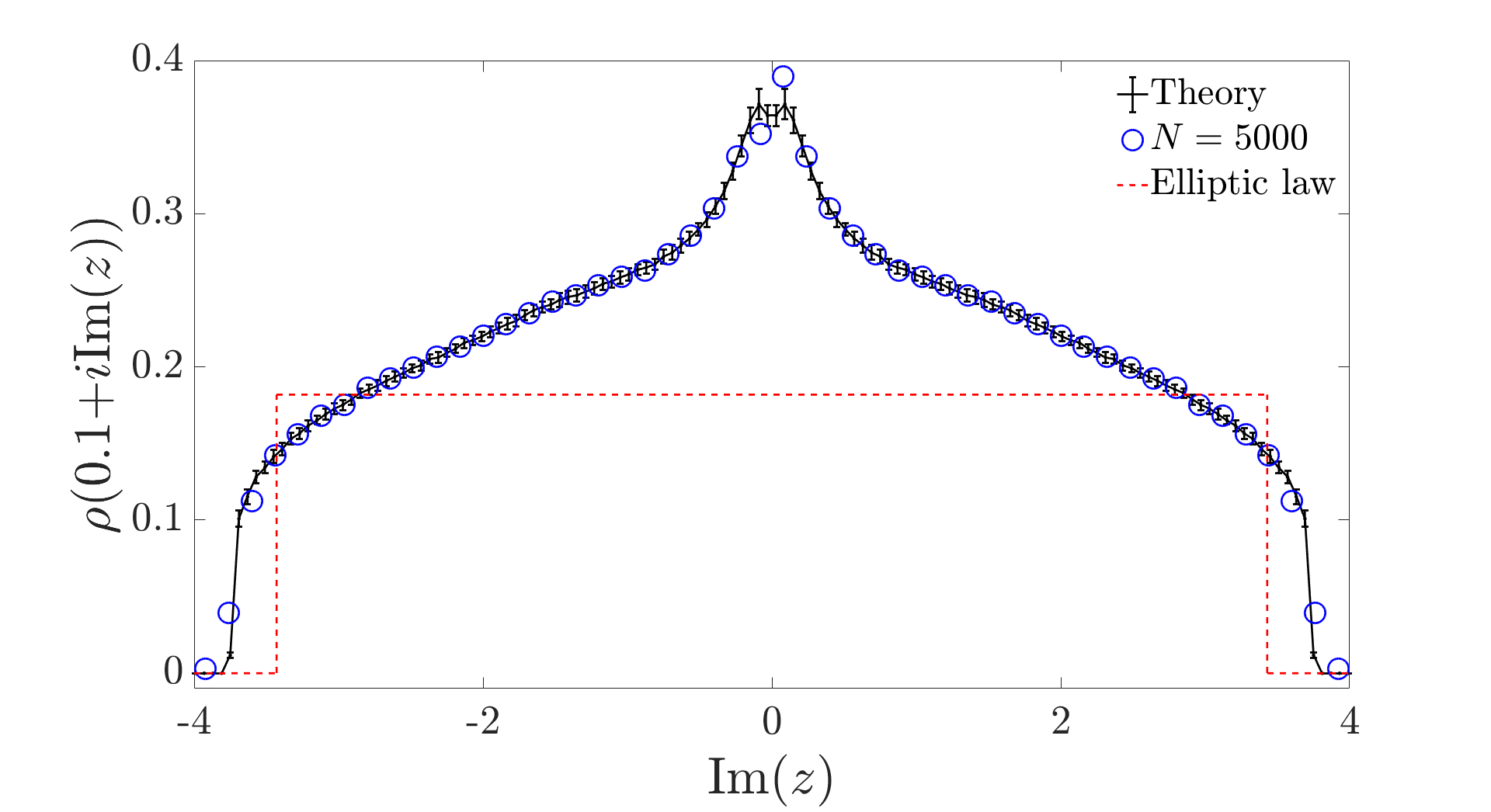}\label{subfig:SPDEnsityCutRe0p1}}
\subfigure[${\rm Re}(z)=0.2$]{\includegraphics[width=0.45\textwidth]{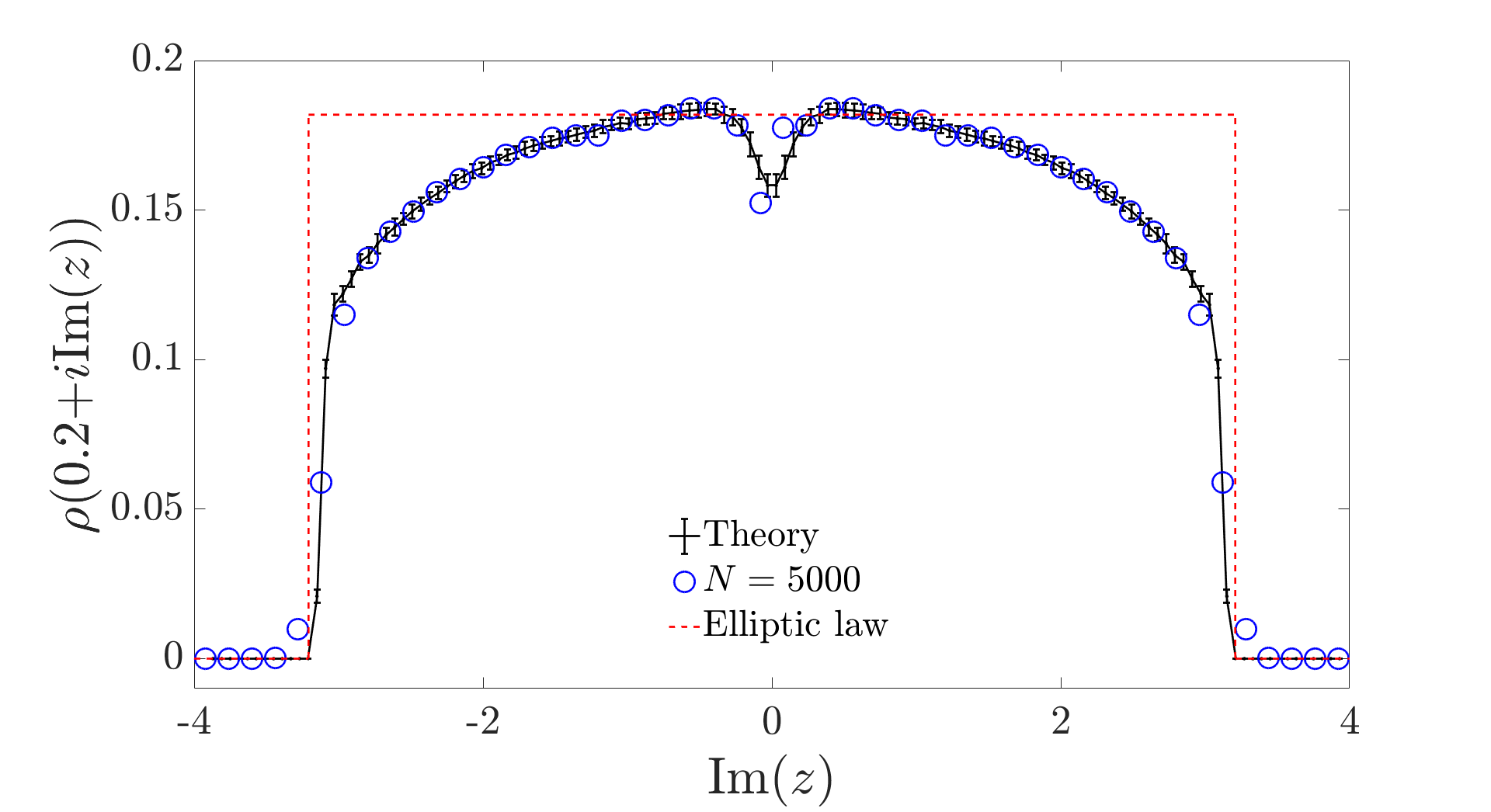}\label{subfig:SPDEnsityCutRe0p2}}
\subfigure[${\rm Re}(z)=0.3$]{\includegraphics[width=0.45\textwidth]{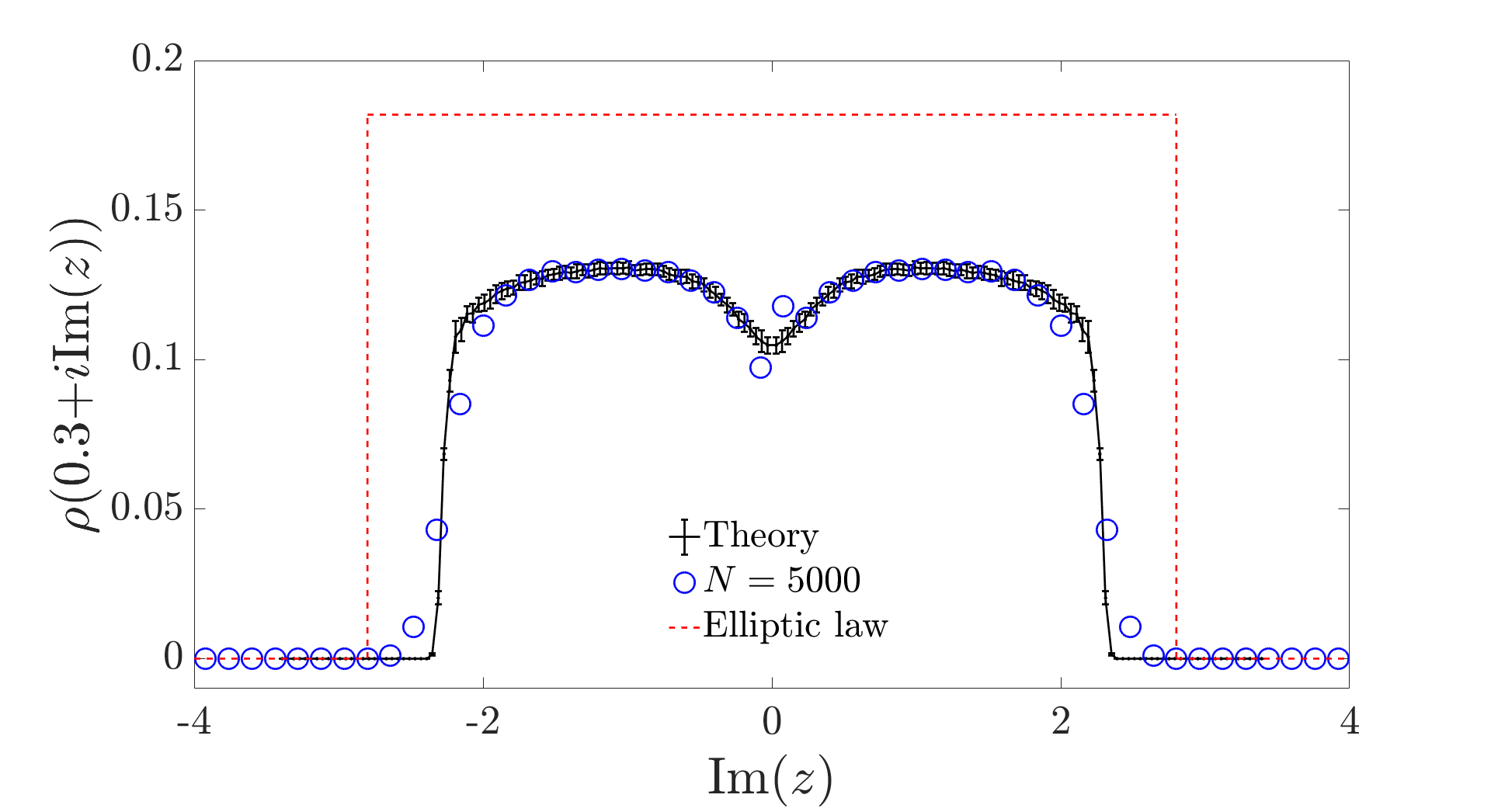}\label{subfig:SPDEnsityCutRe0p3}}
  \caption{The spectral distribution $\rho$ along   cuts  parallel to the imaginary axis for random matrices of Model A with $c=4$.   The present figure is similar to Fig.~\ref{fig:SPDEnsityCutImFixed}, with the  difference that the spectral distribution $\rho$ shown is  for cuts parallel to the imaginary axis.   The values of ${\rm Re}(z)$ are given and the width $\Delta_{{\rm Re}} = 0.02$.   }
 \label{fig:SPDEnsityCutReFixed}
\end{figure*}

Figs.~\ref{fig:SPDEnsityCutImFixed} and \ref{fig:SPDEnsityCutReFixed} show numerical results    for the spectral distribution obtained with the population dynamics algorithm explained above.   In addition, the figures show histograms obtained  from directly diagonalizing  matrices of finite size $N=5000$.   

 We find an excellent agreement between theory and numerical experiments.     Although some care should be taken to interpret the results in Fig.~\ref{fig:SPDEnsityCutImFixed}(b).   While in the population dynamics we can compute $\rho$ exactly along a cut parallel to the imaginary axis,  in direct diagonalization results the spectral distribution $\rho$ is estimated with a histogram of eigenvalues located in a strip of width $\Delta_{\rm Im}$.  For  cuts that are not on the real line, ${\rm Im}(z)\neq 0$, the estimate of $\rho$ improves when $\Delta_{\rm Im}$ decreases, as shown in Fig.~\ref{fig:SPDEnsityCutImFixed}(a).   On the other hand, when the cut is along the real line, ${\rm Im}(z)= 0$, then the estimate of $\rho$  worsens  when $\Delta_{\rm Im}$ decreases.  This is because of finite size effects that are significant for $\rho$ on the real line.  Indeed, on the real line there is an accumulation of eigenvalues, as one can clearly observe in Fig.~\ref{fig:BoundaryFiniteNc2_4Poiss}.   The number of eigenvalues that is real scales as $O(\sqrt{N})$~\cite{edelman1994many, Izaak_Metz_2019}, and is thus a subleading contribution to $\rho$.  However, if we set ${\rm Im}(z)= 0$ and the width $\Delta_{\rm Im}$ is smaller than the typical separation between eigenvalues, then the strip contains only the  $O(\sqrt{N})$ of real eigenvalues, which do not follow the statistics given by  $\rho$.   Therefore, for ${\rm Im}(z)= 0$ it is necessary to consider a  $\Delta_{\rm Im}$ that is small but not too small.  

Figs.~\ref{fig:SPDEnsityCutImFixed} and \ref{fig:SPDEnsityCutReFixed} also compare $\rho$ with the elliptic law given by Eqs.~(66-68), which here amounts to  $\sigma^2= c$ and $\tau  = -3c/4$.  
While the boundary of the spectrum is well predicted by the elliptic law, a feature already observed in Fig.~\ref{fig:BoundaryFiniteNc2_4Poiss}, this is not the case for the spectral distribution $\rho$.   

From Fig.~\ref{fig:SPDEnsityCutImFixed}(b), we observe that the  spectral distribution of antagonistic matrices on Erd\H{o}s-R\'{e}nyi graphs diverges for $z\rightarrow 0$.   Interestingly, this divergence is also observed  in the adjacency matrices of nondirected Erd\H{o}s-R\'{e}nyi graphs, see Ref.~\cite{Reimer_sparse}, and in the adjacency matrices of directed Erd\H{o}s-R\'{e}nyi graphs, see Figure in Ref.~\cite{Izaak_Metz_2019}.    On the other  hand,  the divergence does not occur in  regular graphs.   Hence, the divergence of the spectral distribution for $z\rightarrow 0$ is a generic feature due to  network topology and  is independent of the nature of the interactions $J_{ij}$.    It would be interesting to have a precise understanding of the origin of the peak.

\section{Limiting cases}\label{App:Limiting} 
In Sec.~IV,  we have  derived an exact formula for the boundary of the support set $\mathcal{S}$  of the spectral distribution $\rho$ of random matrices in the general model defined in Sec.~II A.   In particular, we have shown that the boundary of $\mathcal{S}$ is given by  the edge of stability  of Eq.~(\ref{eq:StabBoundary}) at the trivial solution given by Eq.~(\ref{eq:q0Delta}).    Here, we show that the boundary of   $\mathcal{S}$ obtained from the stability analysis of Eq.~(\ref{eq:StabBoundary}) corresponds with  results obtained previously in the literature for the limiting cases of    oriented random matrices~\cite{neri2019spectral}, for which $J_{ij}J_{ji} = 0$, and of dense matrices~\cite{Sommers_et_al, Nguyen,gotzeelliptic, allesina2012}, for which~$c\rightarrow \infty$.

\subsection{Oriented ensemble}\label{sec:App_OrientedBoundary}
In the oriented ensemble, $p(u,l)$ is of the form given by  Eq.~(\ref{eq:oriented}), such that, $J_{ij}J_{ji}=0$ for each pair of indices $i$ and $j$.   We show in this appendix that for oriented matrices the boundary of the continuous part of the spectrum is given by values of $z\in\mathbb{C}$ for which 
\begin{equation}\label{eq:boundary_oriented}
\left|z \right|^2=\frac{\left\langle k (k-1) \right\rangle_{p_{\rm deg}}  }{2c}\left\langle l^2 \right\rangle_{\tilde{p}},
\end{equation}    
where $\tilde{p}$ is the distribution that appears on the right-hand side of  Eq.~(\ref{eq:oriented}).
First, we show that Eq.~(\ref{eq:boundary_oriented})   determines the  edge of stability of Eqs.~(\ref{eq:StabBoundary}) at the solution~Eq.~(\ref{eq:q0Delta}), and  subsequently we show the correspondence with the results for the boundary of $\mathcal{S}$ in Ref.~\cite{neri2019spectral}.

\subsubsection{Derivation of Eq.~(\ref{eq:boundary_oriented}) from the general theory in  Sec.~IV}   

We show that the edge of stability of  Eq.~(\ref{eq:StabBoundary}) at the trivial solution  Eq.~(\ref{eq:q0Delta}) 
 is determined by  Eq.~(\ref{eq:boundary_oriented}). 

In the oriented ensemble, the denominators in the delta distributions of  Eq.~(\ref{eq:StabBoundary}) simplify since  $u_\ell l_\ell =0 $.   As a consequence, Eq.~(\ref{eq:StabBoundary}) reads
\begin{equation}
Q(g,h)=\delta\left(g+\frac{1}{z}  \right) R(h) ,   \label{eq:CavityToPop1Eta0withEps_Oriented}
\end{equation}
where $R$ solves
\begin{equation}
R(h) = \sum_{k=1}^{\infty}\frac{kp_{\rm deg}(k)}{c}\int \prod_{\ell=1}^{k-1} \mathrm{d}\epsilon_\ell R(h_\ell)\int  \prod_{\ell=1}^{k-1}\mathrm{d}u_\ell  \mathrm{d}l_\ell \: p^{\rm O}(\JU_\ell,\JD_\ell)\:
 \delta\left(h -\frac{ \sum_{\ell=1}^{k-1} h_\ell  l_\ell^2}{\left|z \right|^2}   \right),  \label{eq:R}
\end{equation}
where  $p^{\rm O}$ is the distribution defined in Eq.~(\ref{eq:oriented}).
Eq.~(\ref{eq:CavityToPop1Eta0withEps_Oriented}) implies that  for oriented matrices the  variables $g$ and $h$ decouple.     Hence, it suffices to study the stability of   Eq.~(\ref{eq:R}) at
the trivial solution $R_0(h) = \delta(h)$.

Evaluating the average value of $h$, we readily obtain 
\begin{eqnarray}
\left\langle h\right\rangle_{R}=\int \mathrm{d}g \mathrm{d}h \; h\; Q(g,h) = \int {\rm d}h   R(h) h 
=\frac{\left\langle k (k-1) \right\rangle_{p_{\rm deg}}  }{2c}\left\langle  l^2 \right\rangle_{\tilde{p}} \frac{\left\langle h\right\rangle_{R}}{\left|z \right|^2},\label{eq:mean_epsilon_oriented} 
\end{eqnarray} 
where $\tilde{p}$ is the distribution appearing on the right-hand-side of  Eq.~(\ref{eq:oriented}).
Hence, the edge of stability is given by the values of $z$ for which  Eq.~(\ref{eq:boundary_oriented}) holds, which is what we were meant to show.

\subsubsection{Derivation of Eq.~(\ref{eq:boundary_oriented}) from the theory in  Ref.~\cite{neri2019spectral}}

We derive the result Eq.~(\ref{eq:boundary_oriented}) from the results obtained in Ref.~\cite{neri2019spectral}.   
Reference~\cite{neri2019spectral} considers random matrices  $\mathbf{A}$  of the form 
\begin{equation}
\mathbf{A} = \tilde{\mathbf{J}}\circ \tilde{\mathbf{C}} \label{eq:orientedModelx}
\end{equation}
where $\tilde{\mathbf{C}}$ is the adjacency matrix of a random, directed graph with a prescribed joint degree distribution $p_{\rm deg}(k_{\text{in}},k_{\text{out}})$ of indegrees $k_{\text{in}}$ and outdegrees $k_{\text{out}}$, and where $\tilde{\mathbf{J}}$ is a random matrix with real-valued i.i.d.~entries drawn from a distribution $\tilde{p}(x)$.  

According to Ref.~\cite{neri2019spectral}, in the limit $N\rightarrow \infty$ the boundary of the continuous part of the spectrum of $\mathbf{A}$ is given by the values $z\in\mathbb{C}$ for which 
\begin{equation}
    |z|^2 = \frac{\langle k_{\rm in}k_{\rm out} \rangle}{\tilde{c}} \langle x^2  \rangle_{\tilde{p}},  \label{eq:boundaryPrevious}
\end{equation}
where $\tilde{c}$ is the mean indegree (or outdegree)
\begin{equation}
    \tilde{c}=\left\langle k_{\text{in}} \right\rangle_{p_{\rm deg}}=\left\langle k_{\text{out}} \right\rangle_{p_{\rm deg}}.  \label{eq:meanout}
\end{equation}

Note that  Eq.~(\ref{eq:orientedModelx}) considers a random, directed graph $\tilde{\mathbf{C}}$  with symmetric couplings $ \tilde{\mathbf{J}}$, while  Eq.~(13) with $p = p^{\rm O}$ considers a random, nondirected graph $\mathbf{C}$ with asymmetric couplings $\mathbf{J}$.   Both models are related through the correspondence 
\begin{equation}
p_{\rm deg}(k_{\text{in}},k_{\text{out}})=\sum_{k=0}^\infty p_{\rm deg}(k) \frac{1}{2^k} \sum_{n=0}^k  {k\choose n}
\delta_{k_{\text{in}},n} \delta_{k_{\text{out}},k-n} \label{eq:degreeDistriJoint}
\end{equation} 
between the degree distributions of the directed and nondirected graphs.

Using Eq.~(\ref{eq:degreeDistriJoint}), one can show that 
\begin{equation}
\tilde{c} = \frac{c}{2} \label{eq:ctilde}
\end{equation}   
and 
\begin{equation}
\langle k_{\rm in}k_{\rm out} \rangle_{p_{\rm deg}}  = \frac{1}{4} \left\langle k (k-1)\right\rangle_{p_{\rm deg}} . \label{eq:kinkout}
\end{equation}

Substituting  Eqs.~(\ref{eq:ctilde}) and (\ref{eq:kinkout})  into Eq.~(\ref{eq:boundaryPrevious}), we obtain   Eq.~(\ref{eq:boundary_oriented}), which we meant to derive.

\subsection{Large connectivity limit}
\label{sec:App_Large_cLimit}
We take the limit $c\rightarrow \infty$, with $c/N\approx 0$,   and find the law given by Eq.~(\ref{eq:elliptic1}), which is reminiscent of the elliptic law for i.i.d.~matrices with $c=N-1$.   In order to obtain a bounded support set, we rescale     $\sigma$ and $\tau$ as in Eq.~(\ref{eq:tau_elliptic_main}).

\subsubsection{Support set} \label{sec:App_Large_cLimit_Boundary}
We first derive an expression for the support set $\mathcal{S}$.   We show that  the edge of stability of the  Eq.~(\ref{eq:StabBoundary})  at the trivial solution Eq.~(\ref{eq:q0Delta}) provides us with the boundary of the elliptic law in Eqs.~(\ref{eq:elliptic1}).   Using the law of large numbers, we can identify the  sums inside the delta distributions on the right-hand-side of  Eq.~(\ref{eq:StabBoundary}) with their mean values.   As a consequence, to the leading order $1/c$, the distribution $Q$ takes the form 
 \begin{eqnarray}
 Q(g,h) = \delta(g - \hat{g})\delta(h - \hat{h})
 \end{eqnarray}
 where $\hat{g}$ and $\hat{h}$ satisfy the self-consistent equations
 \begin{equation}
     \hat{g} =- \frac{1}{z +  \hat{g} \tau }\label{eq:g}
 \end{equation}
 and 
 \begin{equation}
     \hat{h} =  \hat{h} |\hat{g}|^2 \sigma^2.
 \end{equation}  
The edge of stability of the previous equation is given by
  \begin{equation}
  |\hat{g}(z)|^2  = \frac{1}{\sigma^2}.  \label{eq:modg}
 \end{equation} 
 
In order to obtain  $|\hat{g}|^2$, we first consider the two equations 
\begin{eqnarray}
 \hat{g}+\overline{\hat{g}} = - |\hat{g}|^2 \left( z+\overline{z}+(\hat{g}+\overline{\hat{g}})\tau\right)\label{eq:1}
\end{eqnarray}
and 
\begin{eqnarray}
 \hat{g}-\overline{\hat{g}} = - |\hat{g}|^2 \left( z-\overline{z}+(\hat{g}-\overline{\hat{g}})\tau\right),\label{eq:2}
\end{eqnarray}
which are readily obtained from Eq.~(\ref{eq:g}).
Using Eq.~(\ref{eq:modg}) in Eqs.~(\ref{eq:1}) and (\ref{eq:2}), we obtain 
\begin{eqnarray}
 2{\rm Re}(\hat{g}) = \hat{g}+\overline{\hat{g}} = -\frac{2{\rm Re}(z)}{\sigma^2 + \tau} \label{eq:Reg}
\end{eqnarray}
and
\begin{eqnarray}
 2{\rm Im}(\hat{g})= \hat{g}-\overline{\hat{g}} = \frac{2{\rm Im}(z)}{\sigma^2 - \tau},   \label{eq:Img}
\end{eqnarray}
and thus also
\begin{eqnarray}
|\hat{g}|^2 = \left(\frac{{\rm Re}(z)}{\sigma^2 + \tau}\right)^2  + \left(\frac{{\rm Im}(z)}{\sigma^2 - \tau}\right)^2 .  \label{eq:g2}
\end{eqnarray}
Lastly, the stability condition Eq.~(\ref{eq:modg}) together with  (\ref{eq:g2}) provides us with  the boundary 
 \begin{eqnarray}
 \left(\frac{{\rm Re}(z)}{\sigma^2 + \tau}\right)^2  + \left(\frac{{\rm Im}(z)}{\sigma^2 - \tau}\right)^2 = \frac{1}{\sigma^2}
\end{eqnarray}
for the support set of the spectral distribution of a highly connected random matrix, which is equivalent to Eq.~(68).

\subsubsection{Spectral distribution}
We compute the  spectral distribution of highly connected matrices.    In this case, we rely on the Eqs.~(\ref{eq:rec1}-\ref{eq:specrec}).   In the limit $c\rightarrow\infty$, we can  apply the law of large numbers inside the Dirac distributions of Eqs.~(\ref{eq:rec1}-\ref{eq:rec2}), leading to
\begin{equation}
 \tilde{q}(\mathsf{g}) = q(\mathsf{g}) = \delta(\mathsf{g} - \hat{\mathsf{g}}),
\end{equation}
where $\hat{\mathsf{g}}$ solves the selfconsistent equation (setting $\eta=0$) 
\begin{eqnarray}
\hat{\mathsf{g}} = \left(\begin{array}{cc}\hat{g}_{11} & \hat{g}_{12} \\ \hat{g}_{21} &\hat{g}_{22} \end{array}\right) &=&  \left(\begin{array}{cc}   -\hat{g}_{22}\sigma^2 & z  - \hat{g}_{21} \tau \\  \overline{z} - \hat{g}_{12}\tau &  -\hat{g}_{11}\sigma^2\end{array}\right)^{-1}  \nonumber \\
&=& \frac{1}{ \hat{g}_{11}\hat{g}_{22}\sigma^4 - (\overline{z}-\hat{g}_{12}\tau)(z-\hat{g}_{21}\tau)} \left(\begin{array}{cc} - \hat{g}_{11}\sigma^2 & -z  + \hat{g}_{21} \tau \\  -\overline{z} +\hat{g}_{12}\tau & -\hat{g}_{22}\sigma^2\end{array}\right)\label{eq:g_inverse_denseLimit}
\end{eqnarray}
and the spectral distribution is given by
\begin{equation}
    \rho(z) = \frac{1}{\pi} \frac{{\rm d}}{{\rm d}\overline{z}}\hat{g}_{21}. \label{eq:rhoDefx}
\end{equation}

Equation~(\ref{eq:g_inverse_denseLimit}) implies that
\begin{equation}
    \hat{g}_{11}\hat{g}_{22} =  \frac{    \hat{g}_{11}\hat{g}_{22}  \sigma^4}{\left[ \hat{g}_{11}\hat{g}_{22}\sigma^4 - (\overline{z}-\hat{g}_{12}\tau)(z-\hat{g}_{21}\tau) \right]^2}
\end{equation}
such that either  
\begin{equation}
\hat{g}_{11}=\hat{g}_{22}=0 \label{eq:trivialSolg}
\end{equation}
or 
\begin{equation}
  \left[ \hat{g}_{11}\hat{g}_{22}\sigma^4 - (\overline{z}-\hat{g}_{12}\tau)(z-\hat{g}_{21}\tau) \right]^2=  \sigma^4.\label{eq:nontrivialSolg}
\end{equation} 
Equation (\ref{eq:trivialSolg}) is the trivial solution and Eq.~(\ref{eq:nontrivialSolg})  is the nontrivial solution.     

 In Sec.~\ref{sec:App_Large_cLimit_Boundary}, we have shown that  the trivial solution Eq.~(\ref{eq:trivial}) is stable for all $z$ for which 
 \begin{eqnarray}
 \left(\frac{{\rm Re}(z)}{\sigma^2 + \tau}\right)^2  + \left(\frac{{\rm Im}(z)}{\sigma^2 - \tau}\right)^2 > \frac{1}{\sigma^2} \label{eq:ellipse2},
\end{eqnarray} 
while the nontrivial solution holds for 
 \begin{eqnarray}
 \left(\frac{{\rm Re}(z)}{\sigma^2 + \tau}\right)^2  + \left(\frac{{\rm Im}(z)}{\sigma^2 - \tau}\right)^2 \leq \frac{1}{\sigma^2} \label{eq:ellipse3}.
\end{eqnarray}

   In what follows, we first compute the spectral distribution for the trivial solution and then we compute it for the nontrivial solution.

\subparagraph{Trivial solution.}
  For the trivial solution, Eq.~\eqref{eq:g_inverse_denseLimit} reduces to the  two equations
\begin{eqnarray}
\hat{g}_{21}&=\displaystyle \frac{1}{  z-\hat{g}_{21}\tau} \\
\hat{g}_{12}&=\displaystyle  \frac{1}{  \overline{z}-\hat{g}_{12}\tau},
\end{eqnarray}
which admit two complex solutions
\begin{equation}
    \hat{g}_{21}=\displaystyle \frac{z\pm \sqrt{z^2-4\tau}}{2\tau}. \label{eq:trivial}
\end{equation}    
For $|z|>2\sqrt{|\tau|}$, this is an analytical function in $z$, and therefore
\begin{equation}
\frac{{\rm d}}{{\rm d}\overline{z}}    \hat{g}_{21} = 0 .
\end{equation}
Since for all $z$  for which Eq.~(\ref{eq:ellipse2}) holds, it also holds that $|z|>2\sqrt{|\tau|}$,  we obtain 
that
 \begin{eqnarray}
 \rho(z) = 0 \quad {\rm if}  \quad  \left(\frac{{\rm Re}(z)}{\sigma^2 + \tau}\right)^2  + \left(\frac{{\rm Im}(z)}{\sigma^2 - \tau}\right)^2 > \frac{1}{\sigma^2}.
\end{eqnarray}

\subparagraph{Non-trivial solution.}
For the nontrivial solution Eq.~(\ref{eq:nontrivialSolg}) we obtain
\begin{equation}
\hat{\mathsf{g}} = \left(\begin{array}{cc}\hat{g}_{11} & \hat{g}_{12} \\ \hat{g}_{21} &\hat{g}_{22} \end{array}\right) = \pm\frac{1}{\sigma^2} \left(\begin{array}{cc} - \hat{g}_{11}\sigma^2 & -z  + \hat{g}_{21} \tau \\  -\overline{z} +\hat{g}_{12}\tau & -\hat{g}_{22}\sigma^2\end{array}\right)
\end{equation}
and therefore
\begin{eqnarray}
  \pm\hat{g}_{21}\sigma^2&=  -\overline{z} +\hat{g}_{12}\tau, \\
    \pm \hat{g}_{12}\sigma^2&=   -z +\hat{g}_{21}\tau.
\end{eqnarray}
From these equations we obtain  a closed eqution for $\hat{g}_{21}$, viz., 
\begin{eqnarray}
\hat{g}_{21} &=  \mp \displaystyle \frac{\sigma^2}{\sigma^4- \tau^2}\overline{z} -\displaystyle \frac{ \tau}{\sigma^4-\tau^2}z,
\end{eqnarray}
and 
accordingly
\begin{eqnarray}
    \rho(z)&= \frac{1}{\pi}\frac{{\rm d}}{{\rm d}\overline{z}} \hat{g}_{21}  = \ \displaystyle\frac{1}{\pi} \frac{\sigma^2}{\sigma^4- \tau^2},
\end{eqnarray}
if we select the positive solution.

\section{Giant components in random graphs}\label{App:LCC} 
We  revisit   percolation theory  for nondirected random graphs \cite{molloy_reed_1998} and directed random graphs   \cite{PhysRevE.64.025101}, and then apply percolation theory to the random graph ensembles with degree distributions  givenm by Eq.~(\ref{eq:p_k_linearcomb}).  For a nondirected graph, $p_{\rm deg}$ is the degree distribution, while for a directed graph we obtain the joint distribution of indegrees and outdegrees from  Eq.~(\ref{eq:degreeDistriJoint}).

\subsection{Largest connected component in nondirected graphs}\label{sec:LLC}  
Let $G = (V,E)$ be a graph with $V$ a set of vertices and $E$ a set of nondirected edges.   We say that a subgraph $G' = (V',E')$ of $G$ is connected if for each pair of vertices $i\in V'$ and $j\in V'$ there exists a path of edges that belong to $E'$ that connect  $i$ to $j$.   The largest connected component is the largest subgraph $G'$ of $G$ that is connected, i.e., both the order   $|V'|$ and the size $|E'|$ of the subgraph or maximal.

The relative order of the largest connected component is defined by  
\begin{equation}
f(G) = \frac{|V'|}{N}. 
\end{equation}

We consider now nondirected, random graphs with a prescribed degree distribution $p_{\rm deg}(k)$.   We denote the generating function of  $p_{\rm deg}(k)$  by 
\begin{equation}
\tilde{M}(x)=\sum_{k=0}^\infty x^k p_{\rm deg}(k) 
\end{equation}
and we will also use the generating function 
\begin{equation}
M(x)=\sum_{k=0}^\infty x^k  \frac{kp_{\rm deg}(k)}{c} =   \frac{\partial_x \tilde{M}(x)}{c},
\end{equation} 
where $c$ is the mean degree of $p_{\rm deg}(k)$.

   In the limit $N\rightarrow \infty$, the relative order $f(G)$  converges with probability one to a deterministic value $f$, which is given by~\cite{molloy_reed_1998}, 
\begin{equation}\label{eq:GiantComp}
1-f=\tilde{M}(y),
\end{equation}
where $y$ is the smallest nonnegative solution of 
\begin{equation}\label{eq:GiantCompH}
y^2=M(y).
\end{equation}

Solving  Eqs.~(\ref{eq:GiantComp}-\ref{eq:GiantCompH}) one finds that
\begin{equation}
f>0 \quad {\rm if} \quad
\sum^{\infty}_{k=0}p_{\rm deg}(k)k(k-2)>0
\end{equation}
and that
\begin{equation}
f=0 \quad {\rm if} \quad
\sum^{\infty}_{k=0}p_{\rm deg}(k)k(k-2)<0.
\end{equation}

Hence, the condition 
\begin{equation}
\sum^{\infty}_{k=0}p_{\rm deg}(k)k(k-2)=0
\end{equation}  
determines the percolation transition in undirected, random graphs.

%Lastly, to understand the results for the leading eigenvalue in Figs.~\ref{fig:leading}(a) and \ref{fig:leading}(b) we analyze the largest connected components in those graphs.     Naively, we  expect that the size of the largest connected component directly relates to system stability, as local perturbations can only reach the nodes within a given connected component.     However, this does not turn out to be the case as we show in  Figs.~\ref{fig:leading}(c) and \ref{fig:leading}(d).   

\subsection{Largest strongly connected component in directed graphs}\label{sec:LLCd}
Let $G = (V,E)$ be a directed graph with $V$ a set of vertices and $E$ a set of nondirected edges.   We say that a subgraph $G' = (V',E')$ of $G$ is strongly connected if for each pair of vertices $i\in V'$ and $j\in V'$ there exists a path starting in node $j$ and ending in node $i$  that follows the edges in $E'$, and there exists also a reverse path  that starts in node $i$ and ends in node $j$.      The largest strongly connected component is the largest subgraph $G'$ that is  strongly connected.  

\begin{figure*}[htbp]
\centering
    \subfigure[nondirected]{\includegraphics[width=0.44\textwidth]{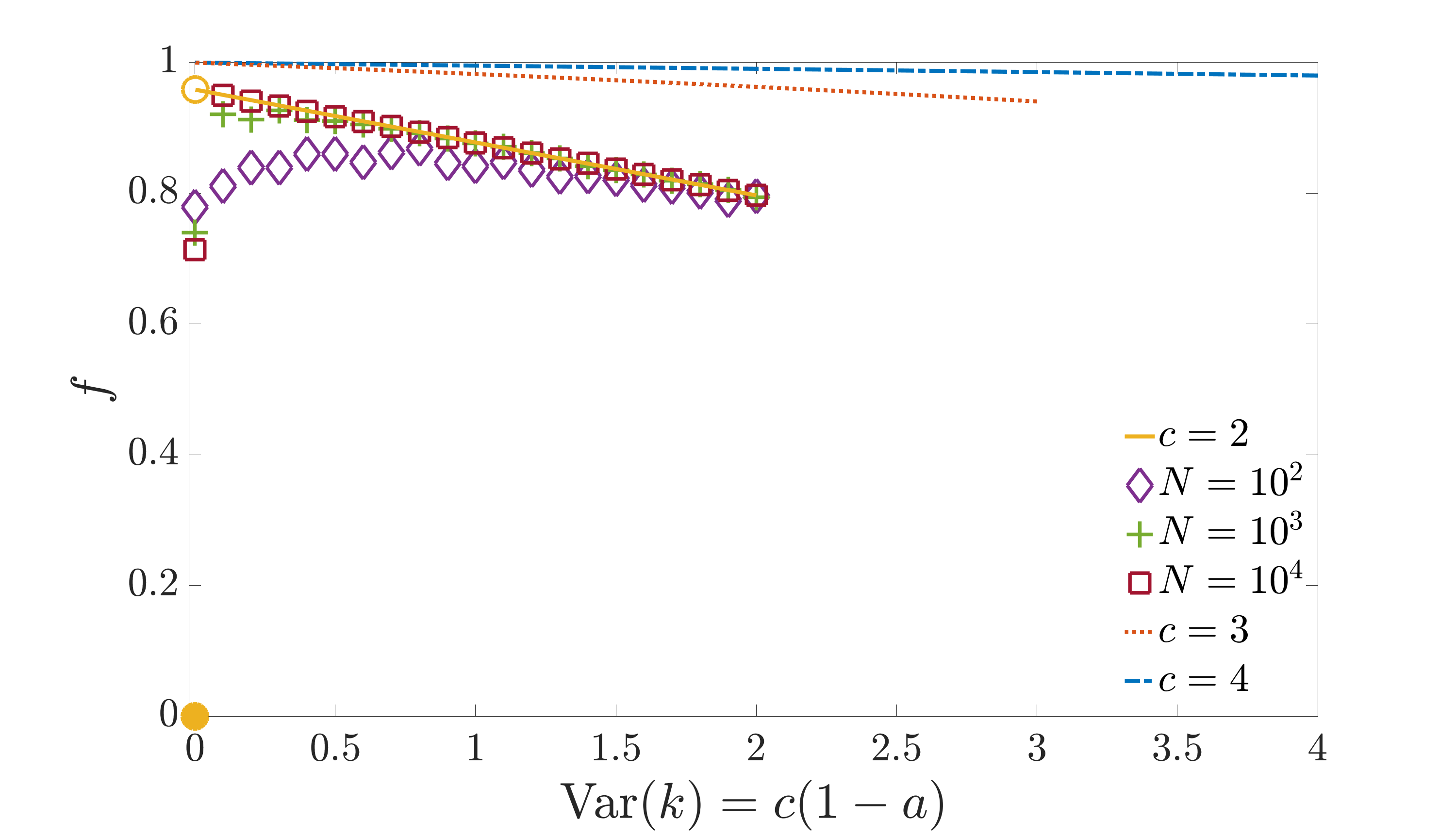}
    \label{subfig:perc1}}
         \subfigure[directed]{\includegraphics[width=0.44\textwidth]{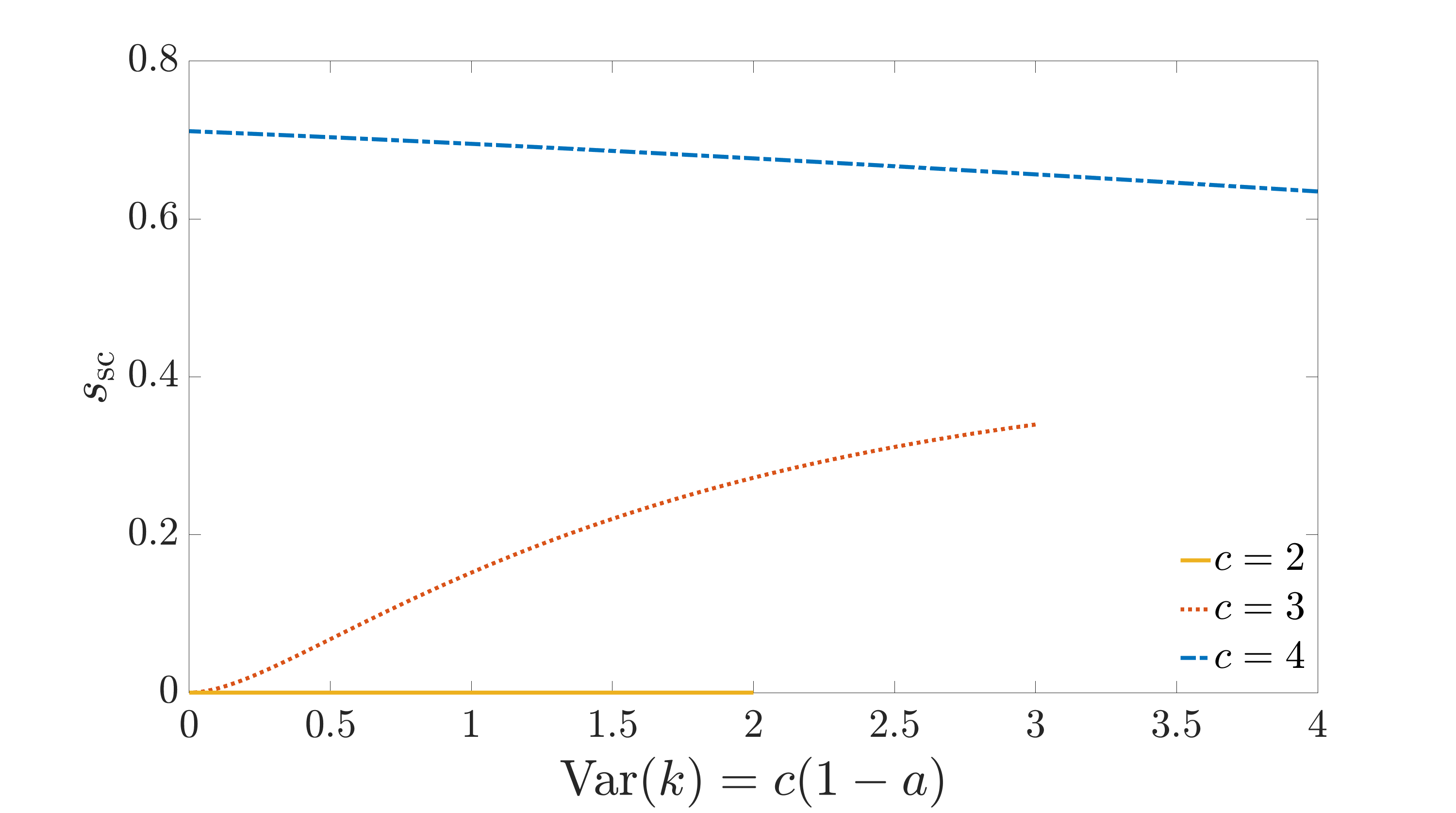}
    \label{subfig:perc2}}
    \caption{Panel (a): Relative order $f$ of the largest connected  component  of random graphs with the prescribed degree distribution given by  Eq.~(73), just as in  Fig.~10.  Lines correspond to theoretical results for infinitely large $N$ obtained from solving Eqs.~(\ref{eq:GiantComp}-\ref{eq:GiantCompH}), while the markers are simulation results for finite $N$ and $c=2$; we have included simulations in order to verify the peculiar discontinuity of $f$ at $a=1$.  Panel (b):    Relative order $s_{\rm sc}$ of the largest strongly connected component of random, directed graphs with a joint degree distribution $p_{\rm deg}(k_{\rm in},k_{\rm out})$ given by Eq.~(\ref{eq:degreeDistriJoint}), just as in  Fig.~10. Lines  are theoretical values  for infinitely large $N$ obtained from solving Eqs.~(\ref{eq:sc}-\ref{eq:bb}).
    }\label{fig:leadingAPPG2}
\end{figure*}

 We define the relative order of the largest strongly connected component as 
\begin{equation}
s_{\rm sc}(G) = \frac{|V'|}{N}. 
\end{equation}

Let us consider directed, random graphs with a prescribed degree distribution $p_{\rm deg}(k_{\rm in},k_{\rm out})$ of indegrees and outdegrees.   Then, 
it holds that~\cite{PhysRevE.64.025101, neri2019spectral}
\begin{eqnarray}
s_{\rm sc} =   s_{\rm in} +s_{\rm out}  + s_{\rm t} - s_{\rm wc} ,  \label{eq:sc}
\end{eqnarray} 
where $s_{\rm in}$, $s_{\rm out}$, $s_{\rm t}$ and $s_{\rm wc}$ are the fraction of nodes that belong to the  incomponent, outcomponent, tendrils and the weakly connected component, respectively.   It holds that  
 \begin{eqnarray}
s_{\rm out} &=& 1- \sum^{\infty}_{k_{\rm in}=0}a^{k_{\rm in}}\sum^{\infty}_{k_{\rm out}=0} p_{{\rm deg}}(k_{\rm in}, k_{\rm out}), \label{eq:sin} 
\end{eqnarray}   
\begin{eqnarray}
s_{\rm in} &=& 1- \sum^{\infty}_{k_{\rm out}=0} b^{k_{\rm out}} \sum^{\infty}_{k_{\rm in}=0} p_{{\rm deg}}(k_{\rm in},k_{\rm out}), \label{eq:sout}
\end{eqnarray}   
and 
  \begin{eqnarray}
s_{\rm t} - s_{\rm wc} =   \sum^{\infty}_{k_{\rm in}=0}\sum^{\infty}_{k_{\rm out} =0}p_{{\rm deg}}(k_{\rm in},k_{\rm out} )\:a^{k_{\rm in}} b^{k_{\rm out} } -1,
\end{eqnarray} 
where $a$ and $b$ solve 
the equations
\begin{eqnarray}
a &=& \sum^{\infty}_{k_{\rm in}=0}a^{k_{\rm in}}\sum^{\infty}_{k_{\rm out}=0}\frac{k_{\rm out} \: p_{{\rm deg}}(k_{\rm in},k_{\rm out})}{\tilde{c}},  \label{eq:a} 
\end{eqnarray}   
and
\begin{eqnarray}
b &=&\sum^{\infty}_{k_{\rm out}=0} b^{k_{\rm out}} \sum^{\infty}_{k_{\rm in}=0}\frac{k_{\rm in} \: p_{{\rm deg}}(k_{\rm in},k_{\rm out})}{\tilde{c}}, \label{eq:bb}
\end{eqnarray} 
where $\tilde{c}$  is the mean outdegree defined in  Eq.~(\ref{eq:meanout}).

Solving  Eqs.~(\ref{eq:GiantComp}-\ref{eq:GiantCompH}), we obtain that   \cite{PhysRevE.64.025101, neri2019spectral} 
\begin{equation}
s_{\rm sc}>0 \quad {\rm if} \quad
\frac{\sum^{\infty}_{k_{\rm in}=0} \sum^{\infty}_{k_{\rm out}=0}  p_{{\rm deg}}(k_{\rm in}, k_{\rm out}) k_{\rm in}k_{\rm out}}{\tilde{c}} >1 ,
\end{equation}
and  
\begin{equation}
s_{\rm sc}=0 \quad {\rm if} \quad
\frac{\sum^{\infty}_{k_{\rm in}=0}\sum^{\infty}_{k_{\rm out}=0}   p_{{\rm deg}}(k_{\rm in}, k_{\rm out}) k_{\rm in}k_{\rm out}}{\tilde{c}} <1. 
\end{equation} 
Hence, the condition 
\begin{equation}
\frac{\sum^{\infty}_{k_{\rm in}=0}\sum^{\infty}_{k_{\rm out}=0}   p_{{\rm deg}}(k_{\rm in}, k_{\rm out}) k_{\rm in}k_{\rm out}}{\tilde{c}} =1  \label{eq:critCondSC}
\end{equation}  
determines the percolation transition of the strongly connected component in directed, random graphs.    

\subsection{Random graphs with degree distribution given by Eq.~(\ref{eq:p_k_linearcomb}) }

\subsubsection{Nondirected} 
 Random graphs with the prescribed degree distribution given by Eq.~(\ref{eq:p_k_linearcomb}) percolate when 
\begin{equation}
a =  c - 1.  \label{eq:aLargest}
\end{equation}  
In the special case of $a=0$, corresponding to  the Erd\H{o}s-R\'{e}nyi ensemble with the Poisson degree distribution, 
the percolation transition takes place at 
$c=1$.    In Panel (a) of Fig.~\ref{fig:leadingAPPG2}, we plot for the relative order $f$ of the largest connected component  as a function of the variance ${\rm var}(k)$ of the degree distribution  for random graphs with a prescribed degree distribution given by Eq.~(\ref{eq:p_k_linearcomb}).
We obtain that for infinitely large graphs $f$ decreases monotonically as a function of ${\rm var}(k)$, even for $c=2$.

\subsubsection{Directed}
We consider a  random, directed graph with a prescribed degree distribution $p_{\rm deg}(k_{\rm in},k_{\rm out})$ obtained by plugging  Eq.~(\ref{eq:p_k_linearcomb}) in Eq.~(\ref{eq:degreeDistriJoint}).   Such a random graph can be constructed by adding unidirectional links on a nondirected graph with degree distribution $p_{\rm deg}(k)$, as we discussed in Sec.~II A.  
   Using Eqs.~(\ref{eq:ctilde}) and (\ref{eq:kinkout}), we find that  critical condition (\ref{eq:critCondSC})  reads 
\begin{equation}
\frac{1}{2c} \sum^{\infty}_{k=0} k (k-1)  p_{\rm deg}(k)  =1.
\end{equation}   
If $p_{\rm deg}(k)$ is given by Eq.~(\ref{eq:p_k_linearcomb}), then the strongly connected component of the graph percolates when 
\begin{equation}
a=c-2.
\end{equation}  
In the special case of an Erd\H{o}s-R\'{e}nyi ensemble ($a=0$)   with Poisson degree distribution  the strongly connected component percolates at   $c =2$.
In Panel (b) of Fig.~\ref{fig:leadingAPPG2}, we plot the relative order $s_{\rm sc}$ as a function of ${\rm var}(k)$ for  directed graphs $p_{\rm deg}(k_{\rm in},k_{\rm out})$ given by Eq.~(\ref{eq:degreeDistriJoint}).  We find that $s_{\rm sc}$ increases as a function of ${\rm var}(k)$ for $c=3$ and decreases as a function of ${\rm var}(k)$ for $c=4$.

\section{Eigenvalues of the adjacency matrices of tree graphs with predator-prey interactions}\label{App:last} 
Consider a square matrix $\mathbf{A}$ of dimensions $n\times n$ that has the following two properties: 
\begin{itemize}
\item  the graph represented by  $\mathbf{A}$, in the sense that there exists a nondirected edge between two nodes $i$ and $j$ when either $A_{ij}\neq 0$ or $A_{ji}\neq 0$,  is a tree;
\item it is antagonistic, i.e., $A_{jk}A_{kj}<0$ for all pairs $j,k\in \left\{1,2,\ldots,N\right\}$ with $j\neq k$.  
\end{itemize}

We call such matrices the adjacency matrices of  antagonistic, tree graphs.     We make the following claim: if $\lambda_j$ is an eigenvalue of $\mathbf{A}$, then  ${\rm Re}[\lambda_j] = 0$.   In other words, all eigenvalues of the adjacency matrices of  antagonistic, tree graphs lie on the imaginary axis.

First, we have  verified this numerically for a large number of examples by numerically diagonalising antagonistic tree matrices.     Second,  we present  mathematical evidence in favor of this claim.    We will show that for  antagonistic tree matrices it holds that  
\begin{equation}
{\rm sign}\left(\sum^N_{j=1}\lambda^{2\ell}_j\right) =  (-1)^{\ell} \label{eq:sign2}
\end{equation} 
and 
\begin{equation}
\sum^N_{j=1}\lambda^{2\ell+1}_j  =  0 \label{eq:sign3}
\end{equation}   
 for all values $\ell\in\mathbb{N}$.   The most direct explanation of these two relations is that  ${\rm Re}[\lambda_j] = 0$ for all eigenvalues $\lambda_j$.  Eq. (\ref{eq:sign2}) is then a direct consequence of ${\rm i}^{2\ell} = (-1)^{\ell}$ and Eq.~(\ref{eq:sign3}) is a direct consequence of the fact that if $\lambda_j$ is an eigenvalue, then also its complex conjugate $\lambda^\ast_j$ is an eigenvalue.

Let us derive the Eqs.~(\ref{eq:sign2}) and (\ref{eq:sign3}).  First, it holds that 
\begin{equation}
{\rm Tr}[\mathbf{A}^{\ell}] = \sum^N_{j=1}\lambda^{\ell}_j. \label{eq:lambdal}
\end{equation}
This is a direct consequence of the fact that the diagonal elements of the Jordan canonical form of $\mathbf{A}^\ell$ are the eigenvalues $\lambda^{\ell}_j$.  Second, it holds that
\begin{equation}
{\rm Tr}[\mathbf{A}^{\ell}]  = \sum_{ (i_1,i_2,\ldots,i_\ell)\in\mathcal{P}_\ell(\mathbf{A})}A_{i_1i_2}A_{i_2i_3} \ldots A_{i_{\ell}i_1}
\end{equation} 
where the sum runs over the set $\mathcal{P}_\ell$ of all closed paths  of length $\ell$ in the graph.    Since the graph is a  tree, it holds that 
\begin{equation}
{\rm Tr}[\mathbf{A}^{2\ell+1}]  = \ 0
\end{equation} 
for all natural numbers $\ell$.  On the other hand, since the graph is an antagonstic tree, it holds that 
\begin{equation}
{\rm sign}\left({\rm Tr}[\mathbf{A}^{2 \ell}] \right) =  (-1)^{\ell}. \label{eq:sign}
\end{equation} 
Combining equations Eq.~(\ref{eq:lambdal}) and Eq.~(\ref{eq:sign}), we obtain Eqs.~(\ref{eq:sign2}) and (\ref{eq:sign3}).

\section*{References}
\bibliographystyle{ieeetr} % plain, alpha, ieeetr
\bibliography{Biblio_aps}

%\section{Numerical evaluation of the leading eigenvalue}

\end{document}